%% file: apssamp.tex
\documentclass[%
 reprint,
 amsmath,amssymb,
 aps,
 pra,
floatfix,
superscriptaddress]{revtex4-2}

\usepackage{graphicx}
\usepackage{dcolumn}
\usepackage{bm}
\usepackage{booktabs}
\usepackage[english]{babel}
\usepackage{hyperref}
\input{extra_settings} 

\begin{document}

\preprint{APS/123-QED}

\title{Optimising finite-time quantum information engines using Pareto bounds}

\author{Rasmus Hagman}
\affiliation{Department of Microtechnology and Nanoscience (MC2), Chalmers University of Technology, S-412 96 Göteborg, Sweden}
 
\author{Jonas Berx}%
\affiliation{%
 Niels Bohr International Academy, Niels Bohr Institute, University of Copenhagen - Blegdamsvej 17,
2100 Copenhagen, Denmark
}%

\author{Janine Splettstoesser}
\affiliation{Department of Microtechnology and Nanoscience (MC2), Chalmers University of Technology, S-412 96 Göteborg, Sweden}

\author{Henning Kirchberg}
\email[]{henning.kirchberg@chalmers.se}
\affiliation{Department of Microtechnology and Nanoscience (MC2), Chalmers University of Technology, S-412 96 Göteborg, Sweden}

\date{\today}

\begin{abstract}
Information engines harness measurement and feedback to convert energy into useful work. In this study, we investigate the fundamental trade-offs between ergotropic output power, thermodynamic efficiency and information-to-work conversion efficiency in such engines, explicitly accounting for the finite time required for measurement. As a model engine, we consider a two-level quantum system from which work is extracted via a temporarily coupled quantum harmonic oscillator that serves as the measurement device. This quantum device is subsequently read out by a classical apparatus. We compute trade-offs for the performance of the information engine using Pareto optimisation, which has recently been successfully used to optimise performance in engineering and biological physics. Our results offer design principles for future experimental implementations of information engines, such as in nano-mechanical systems and circuit QED platforms.
\end{abstract}

\maketitle


\input{sections/section_intro}

\input{sections/section_model}

\input{sections/section_results}

\input{sections/section_conclusions}

\begin{acknowledgments}
\input{sections/section_acknowledgements}
\end{acknowledgments}
\input{sections/availability}

\appendix
\input{appendices/mutual_information}
\input{appendices/free_energy}
\input{appendices/mutual_information_bound}
\input{appendices/measurement_work}
\input{appendices/probabilities}
\input{appendices/zeno}
\input{appendices/power_vs_time}
\input{appendices/low_meter_temperature}
\input{appendices/Pareto_formalism}
\input{appendices/paretoovertime}


\bibliography{sources}

\end{document}

%% file: extra_settings.tex
\usepackage[capitalise]{cleveref}
\usepackage{lipsum}
\usepackage{glossaries}
\usepackage{tikz}
\usepackage{pgfplots}
\pgfplotsset{compat=1.18}
\usetikzlibrary{arrows.meta, shapes.misc}
\usetikzlibrary{fit,backgrounds,positioning}
\usepackage{siunitx} 
\usepackage{braket}
\glsdisablehyper

\newacronym{ie}{IE}{information engine}
\newacronym{tls}{TLS}{two-level system}
\newacronym{qho}{QHO}{quantum harmonic oscillator}
\newacronym{cop}{COP}{coefficient of performance}

\newcommand{\tr}[2][]{\operatorname{tr}_{\text{#1}}\left\{#2\right\}}
\newcommand{\dm}[1][]{\hat{\rho}_{\text{#1}}}
\newcommand{\Ham}{\hat{H}}
\newcommand{\HamS}{\hat{H}_{\text{S}}}
\newcommand{\HamM}{\hat{H}_{\text{M}}}
\newcommand{\HamI}{\hat{V}_{\text{I}}}
\newcommand{\Wmeas}{W_{\text{meas}}}
\newcommand{\Wext}{W_{\text{ext}}}
\newcommand{\Wnet}{W_{\text{net}}}
\newcommand{\tm}{t_{\text{m}}}
\newcommand{\kb}{k_{\text{B}}}
\newcommand{\Temp}[1][]{T_{\text{#1}}}
\newcommand{\Tfrac}{\frac{\Temp[M]}{\Temp[S]}}
\newcommand{\geff}{g_\text{eff}}

%% file: sections/section_intro.tex
\section{\label{sec:introduction}Introduction}

Traditional heat engines convert thermal energy into mechanical work by operating between two heat reservoirs at different temperatures. In recent years, heat engines have been analysed in quantum and nanoscale systems, both theoretically and experimentally~\cite{Benenti2017Jun,QTDBook2018,Kirchberg2022,Kirchberg2023,Campbell2025Apr}. Information engines produce work not only by using heat as a resource, but also by exploiting information and feedback~\cite{Maruyama2009Jan,Sagawa2010,Mandal2013}. In contrast to standard heat engines, they can even extract energy from a single heat reservoir by processing information. It is the fundamental link between information and thermodynamics, exemplified by Maxwell's demon, that makes information engines possible~\cite{Szilard1929,Bennett1982,junior2025}. The possibility to exploit information to convert energy to work has been studied theoretically~\cite{bresque2021,Fadler2023,Sanchez2019} and it has been successfully demonstrated in  experiments in the classical and quantum regime~\cite{Raizen2009,Toyabe2010,Koski2014,Pruchyathamkorn2024}. To analyse the performance of such an information engine, information has to be quantified as a resource, which is crucial to determine the efficiency of the engine~\cite{Paneru2018Jan,Monsel2025Jan}. Even fluctuation relations including information~\cite{Sagawa2010,Sagawa2012Nov,Potts2018Nov} and the precision of information engines~\cite{Paneru2020Feb,Monsel2025Jan} have been analysed.

A fundamental question arises regarding the duration of a measurement and, consequently, how quickly information can be acquired to perform feedback on the system to extract work~\cite{busshard2010,bresque2021,Kirchberg2025}. 
This timescale may be particularly important in quantum information engines. Due to their small size, the \textit{operation} time can be very short, making the measurement duration comparable to the engine's cycle time.

Consequently, the duration of the measurement becomes a key factor in the overall performance of the engine. In particular, the energy cost arising from the system-meter coupling but also the potential extracted work through measurement depend on the measurement time, which thus influences, \textit{inter alia}, the efficiency of converting information into work and the operational power (i.e., the work extracted per measurement time).

This raises an important question when operating an information engine: What are the constraints on measurement time and how does it impact performance?

\begin{figure}[bt]
    \centering
    \input{figures/model_diagram}
    \caption{Schematic of the information engine, showing a concrete  model in which the \textit{system} is a two-level system (\gls{tls}) and the \textit{meter} is a quantum harmonic oscillator \gls{qho}. Coupling them via the time-dependent interaction $\hat{V}_{\rm I}$ costs the measurement work $W_\mathrm{meas}$ and goes along with information transfer $I$. In the $j$-th cycle, the \gls{qho} is projectively measured in its energy eigenstate $n_j$ with a classical measurement device with a corresponding entropy flow $\mathcal{S}$. The cycle is used to extract energy $\Wext$. After each cycle the \gls{tls} and the \gls{qho} are coupled to thermal baths at temperatures $\Temp[S]$ and $\Temp[M]$.} 
    \label{fig:ie_model}
\end{figure}
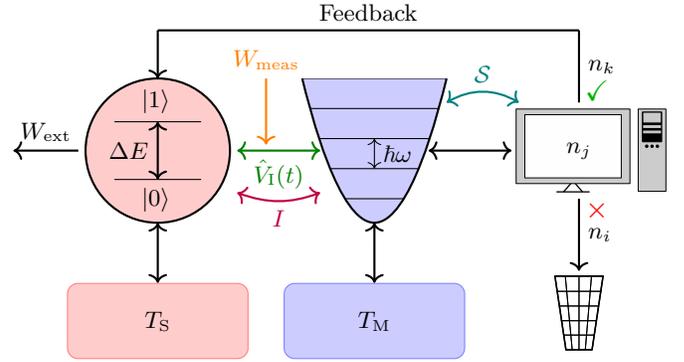

In this paper, we address this challenge and treat the measurement time as a crucial parameter, as it intrinsically links information acquisition, energy cost and extractable work. We thereby extend a previous study~\cite{Kirchberg2025} and complement earlier work on finite-time cyclic information engines that consider bounds on their performance metrics with respect to the dissipation time into the heat reservoir~\cite{Fadler2023}. 
Specifically, we present a comprehensive thermodynamic analysis of a quantum information engine, such as 
illustrated in Fig.~\ref{fig:ie_model}, where we examine the time-dependent coupling between the quantum system and the quantum meter, focusing on their joint evolution. By placing the Heisenberg cut between the quantum meter and the classical apparatus, the study preserves quantum coherent evolution during measurement and enables analysis of measurement time and its associated energy costs. This is in line with  von Neumann’s framework~\cite{VonNeumannBook}, according to which a quantum measurement begins with a "premeasurement," which involves an interaction between the target system and a secondary system known as the ``quantum meter." The quantum meter is then collapsed by a classical measurement apparatus, rendering the measurement outcome objective~\cite{Korbicz2017}. 

For this analysis, we here introduce the experimentally relevant system-meter configuration shown in Fig.~\ref{fig:ie_model}: a two-level system measured by a quantum harmonic oscillator. This setup is directly relevant to potential experimental platforms for quantum information engines such as
nitrogen-vacancy centres, where spin degrees of freedom can be coupled and decoupled to a nano-mechanical oscillator by switching an external magnetic field on (coupling) and off (decoupling)~\cite{Arcizet2011}. Another example arises in quantum computing, where the state of a qubit is read out by coupling it to a microwave resonator~\cite{Blais2004,Cottet2018}.

A key feature of our study is the identification of distinct thermodynamic regimes of operation. Depending on the measurement duration, the engine can function either as a heat engine—producing net work—or as a heat valve—transferring heat between reservoirs through work input. 

To determine optimal engine configurations, a set of selected performance metrics must be optimised simultaneously, i.e., achieving both high power and high efficiency in converting information and energy into work via measurement.
This is a challenging task, as all performance metrics are interdependent through the underlying engine parameters; achieving maximum power and maximum efficiency simultaneously is not possible. This interdependence gives rise to fundamental trade-offs—known as Pareto fronts—between the metrics~\cite{Emmerich2018}.
Pareto optimality is well-established in economics~\cite{Fudenberg1991-jb} and engineering~\cite{Tomoiaga2013-bi}, and has recently been applied in fields such as stochastic thermodynamics~\cite{berx_2024_2,berx2024,forao2025} and biological physics~\cite{Sheftel2013,Berx_2025}. However, to our knowledge, it has not yet been employed in the optimisation of quantum information engines.
In this paper, we apply multiobjective optimisation to explore these Pareto trade-offs in the context of information engines for the first time. We anticipate that this approach will provide practical guidance for jointly tuning experimental parameters to realise information engines operating in desired regimes.

The paper is organized as follows. In Sec.~\ref{sec:method}, we introduce the general model for the information engine and describe its operational cycle. In Sec.~\ref{sec:method_example}, we illustrate the implementation of the engine using a two-level system coupled to a harmonic oscillator, where time-dependent information acquired through measurement is subsequently converted into work. Sec.~\ref{sec:results} presents a detailed performance analysis, including the efficiencies of information-to-work and energy-to-work conversion, as well as the power output, all as functions of the measurement time. In \cref{sec:pareto} we analyse the Pareto-optimal trade-offs between the performance metrics—such as achieving maximum power at maximum efficiency—and determine optimal combinations of system-meter parameters and measurement duration. Finally, Sec.~\ref{sec:conclusion} summarizes our main results, outlines guidelines for experimental setups and lists potential research avenues for extending our Pareto optimisation in information-driven processes.

%% file: figures/model_diagram.tex
\begin{tikzpicture}[scale=0.8]
    \pgfmathsetmacro{\circrad}{1.2} 
    \pgfmathsetmacro{\xqho}{3*\circrad}  
    \pgfmathsetmacro{\xproj}{2*\xqho} 
    \pgfmathsetmacro{\qhospacing}{0.5} 
    \pgfmathsetmacro{\qholevels}{5} 

    \fill[red!20, thick] (0,0) circle (\circrad);
    \draw[thick] (0,0) circle (\circrad);
    \draw (-0.6*\circrad,0.4*\circrad) -- (0.6*\circrad,0.4*\circrad) node[above,xshift=-0.5*\circrad cm] {$\ket{1}$};
    \draw (-0.6*\circrad,-0.4*\circrad) -- (0.6*\circrad,-0.4*\circrad) node[below,xshift=-0.5*\circrad cm] {$\ket{0}$};
    \draw[<->, thick] (0,0.4*\circrad) -- node[left] {$\Delta E$} (0,-0.4*\circrad);
    \node[draw=red!50, fill=red!20, rounded corners, below=0.8cm, minimum width=2*\circrad cm, minimum height=1cm, align=center] (TLSbox) at (0,-1.2) {$\Temp[S]$};
    \draw[<->, thick] (0,-\circrad) -- (TLSbox.north);

    \draw[->, thick] (-1.1*\circrad,0) -- node[above] {$\Wext$} (-2*\circrad,0);

    \begin{scope}[shift={(\xqho,0)}]
        \begin{scope}[yscale=-1]
            \fill[blue!20] (-\circrad,-\circrad) parabola bend (0,\circrad) (\circrad,-\circrad);
            \begin{scope}
                \clip (-\circrad,-\circrad) parabola bend (0,\circrad) (\circrad,-\circrad);
                \foreach \n in {0,...,\qholevels} {
                    \draw (-\circrad,\n*\qhospacing-\circrad) -- (\circrad,\n*\qhospacing-\circrad);
                }
                \draw[<->] (0.0, 2*\qhospacing-\circrad) -- node[right] {$\hbar\omega$} (0.0, 3*\qhospacing-\circrad);
            \end{scope}
            \draw[thick, black] (-\circrad,-\circrad) parabola bend (0,\circrad) (\circrad,-\circrad);
        \end{scope}
    \end{scope}
    \node[draw=blue!50, fill=blue!20, rounded corners, below=0.8cm, minimum width=2*\circrad cm, minimum height=1cm, align=center] (QHObox) at (\xqho,-1.2) {$\Temp[M]$};
    \draw[<->, thick] (\xqho,-\circrad) -- (QHObox.north);

    \draw[<->, thick, green!50!black] (1.1*\circrad,0) -- node[below] {$\hat{V}_{\rm I}(t)$} (0.75*\xqho,0);
    \draw[->, thick, orange] (0.5*\xqho,\circrad) node[above]{$\Wmeas$}--  (0.5*\xqho,0.1);

    \draw[<->, thick, purple] (1.1*\circrad,-0.7) to[out=-20, in=-160] node[below] {$I$} (0.75*\xqho,-0.7);
    
    \node[inner sep=0pt] at (\xproj,0) {\includegraphics[width=2.0cm]{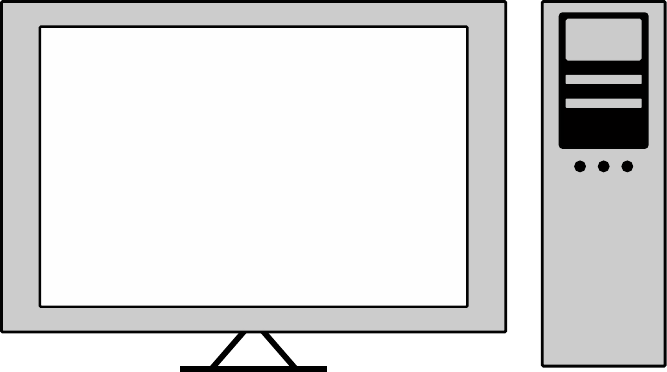}};
    \node[] at (\xproj-0.2, 0) {${n_j}$};
    
    \node[inner sep=0pt] at (\xproj-0.2,-2.7) {\includegraphics[height=1cm]{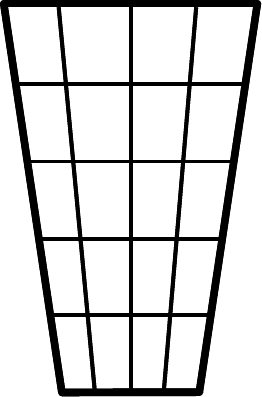}};

    \draw[<->, thick] (1.25*\xqho,0) -- (\xproj-1.1*\circrad,0);

    \draw[<->, thick, blue!50!green] (\xqho+\circrad, 0.8) to[out=30, in=150] node[midway, above] {$\mathcal{S}$} (\xproj-\circrad, 0.8);
    
    \draw[thick] (\xproj-0.2, 2) -- (0, 2) node[above,midway] {Feedback};
    \draw[thick] (\xproj-0.2, 0.8) -- (\xproj-0.2, 2) node[midway,right]{${n_k}$};
    \node[scale=1.2, text=green!70!black] at (\xproj+0.1, 1.0) {\checkmark};
    \draw[thick, ->] (0,2) -- (0,1.2);
    \draw[thick, ->] (\xproj-0.2, -0.8) -- (\xproj-0.2, -2) node[midway,right]{$n_i$};
    \node[scale=1.2, text=red] at (\xproj+0.1, -1.0) {\texttimes};

\end{tikzpicture}

%% file: sections/section_model.tex
\section{\label{sec:method} Information-engine principles}

\subsection{General model}\label{sec:general_model}

The core of the quantum information engine consists of a system, from which work is extracted, and a meter obtaining information about the system; see the sketch in Fig.~\ref{fig:ie_model} for the concrete implementation introduced in Sec~\ref{sec:method_example}. The system and the meter are coupled for a certain time period --- the measurement time $t_\mathrm{m}$~\cite{Kirchberg2025}. 
Due to this coupling, the meter obtains information about the system, which is subsequently used to extract work.
This system-meter setup is captured by the Hamiltonian
\begin{equation}\label{eq:ie_model_general}
    \Ham = \HamS + \HamM + \HamI(t)
\end{equation}
where $\HamS$ is the Hamiltonian of the system, $\HamM$ is the Hamiltonian of the meter, and $\HamI(t)$ describes their time-dependent interaction. This interaction part of the Hamiltonian is finite in a time interval $t\in(0,t_\mathrm{m})$ and zero otherwise.

In time intervals in which system and meter are uncoupled from each other, the system interacts with a heat bath at temperature $\Temp[S]$ while the meter interacts  with a heat bath at temperature $\Temp[M]$. We refer to these two baths as the system bath and the meter bath.
\subsection{Engine cycle}
\label{sub:enging_cycle}

The engine cycle consists of the following steps, see also~\cite{Kirchberg2025} where an analogous cycle was analysed.
\paragraph{Initialization:}
The system (S) and the meter (M) are initially uncoupled and in thermal states
\begin{align}
\label{eq:sys_in}
\hat{\rho}_\mathrm{S}(0) = \frac{\exp (-\HamS/\kb\Temp[S])}{\tr{\exp (-\HamS/\kb\Temp[S])}}\ , 
\\ 
\hat{\rho}_\mathrm{M} (0) = \frac{\exp (-\HamM/\kb\Temp[M])}{\tr{\exp (-\HamM/\kb\Temp[M])}}\ ,
\end{align}
specified by the temperatures $\Temp[S]$ and $\Temp[M]$ of the respective baths they are initially coupled to. Here, $\kb$ is the Boltzmann constant. The total initial state is hence captured by $\dm(0) = \dm[S](0)\otimes\dm[M](0)$.

\paragraph{Unitary evolution:} 
Next, the system and meter are decoupled from their respective baths. Then, for the period of time $0 < t < \tm$, which we refer to as the \textit{measurement time}, the system and meter are coupled to each other. This coupling is reflected by $\HamI(t)$ being non-zero for times $0 < t < \tm$. Importantly, since $[\HamI(t), \HamM] \neq 0$,  the states of the system and of the meter become correlated during the time period $\tm$.
Assuming instant switching of the coupling Hamiltonian at times $t=0$ (``on") and at $t=\tm$ (``off"), the measurement work, namely the work spent to correlate the system and meter states, is
\begin{equation}\label{eq:measurement_work_definition}
    \Wmeas(\tm) \equiv \tr{ [\HamS+\HamM]\big(\dm(\tm)-\dm(0) \big) },
\end{equation}
where $\dm(t) = \hat{U}(t) \dm(0)\hat{U}^\dagger(t)$ with $\hat{U}(t) = \mathcal{T}\exp{\left[-\frac{i}{\hbar}\int_0^tdt' \Ham(t')\right]}$, while $\mathcal{T}$ being the time-ordering operator and $\hbar$ being the reduced Planck constant. Note that the Hamiltonian entering the expression in Eq.~\eqref{eq:measurement_work_definition} is the \textit{total} Hamiltonian at times $t=0$ and $t=t_\mathrm{m}$, where the system-meter coupling is switched off, $\hat{V}_{\rm I}=0$. 

\paragraph{Projective measurement \& information acquisition:}
At time $\tm$ the coupling between system and meter is turned off.
The state of the meter is determined by a projective measurement on the eigenstates $\ket{n}$ of $\HamM$ by a classical meter~\cite{Taranto2023}. This quantum-to-classical transition is the Heisenberg cut~\cite{Heisenberg1949,VonNeumannBook,Atmanspacher01091997}.  After the projective measurement we can formulate the conditional density matrix 
\begin{equation}\label{eq:conditional_density_matrix}
    \hat{\rho}_{\rm S}(\tm|n) = \frac{\braket{n|\dm(\tm)|n}}{\tr[S]{\braket{n|\dm(\tm)|n}}} \equiv \frac{\braket{n|\dm(\tm)|n}}{P(n,\tm)},
\end{equation}
which depends on the measurement outcome $n$.
The conditional probability of finding the system in state $\ket{i}$, given the measurement outcome $n$, is given by the diagonal elements of $\hat{\rho}(\tm|n)$:
\begin{equation}\label{eq:condProb}
    P(i|n,\tm) = \braket{i|\dm(\tm|n)|i}.
\end{equation}
Similarly, the joint probability of the meter being in state $\ket{n}$ and the system being in state $\ket{i}$ at time $\tm$ is given by 
\begin{equation}\label{eq:joint_probs}
    P(i,n,\tm) = \braket{i| \braket{n|\dm(\tm)|n}| i}. 
\end{equation}

Starting from Eq.~\eqref{eq:conditional_density_matrix}, the conditional entropy $S_n(t)$ given the measurement outcome $n$ reads
\begin{equation}
    S_n(t) = -\kb \sum_n \hat{\rho}(\tm|n) \ln \hat{\rho}(\tm|n).
\end{equation}
Averaging over engine cycles, the entropy $S$ is 
\begin{equation}
\label{eq:entropy}
    S(t) = \sum_n P(n,t) S_n(t),
\end{equation}
with $P(n,t)$ as defined in~\cref{eq:conditional_density_matrix}.

Thus, we quantify the information gain, $I(\tm)$, during the measurement process by the entropy difference \cite{Kirchberg2025}, see discussion in \cref{app:mutual_info},
\begin{equation}\label{eq:mutual_info}
    I(\tm) \equiv  S(0) - S(\tm).
\end{equation}

\paragraph{Work extraction:}
Given the measurement outcome in step \textit{c}, useful work can be extracted from the system, specifically when the system is more likely to be found in the excited state. Conversely, no work is extracted if the system is more likely to be found in its ground state, as attempting to do so would lead to a net energy cost rather than a gain, see later discussion in \cref{sec:method_example} (d). 
We determine the extractable work by ergotropy, the maximal work that can be extracted under unitary transformation~\cite{Allahverdyan2004MaximalSystems,Francica2017DaemonicCorrelations} 
\begin{multline}\label{eq:demonic_ergotropy}
    \Wext (\tm|n) = \tr{\dm[S](\tm|n)\HamS } \\
    - \min_{\hat{U}_n} \tr{ \hat{U}_n\dm[S](\tm|n)\hat{U}_n^\dagger\HamS }.
\end{multline}
The first term on the right-hand side of~\cref{eq:demonic_ergotropy} is the energy of the system after measuring the outcome $n$, whereas the second term is the energy of the system after applying a unitary transformation $\hat{U}_n$. Preferably the unitary transformation, $\hat{U}_n$, makes the second term vanishingly small to yield a maximal $\Wext$. In the example case of a two-level system, as discussed later in Sec.~\ref{sec:method_example}, this unitary work extraction can be realized by stimulated emission by a $\pi$-pulse~\cite{Elouard2017}.
Calculating the ensemble average over many engine cycles yields for the maximally extractable work
\begin{multline}\label{eq:work_extraction}
    \Wext (\tm) = \sum_n P(n,\tm) \big[\tr{\dm[S](\tm|n)\HamS } \\
    - \min_{\hat{U}_n} \tr{ \hat{U}_n\dm[S](\tm|n)\hat{U}_n^\dagger\HamS }\big].
\end{multline}

\begin{figure}[!!!!!htbp]
    \centering
    \input{figures/schematic}
    \caption{Energy flow diagram for the information engine operating between two heat baths at temperatures $\Temp[S],$ and $\Temp[M]$ respectively. The engine is bipartite and consists of the system S from which energy $\Wext$ can be extracted upon measurement and information acquisition by a meter M with energetic cost $\Wmeas$. The extracted work is compensated by the heat $\mathcal{Q}_{\rm S}$ from the system's thermal bath while all energy invested for the measurement is released as heat $\mathcal{Q}_{\rm M}$ into the meter bath. }
    \label{fig:info_engine_schematic}
\end{figure}
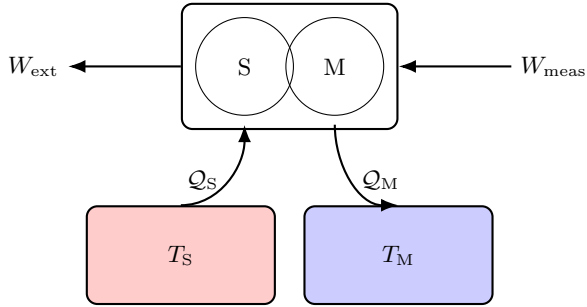

\paragraph{Resetting:}
The system and the meter are coupled to their respective baths, allowing both to rethermalise, thus closing the information-engine cycle. 
We now consider the energy flow during the information-engine cycle as depicted in~\cref{fig:info_engine_schematic}. 
Upon successful extraction of work $\Wext$, Eq.~\eqref{eq:work_extraction}, from the system a given amount of heat $\mathcal{Q}_{\rm S}$ is released into the system to restore it to its initial thermal state, as given in Eq.~\eqref{eq:sys_in}. For the latter process, we assume that all heat from the system bath can be extracted as work $\Wext\equiv \mathcal{Q}_{\rm S}$. The energy invested for the measurement $\Wmeas$, Eq.~\eqref{eq:measurement_work_definition}, (due to coupling and decoupling of system and meter as discussed above) flows as heat into the meter bath, $\Wmeas \equiv \mathcal{Q}_{\rm M}$. 

In principle, we should also consider the cost of resetting the classical meter, as alluded to in Sec.~\ref{sec:introduction}, where the state of the quantum meter is registered. This cost is captured by the Landauer erasure work $W_{\rm reset}=T_{\rm reset} \mathcal{S}$ with $\mathcal{S}$ being the entropy transferred between the quantum meter and the classical meter~\cite{Landauer1961}. 
However, as we describe the projective measurement step with a classical meter, its temperature $T_{\rm reset}$ can be chosen to be arbitrarily small such that we can neglect the erasure work $W_{\rm reset}$~\cite{Elouard2017,Jordan2020,Taranto2023}. 

Next, we consider the \textit{cycle time}, which is the sum of times associated with measurement, projection, work extraction, and resetting. In this study, as alluded to in the introduction, we identify the measurement time $\tm$ under step \textit{b}, the time period for system-meter coupling, as the largest and therefore dominant time scale for the cycle. In order to motivate this, we examine other timescales related to different steps discussed above. 

First, we estimate the time for a projective measurement as described in step \textit{c}. In principle, the quantum meter needs to be coupled and eventually decoupled from another quantum meter and so on, forming an infinite chain of meters, a well-known issue in measurement theory~\cite{MenskyBook}. 
In our model, we use a classical meter to read out the quantum meter where the time can be estimated based on a decoherence mechanism~\cite{Zeh1970}. This related decoherence time can be significantly shorter than $\tm$ due to, for example, strong coupling or high complexity of the classical meter subspace as discussed in~\cite{Taranto2023, Shettell2023}. 

Secondly, the time required for work extraction in step \textit{e} can be estimated, e.g., for a two-level system with energy splitting $\Delta E$ by the duration of stimulated emission under the influence of a $\pi$-pulse, which is on the order of the period time of the system and could be as short as $\hbar/\Delta E$ (ranging from $10^{-12}$ to $10^{-15}$ seconds in molecular systems~\cite{NitzanBook} and $10^{-9}$ to $10^{-10}$ seconds for transmon qubits \cite{Blais2004}). This process can be considered faster than $\tm$. In particular, in the operation of the information engine to produce net useful work, we find for the prototype engine studied in~\cref{sec:method_example} that the system energy is larger than the energy, $\hbar \omega$, associated with the quantum meter. Therefore, we have $(\Delta E/\hbar) \tm > \omega \tm$ which motivates the discussions of the measurement time in terms of $\omega \tm $ in~\cref{sec:results,sec:pareto}.

Thirdly, considering the finite sizes of the system and meter, the reset time in step \textit{f} (thermal relaxation) can be considered rapid when connected to heat reservoirs of infinite size~\cite{Taranto2023} (approximately $10^{-12}$ seconds in condensed molecular systems~\cite{NitzanBook}).

After outlining the detailed operational cycle of the quantum information engine, we now discuss possible performance metrics of the engine, with particular attention to the previously mentioned measurement time which dominates the cycle time.

\subsection{Performance metrics}
\label{subsec:PerformanceMetrics}
As measurement time, information acquisition, energy cost for measurement and the extracted work are interrelated, we consider the following three performance metrics: (A) the ratio of extracted work to information acquired from measurement, which we henceforth name information efficiency, (B) the power, and (C) the efficiency of converting the input energy to net extracted work, i.e., the thermodynamic efficiency. 

(A) Measurement yields information, which in turn bounds the amount of work that can be extracted from the system.
We define the information efficiency, i.e., the conversion of information to extracted work by the ratio
\begin{equation}
\label{eq:eta_info}
    \eta_{\rm info} (\tm) \equiv\frac{\Wext(\tm)}{\Temp[S] I(\tm)},
\end{equation}
where $\Wext$ and $I$ are defined by Eqs.~\eqref{eq:work_extraction} and~\eqref{eq:mutual_info}, respectively. 

Since the maximal extractable work is given by the free energy difference after the projective measurement, as shown in~\cref{app:free_energy} and Ref.~\cite{Kirchberg2025},
\begin{align}
\label{Eq:FreeEnergyGeneral}
    \Delta F(\tm)= \tr{ \HamS\big(\dm(\tm)-\dm(0) \big) } + T_{\rm S} I(t_m),
\end{align}
we have $\Wext (\tm) \leq \Delta F(\tm)$. However, for a non-demolishing measurement, i.e., $[\HamS,\HamI(t)]=0$, we find that the first term in Eq.\ \eqref{Eq:FreeEnergyGeneral} vanishes, so that $\Wext (\tm) \leq \Temp[S]I(\tm) $ (see~\cref{app:free_energy,app:extracted_work_bound}) and the information efficiency, Eq.\ \eqref{eq:eta_info}, is bounded by $0\leq\eta_\text{info}\leq 1$.

(B) The maximum power of the engine is given by
\begin{equation}\label{eq:power}
    \Pi(\tm) \equiv \frac{\Wext(\tm) - \Wmeas(\tm)}{\tm} = \frac{\Wnet(\tm)}{\tm}
\end{equation}
with the net extracted work defined by the difference between the extracted work, Eq.~\eqref{eq:work_extraction}, and the measurement cost, Eq.~\eqref{eq:measurement_work_definition}. The measurement time $\tm$ is considered the cycle time of the information engine as discussed in the previous \cref{sub:enging_cycle} and, thus, enters in the denominator of Eq.~\eqref{eq:power}.
We note that equation~\eqref{eq:power} represents an upper bound on the power output. First, the ergotropy which is used to quantify the extracted work $\Wext$ is a bound as it describes the maximum work extracted under unitary transformation, see Eq.~\eqref{eq:work_extraction}. Second, the measurement time $\tm$ sets a lower bound to the total duration of the engine cycle.

(C) Finally, we consider the efficiency of converting the input energy $\mathcal{Q}_{\rm S}$ to net extracted work $\Wnet=\Wext - \Wmeas$. We recognize that the engine can operate in various regimes, serving different thermodynamic purposes, namely as a heat engine if $\Wnet>0$ or as a heat valve/pump if $\Wnet<0$. For the thermodynamic efficiency, we limit our consideration to the heat engine regime, where $\Wnet \geq 0$. In this regime, heat flows from the bath at temperature $\Temp[S]$ into the bath at temperature $\Temp[M]$, while the engine delivers a net positive amount of work. That is, heat flows from hot to cold, and work is extracted.
As such we define the thermodynamic efficiency as
\begin{equation}\label{eq:efficiency}
    \eta_{\rm HE}(\tm) \equiv \frac{\Wext(\tm) - \Wmeas(\tm)}{\mathcal{Q}_{\rm S}(\tm)} = 1 - \frac{\Wmeas(\tm)}{\Wext(\tm)}.
\end{equation}
Note that we take $ \mathcal{Q}_{\rm S}\equiv \Wext$ in Eq.~\eqref{eq:efficiency}, that is, all energy taken from the system heat bath can be extracted as work, as discussed under step \textit{e} in Sec.~\ref{sub:enging_cycle}.
Since $\Wnet \geq 0$ the efficiency, Eq.\ \eqref{eq:efficiency}, is bounded by $0\leq \eta_{\rm HE}(\tm)\leq 1$.

\section{\label{sec:method_example} Two-level system measured by a harmonic oscillator}
To illustrate the general model introduced in Sec.~\ref{sub:enging_cycle}, we consider a quantum two-level system (\gls{tls}) with energy-level spacing $\Delta E$ as the system S and a quantum harmonic oscillator (\gls{qho}) of frequency $\omega$ and mass $\mathcal{M}$ as the meter, as shown in~\cref{fig:ie_model}. The Hamiltonian of the system and the meter are given by
\begin{subequations}
\label{eq:ie_model_specific}
\begin{eqnarray}
    & \HamS = \Delta E \ket{1}\bra{1},\\
    & \HamM = \frac{\mathcal{M}\omega^2}{2}\hat{x}^2 + \frac{1}{2\mathcal{M}}\hat{p}^2,
\end{eqnarray}
where the energy of the $\ket{0}$ state is set to zero. The interaction Hamiltonian $\HamI(t)$ is switched on for the measurement time only, i.e.,
\begin{equation}\label{eq:interaction_hamiltonian}
    \HamI(t) = \begin{cases}
        g \ket{1}\bra{1}\otimes\hat{p},\quad t\in (0,\tm) \\
        0, \quad t \notin (0,\tm).
    \end{cases}
\end{equation}
\end{subequations}

For this specific example system, the generic information-engine cycle described in \cref{sec:general_model} takes the following concrete form:
\paragraph{Initialization:}\label{paragraph:init_specific}
The initial state of the combined system and meter is 
\begin{align}
\label{eq:InitialStateExample}
    \hat{\rho} (0) &= \dm[S](0)\otimes\dm[M](0) \\ \notag   
    &= \big(a\ket{0}\bra{0} + b\ket{1}\bra{1}\big) \otimes \sum_{n=0}^\infty \frac{1}{Z_M}e^{-\beta_M\HamM}\ket{n}\bra{n},
\end{align}
where $a=(1+e^{-\beta_\text{S}\Delta E})^{-1}$ and $b= (1+e^{+\beta_\text{S}\Delta E})^{-1}$ are the initial populations of the ground and excited states of the \gls{tls} and $\beta_\text{S} = (\kb \Temp[S])^{-1}$ is the inverse temperature of the system bath.
The initial meter state in Eq.\ \eqref{eq:InitialStateExample} is thermal, defined by $\beta_\text{M} = (\kb \Temp[M])^{-1}$ and the partition function $Z_\text{M} = \tr{e^{-\beta_\text{M} \HamM}}$. 

\paragraph{Unitary evolution:}
During the time interval $t\in(0,\,\tm)$, the system and the meter evolve unitarily under the Hamiltonian $\Ham(t)=\HamS+\HamM+\HamI(t)$, resulting in an entangled state.
The state of the full system at time $\tm$ is thus given by
\begin{eqnarray}
    \hat{\rho}(\tm) = e^{-\frac{i}{\hbar} \mathcal{T}\int_0^{\tm} dt' \Ham (t')}\dm(0)e^{\frac{i}{\hbar}\mathcal{T}\int_0^{\tm} dt' \Ham (t')}.
\end{eqnarray}

By evaluating the measurement cost,~\cref{eq:measurement_work_definition}, for the case of a sudden switch of the system–meter interaction~\eqref{eq:interaction_hamiltonian}, we find (see \cref{app:msmt_work})
\begin{equation}
    \label{eq:measurement_work_specific}
    \begin{split}
        \Wmeas(\tm) &\equiv \tr{\dm(\tm)\Ham(\tm)} - \tr{\dm(0)\Ham(0)} \\
     &= bg^2_{\text{eff}}\Big(1-\cos(\omega \tm) \Big), 
    \end{split}
\end{equation}
where $g_{\text{eff}}^2 = g^2\mathcal{M}$ is the effective coupling strength.

\paragraph{Projective measurement \& information acquisition:} 
Using~\cref{eq:joint_probs}, the joint probabilities of finding the \gls{tls} in state $\ket{i}$ and the \gls{qho} in state $\ket{n}$ are given by (see \cref{app:probs})
\begin{align} \label{eq:probability0}
    P(0,n,\tm) &= a \left(1-e^{-\beta_\text{M}\hbar\omega}\right)e^{-\beta_\text{M}\hbar\omega n} \\ \label{eq:probability1}
    P(1,n,\tm) &= \\ b\sum_m P_m &\left|\left(\frac{m!}{n!}\right)^{1/2}\alpha^{n-m}e^{-|\alpha|^2/2}L_m^{(n-m)}\left(|\alpha|^2\right)\right|^2 \notag
\end{align}
where $L_m^{(n-m)}(x)$ are the generalised Laguerre polynomials, and 
\begin{equation}
 \alpha=\alpha(\tm) = \frac{\geff}{\sqrt{2\hbar \omega}}\Big(\sin(\omega \tm)-i\Big(\cos(\omega \tm) -1\Big)\Big)   .
\end{equation}
The corresponding conditional probabilities are then given by $P(i|n,\tm) = P(i,n,\tm)/\sum_iP(i,n,\tm)$ according to Eq.\ \eqref{eq:condProb}.

Thus, we can calculate the information acquisition defined in ~\cref{eq:mutual_info}. As we describe a non-demolishing measurement in this example, $[\hat{H}_{\rm S},\hat{V}_{{\rm I}}]=0$, the work extracted from the information engine is bounded by $\Wext(\tm) \leq T_{\rm S} I(\tm)$ as discussed in Sec.\ \ref{subsec:PerformanceMetrics} (B) and~\cref{app:free_energy,app:extracted_work_bound}.
%
\begin{figure}[bt]
    \centering
    \includegraphics[width=\linewidth]{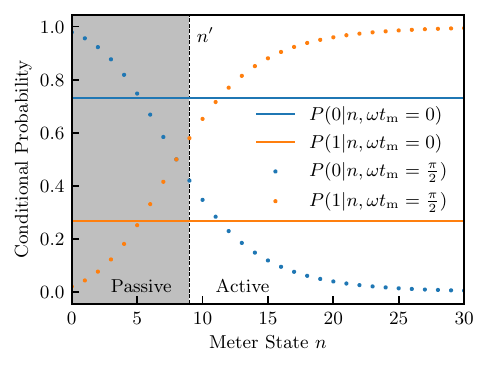}     
    \caption{Conditional probabilities of the two-level system states $P(i|n,\tm=0)$ for $i=0,1$ given a measurement outcome $n$. 
    The solid lines show the conditional probabilities $P(i|n,\tm=0)$ at time $\tm=0$, i.e., when the system and meter are uncorrelated. The dotted lines show $P(i|n,\tm)$ at finite measurement time $\tm$ which depend on the measurement outcome $n$. 
    The dashed vertical line marks the first measurement outcome $n'$ at which $P(1|n,\tm) > P(0|n,\tm)$.
    The chosen parameters are $ \omega \tm = \pi/2 $, $\Delta E =\kb\Temp[S] = g^2_{\text{eff}}$, $\hbar\omega = 0.1\Delta E$, $\Temp[M]/\Temp[S] = 0.3$. }
    \label{fig:cond_probs_v_level}
\end{figure}

\paragraph{Work extraction:}\label{subsec:model_work_extraction}
Given a measurement outcome $n$ of the meter, we consider the work extraction by ergotropy, Eq.\ \eqref{eq:demonic_ergotropy}, i.e., a unitary transformation is performed that maximises the extracted work according to~\cref{eq:demonic_ergotropy}.
For a \gls{tls} this corresponds to population inversion, such that the extracted energy in this case is
\begin{multline}
\label{eq:work_ext_TLS}
    \Wext(\tm|n) = \Delta E
    \Big[ P(1|n,\tm) 
    -P(0|n,\tm) \Big]\times  \\
    \Theta \big(P(1|n,\tm)-P(0|n,\tm)\big),
\end{multline}
where $\Theta(x)$ is the Heaviside function. Averaging over several engine cycles the ergotropy, ~\cref{eq:work_extraction}, reads
\begin{equation}
\label{eq:work_extracted_specific}
    \begin{split}
        \Wext(\tm) &= \sum_nP(n,\tm) \Wext(\tm|n) \\
        &= \Delta E \sum_nP(n,\tm)
        \Big[ P(1|n,\tm) -P(0|n,\tm) \Big] \times \\
        & \Theta \big(P(1|n,\tm)-P(0|n,\tm)\big).
    \end{split}
\end{equation}
To illustrate the behaviour of ergotropy and, thus, of work extraction in this setting, we show in \cref{fig:cond_probs_v_level} the conditional probabilities $P(i|n,\tm)$, as defined in \cref{eq:condProb} and calculated with \cref{eq:probability0,eq:probability1}. Here $i=0,1$ denotes the state of the \gls{tls} and $n$ corresponds to the outcome of a projective measurement on the \gls{qho} at time $\tm$
At the initial time $t = 0$, the \gls{tls} and the \gls{qho} are uncoupled, implying that the conditional probabilities $P(0|n,t=0)=a$ and $P(1|n,t=0)=b$, see Eq.\ \eqref{eq:InitialStateExample}, are independent of the measurement outcomes $n$ as shown by the solid straight lines in \cref{fig:cond_probs_v_level}.
This corresponds to a passive state, characteristic of thermal equilibrium, from which no work can be extracted when averaging over all cycles. 
At a later time $t = \tm$, the interaction between the system and the meter correlates them. 
Consequently, the conditional probabilities $P(i|n,\tm)$ of the TLS depend on the meter outcome $n$. 
For measurement outcomes $n\geq n'$, the conditional probability of finding the \gls{tls} in the excited state exceeds that of the ground state, i.e., $P(1|n,\tm) > P(0|n,\tm)$.
This is a necessary condition for extracting work from the system, i.e., $ \Wext(\tm|n)\neq 0$ in \cref{eq:work_ext_TLS}, by applying a $\pi$-pulse, see the white region in~\cref{fig:cond_probs_v_level}. Averaging over all cycles leads to an overall positive work extraction $\Wext$,~\cref{eq:work_extracted_specific}.

\paragraph{Resetting:}
The \gls{ie} cycle is completed by resetting both the \gls{tls} and the \gls{qho} to their initial states by rethermalisation through interaction with their respective thermal reservoirs.

%% file: figures/schematic.tex
\begin{tikzpicture}[
    every node/.style={font=\small},
    circle style/.style={draw, minimum size=1.75cm, shape=circle, align=center},
    circle style 2/.style={draw, minimum size=1.3cm, shape=circle, align=center},
    arrow style/.style={-{Latex}, thick},
    box style/.style={draw=black, rounded corners, thick, inner sep=5pt},
    box style 2/.style={draw=black, rounded corners, thick, inner sep=5pt, minimum width=2.5cm, minimum height=1.3cm}
]

    \node[circle style 2] (V1) at (2.7,0) {S};
    \node[circle style 2] (V2) at (3.9,0) {M};
    \node[box style, fit=(V1)(V2)] (SM_part) {};
    \node[] (S) at (2.7,-0.66) {};
    \node[] (M) at (3.9,-0.66) {};
    \node[box style 2, fill=red!20] (TS) at (SM_part.south west |- SM_part.south) [below=1cm of SM_part.south west] {$\Temp[S]$}; 
    
    \node[box style 2, fill=blue!20] (TM) at (SM_part.south east |- SM_part.south) [below=1cm of SM_part.south east] {$\Temp[M]$}; 

    \draw[arrow style] (TS.north) to[out=0, in=270] node[pos=0.5, allow upside down, left] {$\mathcal{Q}_\text{S}$} (S);
    \draw[arrow style] (M) to[out=270, in=180] node[pos=0.5, allow upside down, right] {$\mathcal{Q}_\text{M}$} (TM.north);
    
    \draw[arrow style] (SM_part.west) -- ++(-1.5,0) node[left] {$\Wext$};
    
    \draw[arrow style] (SM_part.east) ++(1.5,0) node[right] {$\Wmeas$} -- (SM_part.east);

\end{tikzpicture}

%% file: sections/section_results.tex
\section{\label{sec:results} Performance and measurement time}
In this section, we study the performance metrics introduced in \cref{subsec:PerformanceMetrics} for the \gls{tls} measured by a \gls{qho} to investigate the interrelation between the extractable work, the information acquisition, the measurement time, and the energetic cost for this measurement.

\subsection{Information-to-work conversion}\label{subsec:info_to_work}

\begin{figure}[!htp]
    \centering
    \includegraphics{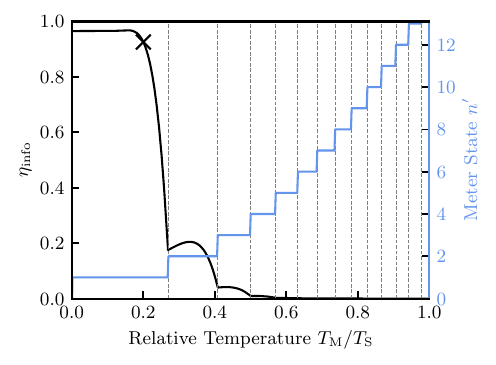}
    \caption{Information efficiency $\eta_\text{info}$ (solid black line),~\cref{eq:eta_info}, on the left vertical axis, and the lowest lying meter level $n'$ (solid blue line) at which the condition $P(1|n,\tm) > P(0|n,\tm)$ is first satisfied on the right vertical axis. Both are plotted as functions of the relative temperature $\Temp[M]/\Temp[S]$. The chosen parameters are $\geff^2/\Delta E= 0.1$, $\Delta E = 4\kb\Temp[S] $, $\omega\tm = \pi/2$, and $\hbar\omega= 1.5\kb\Temp[S]$. The $\times$ denotes a specific operating point, serving as reference in \cref{fig:P_eta_both}(a) and \cref{fig:P_eta_info_both}.}
    \label{fig:info_efficiency_fixed_params}
\end{figure}

The amount of work that can be extracted from the \gls{tls} is bounded by the information gained through measurement by the \gls{qho} (see also~\cref{app:free_energy,app:extracted_work_bound}). 
Consequently, the efficiency of the engine in converting information to work is bounded by unity, $\eta_\text{info}\leq 1$, as shown in \cref{eq:eta_info}. 

\Cref{fig:info_efficiency_fixed_params} shows the information efficiency $\eta_{\rm info}$, \cref{eq:eta_info}, as a function of the relative meter temperature $T_{\rm M}/T_{\rm S}$ for a fixed set of parameters.
At low relative temperatures $\eta_\text{info}$ approaches unity, indicating that nearly all information gained can be converted into work.
This fact can be understood from the temperature-dependence of the population of the meter's states: at low temperatures the ground state---and generally the lowest lying states---have a significantly higher occupation than all states above a certain crossover energy; by contrast, at higher temperatures the meter's state population is more uniformly distributed over a range of relevant meter states. As a result, effects impacting the population distribution become less visible at high temperatures.
Consequently, for low temperatures the meter exhibits an increased sensitivity which allows one to measure the state of the system more accurately.

Importantly, the decrease of $\eta_{\rm info}$ seen in \cref{fig:info_efficiency_fixed_params} with increasing meter temperature is however non-monotonic and not smooth. 
This behaviour can be understood by examining both the numerator and denominator of $\eta_{\rm info}$ independently, given by $\Wext$, \cref{eq:work_extracted_specific}, and $I$, \cref{eq:mutual_info}, respectively.
Consider first $\Wext$. In order to have $\Wext\neq 0$ the condition $P(1|n,\tm)- P(0|n,\tm)\geq 0$ must be fulfilled. Given a fixed meter temperature this condition is met starting from a meter outcome $n\geq n'$ (see vertical dashed line in \cref{fig:cond_probs_v_level} for an example).
With increasing meter temperature $n'$ shifts to larger values in discrete jumps (see consecutive jumps of $n'$ in \cref{fig:info_efficiency_fixed_params}). This explains the sharp kinks in the temperature-dependence of $\eta_{\rm info}$. Moreover, with increasing $T_{\rm M}/T_{\rm S}$ the difference $P(1|n>n',\tm)- P(0|n>n',\tm)$ decreases. Thus, $\Wext$, the numerator of $\eta_\text{info}$, decreases. 

Consider next the information $I$. The latter decreases with increasing $T_{\rm M}/T_{\rm S}$ as less information can be obtained using a hot meter. Immediately following a jump in $n'$, $\Wext$ initially decreases more slowly than the information $I$, resulting in an initial increase in the information efficiency $\eta_\text{info}$.
However, the decline in information is monotonically saturating while the work extraction is continuously decreasing leading to a turnover and eventual decrease in the information efficiency $\eta_{\rm info}$.

\begin{figure}[bt]
    \centering
    \includegraphics{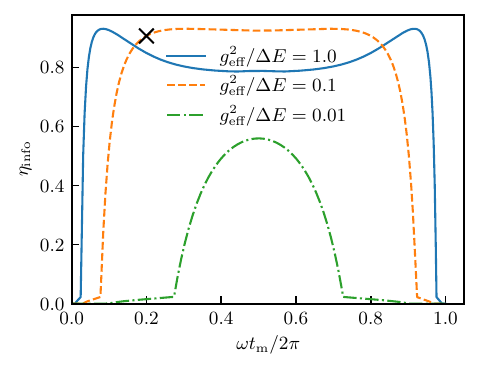}  
    \caption{ Information efficiency, \cref{eq:eta_info}, as a function of time for $\geff^2/\Delta E=0.01,\,0.1,\, 1.0$. The other parameters are $\Temp[M]/\Temp[S] = 0.2$, $\Delta E = 4\kb\Temp[S]$, $\hbar\omega=1.5\kb\Temp[S]$. The $\times$ denotes a specific operating point, serving as reference in \cref{fig:P_eta_both}(a) and \cref{fig:P_eta_info_both}.}
    \label{fig:info_efficiency_time}
\end{figure}

Figures \ref{fig:cond_probs_v_level} and \ref{fig:info_efficiency_fixed_params} display results at a given measurement time $\omega t_\mathrm{m}=\pi/2$. Next, we investigate, for a fixed set of parameters, the information efficiency $\eta_\text{info}$, \cref{eq:eta_info}, as a function of measurement time, shown in  \cref{fig:info_efficiency_time}. 

The information-to-work conversion efficiency, $\eta_\text{info}$, increases with measurement time, reaches a maximum, and then decreases. Note that, given the periodicity of the harmonic oscillator as meter, we observe a symmetry of $\eta_\text{info}$ around $\omega \tm =\pi$.
However, $\eta_\text{info}$ shows a non-smooth evolution as a function of $\tm$. This can be understood from the non-smooth evolution $\Wext$ with $\tm$ entering the numerator of $\eta_\text{info}$. Given the condition $\Wext\geq 0$ we find for short times (i.e., $\geff\tm \ll1$ and $\omega\tm \ll1$) that, see \cref{app:zeno},
\begin{equation}
\label{eq:short_time_kink}
    n' > \frac{\hbar\omega(a-b)}{2\sinh^2\left(\frac{\beta_\text{M}\hbar\omega}{2}\right)(\geff \omega\tm)^2 b}.
\end{equation}
For a given set of system and meter parameters, 
the right hand side of \cref{eq:short_time_kink} continuously decreases with measurement time. 
As $n'$ can only decrease in integer values, $\tm$ must increase until the next integer for $n'$ is found. This discrete change in $n'$ comes with a sudden jump in $\eta_\text{info}$, see \cref{fig:info_efficiency_fixed_params} as discussed above. 

Moreover, the dependence of $\eta_\text{info}$ on measurement time is highly sensitive to the underlying parameters of the information engine—most notably, the system–meter coupling. A stronger coupling leads to a faster increase of the efficiency  $\eta_\text{info}$, occurring at an earlier measurement time, as more information can be transferred between the system and the meter which can be exploited to extract work from the system within a given time interval. 
While $\eta_\text{info}$ peaks at $\tm=\pi/\omega$ for a small system-meter coupling (see $\geff^2/\Delta E=0.01$ in \cref{fig:info_efficiency_time}), the maximum is shifted to smaller values of $\tm$ with increased $\geff^2/\Delta E$. 
Interestingly, for higher system-meter coupling a local minimum in $\eta_\text{info}$ appears. 
Both the information and the related extracted work are increasing from zero with measurement time. With higher system-meter coupling 
the extracted work and the information saturate at higher values. However, the extracted work saturates at a faster rate than the information, leading to an observed local minimum of their ratio described by $\eta_\text{info}$ (see $\geff^2/\Delta E=0.1$ and $\geff^2/\Delta E=1$ in \cref{fig:info_efficiency_time}). 

Overall, the behaviour of $\eta_\text{info}(\tm)$ clearly demonstrates that a finite measurement time within $\omega \tm$ modulo $2\pi $ is essential for converting acquired information into useful work.

\subsection{Power}\label{subsec:power}

The information efficiency discussed above refers to the extracted work, but does not account for the measurement cost. The cost for the measurement impacts, however, the net power,~\cref{eq:power}, namely the \textit{net} work per cycle (or equivalently per measurement) time.

\begin{figure}
    \centering
    \includegraphics[]{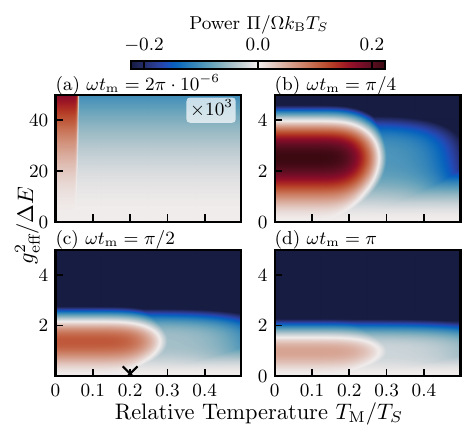}
    \caption{Power output (\cref{eq:power}) as a function of the relative temperature $\Temp[M]/\Temp[S]$ and the effective relative coupling strength $\geff^2/\Delta E$, shown at different measurement times. The fixed parameters are $\hbar\omega = 1.5\kb\Temp[S]$, and $\Delta E = 4 \kb\Temp[S]$. Panels (a)–(d) correspond to $\omega\tm = 2\pi \cdot 10^{-6}$, $\pi/4$, $\pi/2$, and $\pi$, respectively. Red regions indicate positive power output (heat engine regime), while blue regions indicate negative power output (heat valve regime). The data in panel (a) has been multiplied by a factor of $10^3$ to enhance visibility. The $\times$ denotes a specific operating point, serving as reference in \cref{fig:P_eta_both}(a) and \cref{fig:P_eta_info_both}.}
    \label{fig:ergotropy_coupling_heatmap}
\end{figure}

\Cref{fig:ergotropy_coupling_heatmap} shows the power $\Pi$ as a function of the relative meter temperature $\Temp[M]/\Temp[S]$ and the relative system-meter coupling strength $g_{\text{eff}}^2/\Delta E$. We introduce $\Omega=k_B\Temp[S]/\hbar$ as the inverse time scale to make the power unitless.
The power output of the proposed information engine is first evaluated at four different measurement durations as displayed in the heat maps in~\cref{fig:ergotropy_coupling_heatmap}. This parameter manifold reveals distinct regions of positive (red areas in~\cref{fig:ergotropy_coupling_heatmap}) and negative (blue areas) power production. A positive power output requires that the net work satisfies $W_{\rm net} > 0$, as defined in~\cref{eq:power}. Importantly, to achieve a positive power output, a low relative temperature $\Temp[M]/\Temp[S]$ is needed, as can be observed in all panels of~\cref{fig:ergotropy_coupling_heatmap}. As discussed previously, this can be attributed to an enhanced sensitivity of the meter when the difference in state occupations around a given crossover energy is large.
Naturally, in the limit of vanishing system-meter coupling, no power can be extracted anymore, as no information is exchanged between the system and the meter. 

We now consider the impact of the measurement time on the whole parameter manifold in~\cref{fig:ergotropy_coupling_heatmap}.
At very short measurement times ($\omega \tm = 2\pi \cdot 10^{-6}$ in~\cref{fig:ergotropy_coupling_heatmap}(a)), the power output is very small. 
Furthermore, to obtain these small (but finite) power outputs, the relative coupling strength $\geff^2/\Delta E$ needs to be very large, reaching unphysical values for possible information engine setups.
Typical values are $\geff^2/\Delta E \sim 10^{-3}$–$10^{-1}$ for both transmon qubits~\cite{Mallet2009,Walter2017} and optomechanical systems~\cite{Arcizet2011,Pirkkalainen2013}.

As the measurement time increases, see panels \cref{fig:ergotropy_coupling_heatmap}(b)–(c), the magnitude of the power output also increases and the system-meter coupling strength required for positive power output reaches physically reasonable regimes as mentioned before.

However, when further increasing the measurement time, see panel \cref{fig:ergotropy_coupling_heatmap}(d), there is a substantial reduction in the parameter regime for the system-meter coupling yielding positive power. The reason for this is that at longer measurement times, the energetic cost of the measurement process $\Wmeas$, \cref{eq:measurement_work_specific}, which includes the coupling and decoupling of the system and meter, dominates the engine’s performance—ultimately leading to net negative power production.

\begin{figure}[tb]
    \centering
    \includegraphics[]{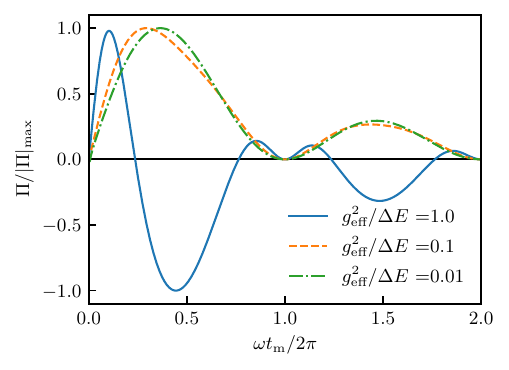}
    \caption{Power output normalised with its maximal absolute value, see \cref{eq:power}, as a function of time for three values of the effective coupling strength, $\geff^2/\Delta E = 0.01,\,0.1,\,1.0$. 
    The chosen parameters are $\Temp[M]/\Temp[S] = 0.1$, $\Delta E = 4\kb\Temp[S]$, $\hbar\omega = 1.5\kb\Temp[S]$.  }
    \label{fig:power_vs_time}
\end{figure}
 
To investigate the measurement-time dependence further, we study next the power output \textit{as a function} of measurement time for a fixed set of parameters, as shown in~\cref{fig:power_vs_time}. As the measurement time increases, more power can initially be extracted from the system as more information about the system is obtained. However, the power $\Pi$ exhibits an oscillatory behaviour, given the quantum harmonic oscillator as meter, reaching multiple local maxima before eventually declining. 
The overall decline of $\Pi$, \cref{eq:power2}, results from the fact that $\tm$ enters in the denominator; the same amount of work is extracted over a longer period of time.
The negative power output emerges for certain measurement times in particular for strong system-meter coupling (solid blue curve in~\cref{fig:power_vs_time}). This results from the substantial contribution of the measurement cost $\Wmeas$,~\cref{eq:measurement_work_specific}, required to couple and decouple the meter with higher values of $\geff^2/\Delta E$.

While a finite measurement time is necessary for obtaining non-zero power output, the optimal duration is highly sensitive to the system and meter parameters, most notably the system-meter coupling strength. All useful work extraction from measurement comes with a finite cost. Both must be optimally tuned by the measurement time to have $\Pi$ positive.
Another illustration of this sensitivity to system and meter parameters is provided in \cref{app:power_vs_time}.

\subsection{Thermodynamic efficiency}\label{subsec:thermo_eff}

\begin{figure}[htp]
    \centering
    \includegraphics[width=\linewidth]{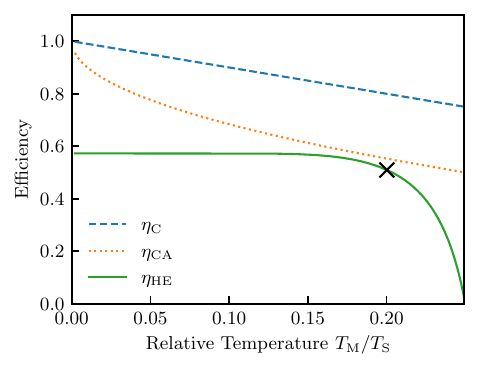}      
    \caption{ Efficiency,~\cref{eq:efficiency}, as a function of relative temperature in the heat engine regime, defined by $\Temp[M] < \Temp[S]$ and $\Wext - \Wmeas > 0$. The Carnot efficiency, $\eta_\mathrm{C}$, and Curzon-Ahlborn efficiency, $\eta_\mathrm{CA}$, are shown for reference. Parameters are fixed as $\Delta E = 4 \kb\Temp[S]$, $\geff^2/\Delta E=0.1 $, $\omega\tm = \pi/2$, and $\hbar\omega = 1.5 \kb\Temp[S]$. The $\times$ denotes a specific operating point, serving as reference in \cref{fig:P_eta_both}(a) and \cref{fig:P_eta_info_both}.}
    \label{fig:efficiency}
\end{figure}

We finally examine the thermodynamic efficiency, $\eta_\text{HE}$, as defined in~\cref{eq:efficiency}, accounting for the net extracted work compared to the absorbed heat. The result is shown in~\cref{fig:efficiency}, where we plot the thermodynamic efficiency as function of the temperature ratio $\Temp[M]/\Temp[S]$, between meter and system. We compare the efficiency to the upper bounds given by Carnot efficiency and the Curzon-Ahlborn efficiency.

Firstly, it is noteworthy that $\eta_\text{HE}$ remains constant over a broad range of $\Temp[M]/\Temp[S]$.
We start by investigating the value of $\eta_\text{HE}$ in the limit of a vanishing meter temperature, $\kb\Temp[M] \ll \hbar\omega$, with the measurement outcome $n'=1$ (as seen in~\cref{fig:info_efficiency_fixed_params}), as threshold to a positive net work extraction, $\Wnet$,~\cref{eq:efficiency} and, thus, $\eta_\text{HE}\geq 0$.
In this limit, we find the thermodynamic efficiency, see~\cref{app:low_temp}, to be
\begin{align}
\label{eq:efficiency_low_temperature}
    \lim_{k_B\Temp[M]/\hbar \omega\to 0}\eta_{\rm HE}  = 1-\frac{ \geff^2(1-\cos{(\omega \tm))}}{\Delta E \bigg[1-e^{-\frac{g_{\rm eff}^2}{\hbar \omega}\big(1-\cos(\omega \tm)\big)}\bigg]}.
\end{align}
Utilising the system and meter parameter indicated in in~\cref{fig:efficiency}, we find $\eta_{\rm HE}=0.57$ in~\cref{eq:efficiency_low_temperature}. The thermodynamic efficiency in ~\cref{eq:efficiency_low_temperature} maximises at $\omega \tm \to 0$ and reads $\eta_{\rm HE}=1-\hbar\omega/\Delta E$ while the power $\Pi(\omega \tm \to 0)$, ~\cref{eq:power}, vanishes. By setting $\Delta E \gg \hbar \omega $, see discussion in~\cref{app:low_temp}, this maximum value can be tuned to unity, $\eta_\text{HE} \to 1$. In particular, in this parameter regime, having a high system temperature, the resulting relative temperature are vanishingly small, $\Temp[M]/\Temp[S] \to 0$, and, $\eta_\text{HE} \equiv  \eta_{\rm CA} \equiv \eta_{\rm C}$. 
 Continuing the discussion for $\eta_\text{HE}$ in~\cref{fig:efficiency}, a decline is observed at $\Temp[M]/\Temp[S] \approx 0.15$, or, when $\kb \Temp[M]$ is within one order of magnitude of $\hbar \omega$. This is attributed to the presence of discrete energy levels $\hbar\omega$ in the meter. When the meter temperature is low, $\kb\Temp[M] < \hbar\omega$, the ground state is more highly occupied, which enhances the change in the meter-state occupations due the system-meter coupling and consequently makes the measurement more precise. With a more precise meter, more information about the system can be obtained to extract its energy as useful work. Therefore, for a fixed measurement time and system temperature, $\eta_\text{HE}$ declines to zero with increasing $\Temp[M]/\Temp[S]$ as the meter becomes less sensitive. 

Secondly, given that the proposed quantum information engine operates between two temperatures, $\Temp[M]\equiv T_\text{cold}$ and $\Temp[S]\equiv T_\text{hot}$, we compare its thermodynamic efficiency with the Carnot efficiency, $\eta_C = 1 - T_\text{cold}/T_\text{hot}$~\cite{carnot1824reflections,Callen1985thermodynamics}, and the Curzon-Ahlborn efficiency, $\eta_{\rm CA} = 1 - \sqrt{T_\text{cold}/T_\text{hot}}$~\cite{Curzon1975,Callen1985thermodynamics}. While the Carnot efficiency represents the theoretical maximum efficiency for heat engines operating in the adiabatic limit of infinite cycle time, the Curzon-Ahlborn efficiency provides a bound for engines operating under finite cycle times and finite power extraction. For the selected parameters, $\eta_\text{HE}$ approaches $\eta_{\rm CA}$, indicating the potential for efficient operation of the proposed quantum information engine at finite measurement time.

\begin{figure}[htp]
    \centering \includegraphics{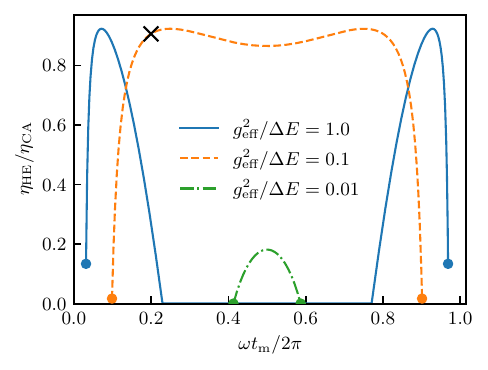}
    \caption{Thermodynamic efficiency, \cref{eq:efficiency}, as a function of time with $\geff^2/\Delta E=0.01,\, 0.1,\,1.0$. The chosen parameters are $\Temp[M]/\Temp[S]=0.2$, $\Delta E = 4\kb\Temp[S]$, $\hbar\omega=1.5\kb\Temp[S]$. The points indicate the times for which $\Wnet<0$ and the information engine is not producing net work. The $\times$ denotes a specific operating point, serving as reference in \cref{fig:P_eta_both}(a) and \cref{fig:P_eta_info_both}.}
    \label{fig:efficiency_time}
\end{figure}

To assess how closely the quantum information engine can approach the Curzon–Ahlborn efficiency, we analyze the ratio $\eta_\text{HE}/\eta_{\rm CA}$ as a function of measurement time in~\cref{fig:efficiency_time}. This ratio increases with measurement time, reaches a maximum, and subsequently decreases. It is important to note that a finite measurement time is needed for the information engine to produce work, i.e., $\Wnet=\Wext-\Wmeas>0$, indicated by the solid points in~\cref{fig:efficiency_time}. This highlights the necessity of a finite, optimised measurement duration for achieving efficient operation in finite-time information engines.
Interestingly, for the chosen set of parameters and a system-meter coupling strength of $\geff^2/\Delta E=0.1$ and $\geff^2/\Delta E=1$, the engine achieves a maximum value of $\eta_\text{HE}/\eta_{\rm CA} \sim 0.95$, indicating that it can operate near the Curzon–Ahlborn efficiency. 
However, for these strong system-meter couplings, the high measurement cost must be accounted for. In particular, for $\geff^2/\Delta E=1$, the measurement cost eventually exceeds the extracted work.
Note that the periodic behaviour of $\eta_\text{HE}/\eta_{\rm CA}$ comes from the harmonic oscillator as meter.

\section{Optimal engine design using Pareto fronts}\label{sec:pareto}

\begin{figure*}[ht!]
    \centering
    \includegraphics[width=0.9\linewidth]{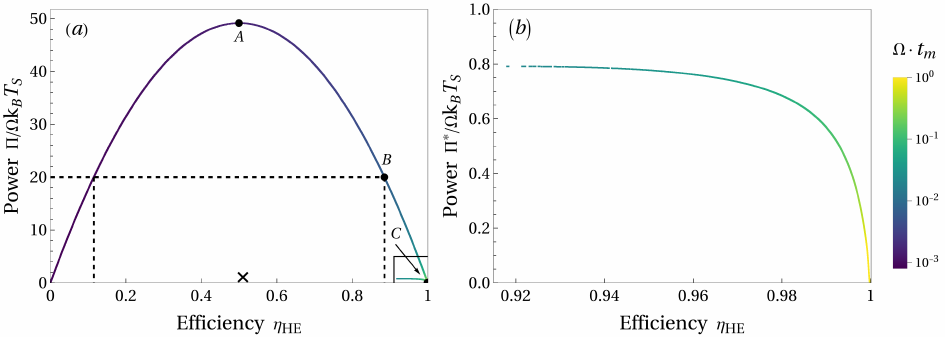}
    \caption{{\bf (a)} Power-efficiency Pareto trade-off for instantaneous work extraction, Eq.~\eqref{eq:power}, with the measurement time $t_m$ shown as a logarithmically scaled colour gradient in units of the characteristic timescale $\Omega^{-1}$. $A$ and $C$ respectively denote the points with highest power and efficiency, while $B$ is the point with highest efficiency for a fixed power given by the horizontal dashed line. The symbol $``\times"$ indicates the suboptimal point associated with the parameters used in Figs~\ref{fig:info_efficiency_fixed_params},~\ref{fig:info_efficiency_time},~\ref{fig:ergotropy_coupling_heatmap},~\ref{fig:efficiency} and~\ref{fig:efficiency_time}, listed in Tbl~\ref{tbl:Pareto_parameters}.{\bf (b)} Power-efficiency Pareto trade-off for $\pi-$pulse work extraction, Eq.~\eqref{eq:power2}, specifically taking into account the time required to extract work via stimulated emission, located in the lower right (boxed) region of panel (a). It shows that the optimal power is significantly lower and the front shape changes drastically. The optimisation parameter ranges are chosen as $\Temp[M]/\Temp[S]\in [10^{-4},1]$, $\Delta E/\kb \Temp[S]\in[10^{-3},10^{2}]$, $\hbar\omega/\kb\Temp[S]\in[10^{-3},10^{2}]$, $\geff^2/\kb\Temp[S]\in[10^{-6},10^{6}]$ and $\omega \tm / 2\pi\in [10^{-3},1]$.}
    \label{fig:P_eta_both}
\end{figure*} 
For our proposed quantum information engine, it is desirable to achieve both a high rate of work extraction—characterised by a large positive power $\Pi$,~\cref{eq:power} —and high efficiency in converting both input energy and information into useful work, respectively given by the thermodynamic efficiency $\eta_\mathrm{HE}$,~\cref{eq:efficiency}, and the information efficiency $\eta_{\rm info}$,~\cref{eq:eta_info}.
However, these performance metrics depend on a shared set of system and meter parameters: the system and meter temperatures $\Temp[S]$ and $\Temp[M]$, system and meter energy level spacings $\Delta E$ and $\hbar \omega$, their coupling strength $\geff^2$ and, of particular interest here, the measurement time $\tm$. Changing the engine's parameters thus influences all performance metrics concurrently and leads to the emergence of trade-offs between them, which need to be optimised in particular for experimental realizations of the proposed engine. 

In this section, we compute a simultaneous optimisation of our metrics by determining Pareto-optimal trade-offs, also called Pareto fronts: a set of engine configurations where improving one metric necessarily compromises at least one other. Standard optimization methods generally either compute only a single optimum e.g., high power, or involve scalarisation of different objectives into a single weighted ``cost function'' which is subsequently minimised~\cite{Erdman2017}. Pareto fronts, however, represent the best possible balance between all metrics simultaneously. While originally developed in the context of economics and engineering, Pareto optimisation has been increasingly used in fields such as biology~\cite{Sheftel2013,berx2024} and stochastic thermodynamics~\cite{forao2025}. 
However, to the best of our knowledge, Pareto optimisation has not previously been applied in the context of quantum information engines.

A more technical description of Pareto-optimal fronts is given in Appendix~\ref{app:Pareto}.
We compute the fronts numerically using a variant of the elitist genetic algorithm NSGA-II~\cite{Deb2008}, which is able to accurately determine non-convex parts of the fronts, in contrast to standard optimization methods. In particular, we will focus on pairwise trade-offs between power, thermodynamic efficiency and information efficiency. 

We consider first the Pareto trade-off between power and thermodynamic efficiency in \cref{fig:P_eta_both}(a). For a given arbitrary power output—e.g., $\Pi = 20 \Omega \kb \Temp[S]$, marked by the horizontal dashed line—the Pareto front defines the range of efficiencies (bounded by vertical dashed lines) over which the quantum information engine can operate. Points on the front with efficiency $\eta_{\rm HE} \leq 0.5$, indicated by point $A$, correspond to the minimum efficiency attainable at a fixed power level, while points with $\eta_{\rm HE} \geq 0.5$, such as point $B$, indicate the maximum achievable efficiency for the same output power. Each point on the Pareto front thus corresponds to a specific set of engine parameters that yield an optimal trade-off between power and efficiency. For example, points $A,\,B,\,C$ in \cref{fig:P_eta_both}(a) illustrate such configurations, with the corresponding parameters listed in Tbl.~\ref{tbl:Pareto_parameters}. 

\begin{table*}[ht]
\centering
\caption{Pareto-optimal engine parameters corresponding to the labelled points $A,\,B,\,C$ and $``\times"$ in Figs.~\ref{fig:P_eta_both} and~\ref{fig:pareto_vs_time}.}
\label{tbl:Pareto_parameters}
\begin{tabular}{@{}l|S[scientific-notation=false,round-mode=figures,round-precision=2]S[scientific-notation=false,round-mode=figures,round-precision=3]|S[scientific-notation=true,round-mode=figures,round-precision=3]S[scientific-notation=true,round-mode=figures,round-precision=3]S[scientific-notation=true,round-mode=figures,round-precision=3]S[scientific-notation=true,round-mode=figures,round-precision=3]S[scientific-notation=true,round-mode=figures,round-precision=3]@{}}
 & $\eta_{\rm HE}$ & $\Pi/\Omega\kb\Temp[S]$ & {$\Temp[M]/\Temp[S]$} & {$\Delta E / \kb \Temp[S]$} & {$\hbar\omega / \kb \Temp[S]$} & {$g_{\text{eff}}^2 / \kb \Temp[S]$} & {$\Omega t_m$}  \\ \midrule
$A$ & 0.499925 & 49.158 & 0.00266406 & 2.21825 & 0.631569 & 40196.1 & 1.58e-3 \\
$B$ & 0.885117 & 20 & 0.00199574 & 2.23663 & 0.145487 & 9357.27 & 0.00687543 \\
$C$ & 1 & 0 & 0.000153107 & 14.4491 & 0.00504849 & 0.00122769 & 1.31 \\
\midrule
$\times$ & 0.5102 & 0.045 & 0.2 & 4 & 1.5 & 0.632456 & 0.166667 \\
\end{tabular}
\end{table*}

Configurations lying below this Pareto front are achievable but suboptimal. One representative suboptimal point below the Pareto front, marked by $``\times"$ in Fig.~\ref{fig:P_eta_both}(a), corresponds to the parameter set we used in the Figs.~\ref{fig:info_efficiency_fixed_params},~\ref{fig:info_efficiency_time},~\ref{fig:ergotropy_coupling_heatmap},~\ref{fig:efficiency}, and~\ref{fig:efficiency_time} of the preceding ~\cref{sec:results}, and Fig.~\ref{fig:power_vs_time_higher_temp} in~\cref{app:power_vs_time}, where this point is also marked with the same symbol. By appropriately tuning the system and meter parameters, this point can be shifted toward the Pareto front, thereby increasing both power output and thermodynamic efficiency. For instance, fixing the efficiency of the point at $\eta_{\rm HE} = 0.51$, as given in Tbl.~\ref{tbl:Pareto_parameters}, the parameters can be judiciously tuned such that the power increases without changing $\eta_{\rm HE}$, approximately up to the point $A$. Conversely, no engine designs above the front are physically possible within our current formulation of the quantum information engine.

\begin{figure}[tp]
    \centering
    \includegraphics[width=0.95\linewidth]{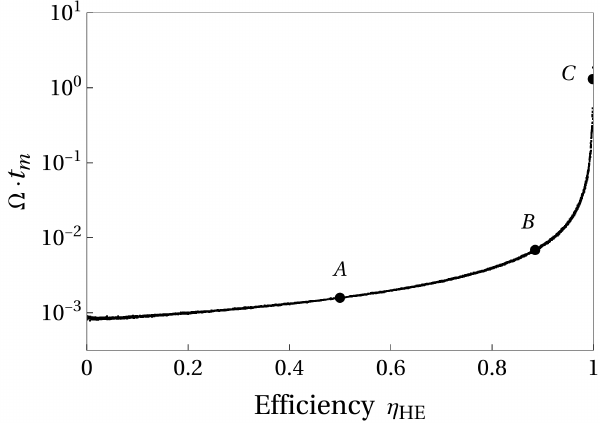}
    \caption{Measurement time along the power-efficiency Pareto front of Fig.~\ref{fig:P_eta_both}(a) plotted against thermodynamic efficiency $\eta_{\rm HE}$. The optimisation parameter ranges are $\Temp[M]/\Temp[S]\in [10^{-4},1]$, $\Delta E/\kb \Temp[S]\in[10^{-3},10^{2}]$, $\hbar\omega/\kb\Temp[S]\in[10^{-3},10^{2}]$, $\geff^2/\kb\Temp[S]\in[10^{-6},10^{6}]$ and $\omega \tm / 2\pi\in [10^{-3},1]$.}
\label{fig:pareto_vs_time}
\end{figure}

Moving along the Pareto front from point $A$, where the power is maximal, to point $C$, where the thermodynamic efficiency is maximal, the engine parameters vary to preserve an optimal combination of maximal power and efficiency. A key parameter in this trade-off is the measurement time, which we highlight using a colour scale in \cref{fig:P_eta_both}(a) and which we investigate in terms of the inverse time scale $\Omega=k_B\Temp[S]/\hbar$. This variation is shown consistently in both \cref{fig:P_eta_both}(a) and in \cref{fig:pareto_vs_time}, where points $A,\,B,\,C$ are marked according to Tbl.~\ref{tbl:Pareto_parameters}. Point $A$ indicates the overall maximum power that is reached at a thermodynamic efficiency $\eta_{\rm HE}\approx 0.5$ where a measurement time $\tm = 1.58\times10^{-3} \Omega^{-1}$ is required. The measurement time increases continuously when going from $A \to C$ along the front.
As expected, when the thermodynamic efficiency of converting input energy to useful work is reaching its global maximal value in point $C$, the maximally achievable power output vanishes. Conversely, when the power output increases the thermodynamic efficiency decreases, i.e., following $C \to A$. This Carnot-like behaviour at $C$, namely maximum achievable efficiency at vanishing output power, can be understood from the fact that, due to the long measurement time and thus the time to complete the cycle of the information engine, the process becomes adiabatically slow. In the present context, a sufficiently long measurement time is required to obtain enough information about the system to transduce all energy into work per cycle.

We consider next the other parameters along the Pareto front. Moving from $A$ to $C$ entails decreasing the relative meter to system temperature, the effective coupling and the oscillator frequency, while simultaneously increasing the TLS energy level spacing and the measurement time, see Tbl.~\ref{tbl:Pareto_parameters}. 
In fact, close to the point $C$, increasing $\Delta E \gg \kb \Temp[S]$ 
leads to a very small population of the excited state of the system which we measure and from where we want to extract work. Similarly, the decrease of $\Temp[M]/\Temp[S]$ leads to meter state occupations that peak sharply in the ground state. The decrease in the effective coupling strength between system and meter leads to a reduced measurement cost, which consequently produces only a very low amount of heat in the meter bath. Together, these combined effects lead to an overall decrease in losses when converting energy to useful work and drive the engine towards an efficiency of $\eta_{\rm HE} = 1$. Note that near point $C$, the Pareto-optimal parameters naturally reproduce the condition $\kb\Temp[M]\ll \hbar\omega$ underlying the approximation in  Eq.~\ref{eq:efficiency_low_temperature}, see~\cref{sec:results} C and~\cref{app:low_temp}, as can be readily confirmed in Tbl.~\ref{tbl:Pareto_parameters}. Additionally, moving toward unit efficiency in Eq.~\ref{eq:efficiency_low_temperature} is also confirmed by the Pareto-optimised parameters, which lead to the required conditions that $\Delta E \gg \hbar \omega$ and $\Temp[M] \ll \Temp[S]$. This agreement highlights that the approximation captures the essential physics of the system in the regime where performance is optimised.

Interestingly, for $\eta_{\rm HE} < 0.5$, descending to the left from point $A$ in Fig.~\ref{fig:P_eta_both}(a), the measurement time $\tm$ continues to decrease along the Pareto front, as shown in Fig.~\ref{fig:pareto_vs_time}. This indicates that a finite, non-zero measurement time is always necessary to balance high power extraction with thermodynamic efficiency.

Note that the power-efficiency Pareto optimisation we discussed intrinsically depends on the exact description of the engine; ``hidden" or neglected processes influencing the performance metrics naturally change the shape of the fronts. One way to model such a neglected process is by explicitly taking into account the time required to extract work via stimulated emission with a $\pi$-pulse, given by $t_{\rm st} = \hbar \pi / \Delta E$, which we previously did not account for. This influences the power that is extracted as follows:
\begin{equation}\label{eq:power2}
    \Pi^*(\tm) = \frac{\Pi(\tm)}{1+t_{\rm st}/\tm}.
\end{equation}

The front obtained by performing the Pareto optimisation of $\Pi^*$ and $\eta_{\rm HE}$ is shown in the bottom right corner of panel (a) in Fig.~\ref{fig:P_eta_both}; an enlarged version is shown in panel (b). As anticipated, the resulting Pareto front of maximised thermodynamic efficiency and modified power $\Pi^*$, Eq.~\eqref{eq:power2}, lies globally below the previously discussed front. However, the newly obtained Pareto front shows the same trend for the measurement time and the system and meter parameters when moving along it.

\begin{figure}[tb]
    \centering
    \includegraphics[width=\linewidth]{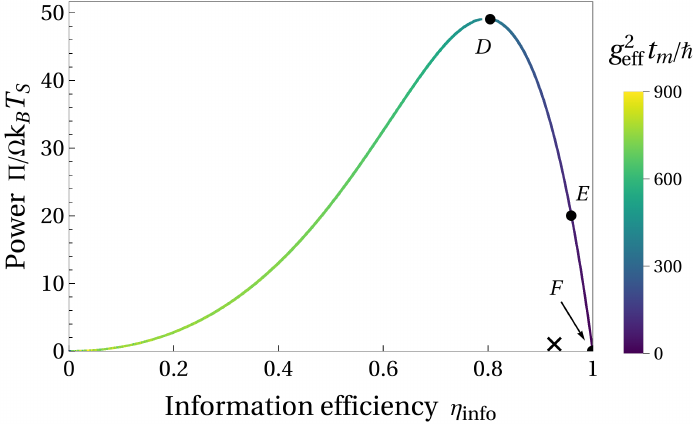}
    \caption{Power-information efficiency Pareto trade-offs for instantaneous work extraction, Eq.~\eqref{eq:power}, with the coupling strength normalized measurement time $g_{\text{eff}}^2 \tm/\hbar$ shown as a colour gradient. $D$ and $F$ respectively denote the points with highest power and information efficiency, while $E$ is the point with highest efficiency for a fixed power $\Pi = 20\Omega\kb\Temp[S]$. The $``\times"$ indicates the suboptimal point associated with the parameters used in Figs~\ref{fig:info_efficiency_fixed_params},~\ref{fig:info_efficiency_time},~\ref{fig:ergotropy_coupling_heatmap},~\ref{fig:efficiency} and~\ref{fig:efficiency_time}, listed in Tbl.~\ref{tbl:Pareto_parameters_info}. The optimisation parameter ranges are $\Temp[M]/\Temp[S]\in [10^{-4},1]$, $\Delta E/\kb \Temp[S]\in[10^{-3},10^{2}]$, $\hbar\omega/\kb\Temp[S]\in[10^{-3},10^{2}]$, $\geff^2/\kb\Temp[S]\in[10^{-6},10^{6}]$ and $\omega \tm / 2\pi\in [10^{-3},1]$.
    \label{fig:P_eta_info_both}}
\end{figure} 

\begin{table*}[ht]
\centering
\caption{Pareto-optimal engine parameters corresponding to the labelled points $D,\,E,\,F$ and $``\times"$ in Fig.~\ref{fig:P_eta_info_both}.}
\label{tbl:Pareto_parameters_info}
\begin{tabular}{@{}l|S[scientific-notation=false,round-mode=figures,round-precision=2]S[scientific-notation=false,round-mode=figures,round-precision=3]|S[scientific-notation=true,round-mode=figures,round-precision=3]S[scientific-notation=true,round-mode=figures,round-precision=3]S[scientific-notation=true,round-mode=figures,round-precision=3]S[scientific-notation=true,round-mode=figures,round-precision=3]S[scientific-notation=true,round-mode=figures,round-precision=3]@{}}
 & $\eta_{\rm info}$ & $\Pi/\Omega\kb\Temp[S]$ & {$\Temp[M]/\Temp[S]$} & {$\Delta E / \kb \Temp[S]$} & {$\hbar\omega / \kb \Temp[S]$} & {$g_{\text{eff}}^2 / \kb \Temp[S]$} & {$\Omega t_m$}  \\ \midrule
$D$ & 0.804465 & 49.058 & 0.0143349 & 2.2971 & 0.64549 & 40613. & 0.00154947 \\
$E$ & 0.959239 & 20 & 0.159166 & 3.1831 & 1.34585 & 18374.7 & 0.000743108 \\
$F$ & 0.998794 & 0 & 0.131394 & 3.55926 & 1.55913 & 645.191 & 0.000645467 \\
\midrule
$\times$ & 0.9266 & 0.045 & 0.2 & 4 & 1.5 & 0.632456 & 0.166667 \\
\end{tabular}
\end{table*}

We study next the trade-off between the power output and information efficiency in Fig.~\ref{fig:P_eta_info_both}. A similar methodology as before can be used to determine the optimal system and meter parameters along the resulting Pareto front. The parameters corresponding to the points $D,\,E,\,F$ and $``\times"$ are given in Tbl.~\ref{tbl:Pareto_parameters_info}, where the point $``\times"$ corresponds once again to the suboptimal set of parameters used in Figs.~\ref{fig:info_efficiency_fixed_params},~\ref{fig:info_efficiency_time},~\ref{fig:ergotropy_coupling_heatmap},~\ref{fig:efficiency},~\ref{fig:efficiency_time} in~\cref{sec:results} and~\ref{fig:power_vs_time_higher_temp} in~\cref{app:power_vs_time}.

In particular, we examine the measurement time weighted by the coupling strength between the system and the meter. 
In fact, $\hbar/ g_{\text{eff}}^2$ can be understood as characteristic transfer time between the system and the meter. In maximising both power output and information efficiency, i.e., following the Pareto front from $D\to F$ in Fig.~\ref{fig:P_eta_info_both}, we observe that $g_{\text{eff}}^2 \tm/\hbar$ approaches zero for $\eta_{\rm info}\to 1$. This configuration allows for the maximal conversion of information into useful work. To be concrete, a small value of $g_{\text{eff}}^2 \tm/\hbar$ implies that a minimal amount of information transferred during $\tm$ can be completely converted into a correspondingly small amount of work. In this regime the power output vanishes. 

As the value of $g_{\text{eff}}^2 \tm/\hbar$ increases, more information can be generated during $\tm$. However, only a portion of this information is converted to work, resulting in a reduced $\eta_{\rm info}$ while producing finite power. Interestingly, $g_{\text{eff}}^2 \tm/\hbar$ continues to increase even when minimising $\eta_{\rm info}$ while maximising $\Pi$, i.e., following the Pareto front $\eta_{\rm info} \to 0$ starting from point $D$ in Fig.~\ref{fig:P_eta_info_both}. 
This underscores the necessity of a finite coupling between the system and the meter to ensure non-zero power output and enable the effective conversion of information into useful work within a finite measurement time $\tm$, before the energetic costs related to large $g_{\text{eff}}$ limit this process.

%% file: sections/section_conclusions.tex
\section{\label{sec:conclusion}Conclusions}

We studied a cyclic information engine in which measurement and subsequent feedback are used to convert energy into useful work. Information is considered as a crucial \textit{resource} that must be exchanged between the system and the meter over a finite time. This \textit{measurement time} can be comparable to the engine's operational time, especially in the nanoscale regime. Importantly, each measurement also incurs an intrinsic cost that depends on the measurement time and must be properly accounted for. In this work, we investigated the interrelation between information acquisition, extractable work, and the energetic cost of measurement — all as functions of the measurement time.

As a prototype, we formulated and analysed a quantum information engine cycle involving a two-level system measured by a quantum harmonic oscillator—a setup inspired by experimental platforms such as nitrogen-vacancy centers or superconducting qubits coupled to resonators. Within this framework, we found that a finite measurement time is essential to transfer sufficient information about the system to the meter to achieve positive power output and high efficiency in the conversion of energy and information into work. We found that lower meter temperatures and stronger system–meter coupling enhance the meter's sensitivity, thereby improving information transfer. While no information can be transferred in the absence of system–meter interaction, excessively strong coupling can become detrimental due to increased measurement costs, ultimately leading—under certain conditions—to zero and even negative net work. This highlights the fundamental fact that information, as the critical resource of information engines, cannot be acquired instantaneously and without cost. Information acquisition through measurement, the related energetic cost and the subsequent work extraction are highly interrelated by the measurement time.

Our results further show that achieving positive power output and high efficiency in the conversion of information and energy into useful work requires careful tuning of the measurement time and of other system and meter parameters. This reveals intrinsic trade-offs between these performance metrics, which we analysed using Pareto fronts—a global optimization framework. Notably, we identified a global maximum in power output at a finite efficiency, occurring at a specific, finite measurement time for the prototype information engine.

Since the Pareto fronts naturally bound the trade-offs between the engine's performance metrics, they identify the optimal combinations of system and meter parameters for maximizing information engine performance. This offers practical guidance for experimentally selecting optimal engine configurations. 

Exploiting measurement and information as a \textit{resource} for controlling and optimising \textit{outputs} can be vital not only in contexts such as energy transduction in nanoscale devices, but also in other research fields such as chemistry to improve reaction yields \cite{Pruchyathamkorn2024}. 
In particular, the consideration of measurement time is crucial to control and optimise time-dependent processes. 
Moreover, at the nanoscale of information driven processes, additional complexity may arise due to fluctuations in quantities such as the power output. This might lead to future investigations of, \textit{inter alia}, the relation between the signal-to-noise ratio, the acquired information and the measurement time.

%% file: sections/section_acknowledgements.tex
We thank Abraham Nitzan, Federico Centrone, Gabriel Landi and Sofia Sevitz for helpful discussions. J.B. and H.K. especially thank Ralf Eichhorn, Bart Cleuren and Supriya Krishnamurthy for organizing the Workshop on Statistical Physics at Nordita in Stockholm, where this collaboration began. We are also grateful to Juliette Monsel for carefully reading our manuscript and providing valuable feedback. Funding from the Knut and Alice Wallenberg Foundation via the Fellowship program (J.S.) and  from the European Research Council (ERC) under the European Union’s Horizon Europe research and innovation program (101088169/NanoRecycle) (H.K., J.S.) is gratefully acknowledged. J.B. is supported by the Novo Nordisk Foundation with grant No. NNF18SA0035142.

%% file: sections/availability.tex
\section*{Code Availability}
The code used to generate the data for this article is available on Zenodo (DOI: 10.5281/zenodo.15646115) under a Creative Commons Attribution 4.0 International (CC BY 4.0) license.

%% file: appendices/mutual_information.tex
\section{Mutual Information }\label{app:mutual_info}
For a non-demolishing measurement, i.e., $[\hat{H}_{\rm S},\hat{V}_{{\rm I}}]=0$, we find the mutual information to be
\begin{align}
\label{know}
I(&\tm)\equiv S(0)-S(\tm) \\
&=-\kb \sum_n \sum_{i=0}^{1}  P(i,n,0) \ln{ P(i|n,0)} \\ \notag &+\kb \sum_n \sum_{i=0}^{1}   P(i,n,\tm)  \ln{P(i|n,\tm)}, \\ \label{know2}
&=-\kb \sum_n\sum_{i=0}^{1}   P(i,n,\tm)  \ln{ P(i|n,0)} \\ \notag &+\kb \sum_n \sum_{i=0}^{1}  P(i,n,\tm) \ln{\frac{P(i,n,\tm)}{P(n,\tm)}},\\ \label{known3}
&=\kb \sum_n \sum_{i=0}^{1}  P(i,n,\tm) \ln{\frac{P(i,n,\tm)}{P(n,\tm)P(i|n,0)}} \\ \notag &=\kb \sum_n \sum_{i=0}^{1}  P(i,n,\tm) \ln{\frac{P(i,n,\tm)}{\sum_iP(i,n,\tm)\sum_n P(i,n,\tm)}} \\ 
& \geq0.
\end{align}
In Eq.\ \eqref{know2}, we have used the fact that $P(i|n,0)$ is independent of $n$ and gives $P(0|n,0)=a$ and $P(1|n,0)=b$ and that we have a non-demolishing measurement, i.e., the system-meter coupling has the form $\HamI \propto \sum_i \ket{i}\bra{i}$ such that one couples to the eigenstates of the system. Therefore we have $\sum_n P(0,n,\tm)=a$ and $\sum_n P(1,n,\tm)=b$ which is equal to the marginal probabilities by tracing out the meter.  Eq.\ \eqref{known3} is the mutual information expression associated with the measurement process.

%% file: appendices/free_energy.tex
\section{Bound on Maximum Extractable Work by Mutual Information}\label{app:free_energy}
Here we investigate the free energy difference of a system-meter setup in three steps: (1) starting in the initial state $\dm[0] = \dm[S]\otimes\dm[M]$, (2) undergoing unitary evolution $\dm(t) = \hat{U}\dm[0]\hat{U}^\dagger$, (3) after projection on the eigenstate of the meter $\braket{n|\dm(t)|n}$.
We assume that the Hamiltonian takes the form $\hat{H}(t) = \HamS + \HamM +\HamI(t)$ where $\HamI(t) \neq 0,\, t\in (0,\tm)$ where in general $[\HamS, \HamI] \neq 0$.

The free energy $F$ is
\begin{equation}
    F = \tr{\dm\hat{H}} - TS(\dm) = \tr{\dm\hat{H}} + \kb T\tr{\dm\ln\dm}.
\end{equation}
The free energy difference going from the initial state $\dm[0]$ to a unitarily evolved state $\dm(\tm)$ is 
\begin{equation}
    \begin{split}
        & F_2 - F_1 \\
        &=\tr{(\HamS+\HamM) (\dm(\tm) - \dm[0])} - T\big(S(\dm(\tm)) - S(\dm[0])\big) \\
        &=  \tr{(\HamS+\HamM) (\dm(\tm) - \dm[0])},
    \end{split}
\end{equation}
where we used the property that unitary evolution is isentropic. The measurement work, defined in \cref{eq:measurement_work_definition}, is
\begin{multline}\label{app_eq:msmt_work}
    \Wmeas = \int_0^{\tm} dt \tr{\dm(t) \dot{\hat{H}}(t)} =\\
    \tr{\big(\dm(\tm) - \dm[0]\big)\big(\HamS+\HamM\big)}
    - \int_0^{\tm} dt \tr{\dot{\hat{\rho}}(t)\hat{H}(t)}\\
    =\tr{\big(\dm(\tm) - \dm[0]\big)\big(\HamS+\HamM\big)}.
\end{multline}
This is the work invested in order to correlate the meter and system with each other.

The free energy of the system after projection on the eigenstates of the meter is given by
\begin{multline}
    F_3(\tm) - F_3(0) =\\
    \sum_n P(n,\tm) \tr{\HamS \frac{\braket{n|\dm(\tm)|n}}{P(n,\tm)}} \\
    -\sum_n P(n,0) \tr{\HamS\frac{\braket{n|\dm[0]|n}}{P(n,0)}} \\
    + \kb T \sum_n P(n,\tm) \tr{\frac{\braket{n|\dm(\tm)|n}}{P(n,\tm)} \ln \frac{\braket{n\dm(\tm)|n}}{P(n,\tm)}} \\
    - \kb T\sum_n P(n,0) \tr{\frac{\braket{n\dm[0]|n}}{P(n,0)} \ln \frac{\braket{n\dm[0]|n}}{P(n,0)}} 
    \\
    = \tr{\HamS \Big(-\dm[S] + \sum_n \braket{n|\dm(\tm)|n}\Big)} \\
    +\kb T \sum_n \tr{\braket{n|\dm(\tm)|n}\big( \ln \braket{n|\dm(\tm)|n} - \ln P(n,\tm) \big)} \\
    -\kb T \tr{\dm[S]\ln\dm[S]}.
\end{multline}
In the final expression, the first term corresponds to the $\HamS$-term in the measurement work (\cref{app_eq:msmt_work}). 
Hence, that part of the energy difference is less interesting as it always has to be paid for through coupling and decoupling.
However, for a non-demolishing measurement, i.e. $[\HamS,\HamI] = 0$, we find that $\dm[S](\tm) = \dm[S](0)$ making the free energy difference
\begin{align}
    \Delta F &= \kb T \sum_n \text{tr}\Big\{ \braket{n|\dm(\tm)|n}\big( \ln \braket{n|\dm(\tm)|n} \nonumber\\
    &- \ln P(n,\tm) \big)\Big\}
    -\kb T \tr{\dm[S]\ln\dm[S]}  \\
    &= \Temp[S] \big(S(\tm) - S(0)\big) = \Temp[S]I(\tm).
\end{align}

%% file: appendices/mutual_information_bound.tex
\section{Maximum Extractable Work by Measurement}\label{app:extracted_work_bound}
This section builds on the original idea and method on bounds on the maximum extractable work using measurement, used in Appendices E-F of Ref.~\cite{Kirchberg2025}, applying them here to a system utilising a \gls{qho} as a meter, in contrast to the free particle used in the original work of Ref.~\cite{Kirchberg2025}. 

Assuming that we follow the scheme laid out in \cref{sec:method_example} we extract work from our system by some unitary transformation leaving the system in a passive state.
However, this passive state does not necessarily correspond to the thermal state given by the temperature $\Temp[S]$ of the thermal bath coupled to the \gls{tls}.
As such, we may imagine that we insert a heat engine and extract work from the rethermalisation process going from some arbitrary passive state to the thermal state specified by the temperature $\Temp[S]$.
The heat capacity is given by
\begin{align}
    C(T)=\kb\beta^2\frac{d^2}{d\beta^2}\ln{Z_\text{S}}
    = \frac{\Delta E^2}{\kb T^2}\frac{e^{-\Delta E/\kb T}}{(1+e^{-\Delta E/\kb T})^2},
\end{align}
where $\beta = (\kb T)^{-1}$ is the inverse temperature and $Z_\text{S} = \sum_i e^{-\beta E_i}$ is the partition function with $E_i$ the $i$th eigenenergy.

Using the conditional probabilities of obtaining $\ket{1}$ or $\ket{0}$ given the measurement outcome $n$ at time $\tm$ we are either in a passive ($P(0|n,\tm)>P(1|n,\tm)$) or active state ($P(1|n,\tm)>P(0|n,\tm)$). If one is in an active state, one can apply a $\pi-$pulse to unitarily extract work from it to end in a passive state.  Once in a passive state, either immediately after measurement or after work extraction, we can define the temperature:
\begin{multline}\label{app:passive_temp}
    \Temp[p](n,\tm) =\\
    \frac{\Delta E}{\kb}\Bigg[
    \ln{\Bigg(\frac{P(1|n,\tm)}{P(0|n,\tm)}\Bigg)^{-1}} \Theta\Big(P(1|n,\tm)-P(0|n,\tm)\Big) \\
    +\ln{\left(\frac{P(0|n,\tm)}{P(1|n,\tm)}\right)^{-1}\Theta\Big(P(0|n,\tm)-P(1|n,\tm)\Big)}
    \Bigg].
\end{multline}

The Carnot efficiency gives the maximum amount of work that can be extracted through heat transfer as
\begin{equation}
    W = Q \left(1-\frac{\Temp[C]}{\Temp[H]}\right) \Rightarrow dW = dQ\left(1-\frac{\Temp[C]}{\Temp[H]} \right),
\end{equation}
where $\Temp[C]$ and $\Temp[H]$ are the temperatures of the cold and hot reservoirs, respectively.
There are two scenarios; (a) the passive state is hotter than the thermal state, that is, $\Temp[p] > \Temp[S]$, or (b) the passive state is colder than the thermal state, that is, $\Temp[p] < \Temp[S]$.

First investigate scenario (a), $\Temp[p] > \Temp[S]$.
In this case $\Temp[C] = \Temp[S]$, and $\Temp[H] =\Temp$ varying from $\Temp[p]$ to $\Temp[S]$ yielding the thermal work
\begin{equation}
    W_\text{th}^\text{(a)} = -\int_{\Temp[p]}^{\Temp[S]} dT\, C(T)\left(1 - \frac{\Temp[S]}{T}\right).
\end{equation}

Next, investigate scenario (b), $\Temp[p] < \Temp[S]$
In this case we have $\Temp[H] = \Temp[S]$, and $\Temp[C] =T$ varying from $\Temp[p]$ to $\Temp[S]$.
The difference between scenarios (a) and (b) is that in scenario (b), part of the heat must be expended to heat up the \gls{tls}:
\begin{equation}
    dQ_{\text{sys}} = dQ\frac{T}{\Temp[S]}
\end{equation}
which raises the temperature by
\begin{equation}
    dT =\frac{dQ_{\text{sys}}}{C(T)} = dQ \frac{T}{\Temp[S]\,C(T)}.
\end{equation}
Hence, the work that can be extracted is 
\begin{equation}
    dW=dQ \left( 1 - \frac{T}{\Temp[S]}\right),
\end{equation}
and 
\begin{equation}
    \begin{split}
    W_\text{th}^\text{(b)} &= \int_{\Temp[P]}^{\Temp[S]} dT C(T)\left(1-\frac{T}{\Temp[S]}\right) \\
         &= \int_{\Temp[p]}^{\Temp[S]} dT C(T) \left(\frac{\Temp[S]}{T} -1\right)\\
    &= -\int_{\Temp[p]}^{\Temp[S]} dT C(T) \left(1 - \frac{\Temp[S]}{T} \right) \\
    &= W_\text{th}^\text{(a)},
    \end{split}
\end{equation}
meaning that in either case the extracted work is the same.
On average then, the extracted work is 
\begin{equation}\label{app:average_thermal_work}
    W_\text{th} = \sum_{n=0}^\infty P(n,\tm) \int_{\Temp[p]}^{\Temp[S]} dT C(T)\left(\frac{\Temp[S]}{T} -1 \right).
\end{equation}
Evaluating the integral
\begin{equation}
    W_\text{th} = \int_{\Temp[p]}^{\Temp[S]} dTC(T)\left(\frac{\Temp[S]}{T} -1 \right)
\end{equation}
by making the variable substitution $y = \Delta E/\kb T$ yields
\begin{equation}
    \label{app:therm_work}
    W_\text{th} = \frac{\Delta E\left( 1-\frac{\Temp[S]}{\Temp[p]} \right) }{1 + e^{\Delta E/\kb\Temp[p]}} - \kb\Temp[S] \ln\left( \frac{1 + e^{-\Delta E/\kb\Temp[p]}}{1 + e^{-\Delta E/\kb\Temp[S]}} \right).
\end{equation}
Inserting the expression for $\Temp[p]$ into \cref{app:therm_work} yields
\begin{multline}
    W_\text{th} = \kb \Temp[S]
    \Big[ P(0|n,\tm)\big(\ln P(0|n,\tm) - \ln b\big)\\ 
    + P(1|n,\tm)\big(\ln P(1|n,\tm) - \ln a\big) \Big] \\
    \times \Theta\big[P(1|n,\tm) - P(0|n,\tm)\big] \\
    +\kb\Temp[S] \Big[ P(0|m,\tm)\big(\ln P(0|n,\tm) -\ln a\big) \\
    + P(1|n,\tm)\big(\ln P(1|n,\tm) - \ln a\big) \Big] \\
    \times \Theta \big[ P(0|n,\tm) - P(1|n,\tm) \big] .
\end{multline}
Adding and subtracting the term 
\begin{multline}
    \kb\Temp[S] \Big[ P(0|m,\tm)\big(\ln P(0|n,\tm) -\ln a\big)+ \\ 
    P(1|n,\tm)\big(\ln P(1|n,\tm) - \ln a\big)\Big] \\
    \Theta\big[P(1|n,\tm) - P(0|n,\tm)\big],
\end{multline}
simplifying, and inserting the resulting expression into the average work defined in \cref{app:average_thermal_work} yields
\begin{multline}
    W_\text{th} = \sum_{n=0}^\infty \kb\Temp[S] \Big\{
    \Big[ P(1|n,\tm) - P(0|n,\tm) \Big] \ln\left(\frac{b}{a}\right)\\
    \times \Theta\big[P(1|n,\tm)-P(0|n,\tm)\big] \\
    + P(0|n,\tm)\big(\ln P(0|n,\tm)-\ln a\big) \\
    + P(1|n,\tm)\big(\ln P(1|n,\tm) - \ln b\big)\Big\} .
\end{multline}

Noting that $\ln{(b/a)} = - \Delta E/\kb\Temp[S]$ and using the ergotropy as defined in \cref{eq:work_extracted_specific} we see that the first term is the average ergotropy, $-W_\text{erg}$. Further, noting that the two last terms make up the mutual information defined in \cref{eq:mutual_info} we can write the average thermal work as
\begin{equation}
    W_\text{th}(\tm)  = -W_\text{erg}(\tm) + \Temp[S] I(\tm).
\end{equation}

%% file: appendices/measurement_work.tex
\section{Evaluation of the Measurement Work}\label{app:msmt_work}
We assume a Hamiltoninan of the form
\begin{equation}
    \hat{H}(t) = \HamS + \HamM + \HamI(t)
\end{equation}
as per the main text where $\HamS$ is the system from which work is to be extracted, $\HamM$ is the meter used to probe the state of the system, and $\HamI$ is the interaction Hamiltonian which in this case has the form
\begin{equation}
    \HamI(t) = \begin{cases}
        g\ket{1}\bra{1}\otimes\hat{p}, \quad t\in (0, \tm) \\
        0,\quad t \notin (0,\tm)
    \end{cases}
    .
\end{equation}
The time-evolution of the system is described by the density matrix $\dm(t) = \exp\big(-\frac{i}{\hbar}\int_0^t dt' \hat{H}(t')\big)\dm(0)\exp\big(-\frac{i}{\hbar}\int_0^t dt' \hat{H}(t')\big)$ and we note that $\dot{\dm}(t) = -\frac{i}{\hbar}[\hat{H}(t),\dm(t)]$.
The system energy is given by $E(t) = \langle\hat{H}(t)\rangle =  \tr{\dm(t)\hat{H}(t)}$ and the time derivative is
\begin{eqnarray}
    &\dot{E}(t) = \tr{\dot{\dm}(t)\hat{H}(t)} + \tr{\dm(t)\dot{\hat{H}}(t)} = \nonumber\\
    &\tr{-\frac{i}{\hbar}\big[\hat{H}(t),\dm(t)\big]\hat{H}(t)} + \tr{\dm(t)\dot{\hat{H}}(t)} \nonumber\\
    &= \tr{\dm(t)\dot{\hat{H}}(t)}
\end{eqnarray}
where we used that $[\hat{H}(t),\dm(t)]=i\dot{\hat{\rho}}/\hbar$.
Thus the energy difference between time $t=0$ and time $t=\tm$ is 
\begin{equation}\label{app:energy_difference}
     E(\tm) - E(0) = \int_0^{\tm} dt \dot{E}(t) = \int_0^{\tm} dt \tr{\dm(t)\dot{\hat{H}}(t)}. 
\end{equation}
Here we may note that $\HamS,\, \HamM$ are both time-independent and thus $\dot{\hat{H}}(t) = \dot{\hat{V}}_\text{I}(t)$.
We write the interaction Hamiltonian as
\begin{align}\label{app:time_derivative}
    \HamI(t) = g \Theta(\tm-t)\Theta(t)\big(\ket{1}\bra{1}\otimes\hat{p}\big)  = 
    g \Theta(\tm-t)\Theta(t)\HamI \\
    \Rightarrow \dot{\hat{V}}_\text{I}(t) = g\Big(-\delta(\tm-t)\Theta(t) + \Theta(\tm-t)\delta(t) \Big)\HamI.
\end{align}
Inserting the time-derivative from \cref{app:time_derivative} into \cref{app:energy_difference} yields
\begin{align}
    &E(\tm) - E(0) =\nonumber\\ 
    &\int_0^{\tm} dt\, \tr{ \dm(t) g\Big( -\delta(\tm-t)\Theta(t) + \Theta(\tm-t)\delta(t) \Big)\HamI } \nonumber\\
    &= \tr{\dm(0)\HamI} - \tr{\dm(\tm)\HamI}
\end{align}
which is the result used in \cref{eq:measurement_work_specific}.
Hence, the energy difference created simply by coupling and decoupling the system and meter is exactly what we defined as $\Wmeas$.

We can further expand this by evaluating each of the traces on their own.
Expressing the initial density matrix as $\dm(0) = \big[a\ket{0}\bra{0} + b \ket{1}\bra{1}\big] \otimes \sum_m P_m\ket{m}\bra{m}$ where $\sum_m P_m = 1$ are the probabilities associated with the eigenstates of the meter.
Thus the first of the traces can be evaluated as
\begin{align}
    \tr{\dm(0)\HamI} &=\sum_{n=0}^\infty\sum_{i=0}^1\bra{n}\bra{i}\Big(a\ket{0}\bra{0}+b\ket{1}\bra{1}\Big) \nonumber\\
    &\otimes \sum_{m=0}^\infty P_m \ket{m}\bra{m} g \ket{1}\bra{1}\otimes\hat{p}\ket{i}\ket{n} \nonumber\\
    &=bg \sum_n\sum_m P_m \braket{n|m}\braket{n|\hat{p}|m} \nonumber\\
    &= bg \sum_m P_m\braket{m|\hat{p}|m} = 0
\end{align}
where we have used the completeness relation to eliminate the sum over $n$. 
The conclusion is that, for this particular model, turning the coupling on has no cost.

The second trace can similarly be evaluated, however this time the density matrix is no longer the initial density matrix but is rather time-evolved.
Performing the same type of polaron transform used in \cref{app:zeno}, we may insert shift operators $\hat{D}^\dagger\hat{D}$ in strategic places, use the completeness relation, and observe that just as for the first trace only the $i=1$ term contributes to yield 
\begin{eqnarray}
    &\tr{\dm(t)\HamI} = bg\sum_m P_m\braket{m|e^{it\hat{H}'/\hbar}\hat{p}e^{-it\hat{H}'/\hbar}|m} \nonumber\\
    &= bg\sum_m P_m\braket{m|\hat{D}^\dagger(t)\hat{D}\hat{p}\hat{D}^\dagger\hat{D}(t)|m}.
\end{eqnarray}
Here we use commutators and the commutation relation $[f(\hat{x}),\hat{p}] = i\hbar\frac{df(x)}{dx}$ to migrate the momentum operator from the centre to the left hand side, and the shift operators should cancel due to being Hermitian.
In practice this becomes
\begin{eqnarray}
    &bg\sum_m P_m\braket{m|\hat{D}^\dagger(t)\hat{D}\hat{p}\hat{D}^\dagger\hat{D}(t)|m} \nonumber \\
    &= bg \sum_m P_m \braket{m|\hat{D}^\dagger(t)(-Mg+\hat{p})\hat{D}\hat{D}^\dagger\hat{D}(t)|m} \nonumber\\
    &= -bg^2M+bg\sum_m P_m \braket{m|\hat{D}^\dagger(t)\hat{p}\hat{D}(t)|m} \nonumber\\
    &= -bg^2M + \sum_m P_m \braket{m|\big(Mg\cos(\omega t) + \hat{p}\big)\hat{D}^\dagger(t)\hat{D}(t)|m} \nonumber\\
    &= -b\geff^2\big(1-\cos(\omega t)\big).
\end{eqnarray}
Here $\geff^2 = g^2M$ and $M$ is the mass of the harmonic oscillator.
All in all the measurement work is then
\begin{equation}
\label{eq:meas_cost}
    \Wmeas = \tr{\dm(0)\HamI} - \tr{\dm(\tm)\HamI} = b\geff^2\big(1-\cos(\omega\tm)\big).
\end{equation}

%% file: appendices/probabilities.tex
\section{Explicit Forms of the Joint Probabilities}\label{app:probs}
The joint probabilities of finding the two-level system in state $\ket{i}$ and the meter in state $\ket{n}$ at time $\tm$ is given by
\begin{equation}
    P(i,n,\tm) = \braket{i|\braket{n| \dm(\tm) |n}|i}
\end{equation}
according to \cref{eq:joint_probs}.
Here we present the calculations that result in explicit forms of these probabilities for our given system in \cref{sec:method_example}.

\subsection{Joint Probability in the Absence of Interaction}
The coupling (\cref{eq:interaction_hamiltonian}) is active only when the \gls{tls} is in the excited state.
We begin with the easier case of examining the joint probability $P(i=0,n,\tm)$ where the coupling is inactive.
With $\hat{H}_0 = \Delta E\ket{1}\bra{1}+\frac{\mathcal{M}\omega^2}{2}\hat{x}^2 + \frac{1}{2\mathcal{M}}\hat{p}^2$ the joint probability is
\begin{equation}\label{app:joint_prob_ground_case}
 \begin{split}
    P(0, n, \tm) &= \bra{0}\bra{n}\dm(\tm)\ket{n}\ket{0}\\
   &= \bra{0}\bra{n}e^{-i\tm\hat{H}_0/\hbar}(a\ket{0}\bra{0}+b\ket{1}\bra{1})\\ &\otimes 
   \sum_m P_m\ket{m}\bra{m}e^{i\tm\hat{H}_0/\hbar}\ket{n}\ket{0}\\
   &= a\sum_m P_m\bra{n}m\rangle\bra{m}n\rangle
   = aP_n \\
   &= a\frac{1}{Z_0}e^{-\beta_M\hbar\omega (n+1/2)} \\
   &= a \left(1-e^{-\beta_\text{M}\hbar\omega}\right)e^{-\beta_\text{M}\hbar\omega n}
\end{split}
\end{equation}
with the thermal distribution $P_n=e^{-\beta_M\hbar\omega (n+1/2)}/\sum_n e^{-\beta_M\hbar\omega (n+1/2)} = e^{-\beta_M\hbar\omega (n+1/2)}/Z_0 $ and $\beta_\text{M}$ being the inverse temperature of the meter.

\subsection{Joint Probability in the Presence of Interaction}
In this case the coupling is active and evaluation is less straightforward.
We again make use of the polaron transformation described in \cref{app:zeno} and write
\begin{equation}
    \begin{split}
    P(1,n,t)&=
    \bra{1}\bra{n}e^{-it\hat{H}_0/\hbar}\hat{D}(\alpha)
    \dm[0]\hat{D}^\dagger(\alpha)e^{it\hat{H}_0/\hbar}
    \ket{n}\ket{1} \\
    &= b\sum_m P_m\bra{n}\hat{D}(\alpha)\ket{m}\bra{m}\hat{D}^\dagger(\alpha)\ket{n}\\
    &= b\sum_m P_m\left|\bra{n}\hat{D}(\alpha)\ket{m}\right|^2.
\end{split}.
\end{equation}
The displacement operator has here been rewritten in the form 
\begin{align}
    &\hat{D}(\alpha) = e^{\alpha(\tm) \hat{a}^\dagger - \alpha^*(\tm)\hat{a}}
    \label{app:displacement_operator_unitary}\\
    &\alpha\equiv \alpha(\tm) = g\sqrt{\frac{\mathcal{M}}{2\hbar \omega }}\Bigg(\sin(\omega  \tm)-i\Big(\cos(\omega \tm) -1\Big)\Bigg).
    \label{app:alpha_parameter_unitary}
\end{align}

The identity from Ref.~\cite{Glauber1969}
\begin{equation}
    \braket{m|\hat{D}(\alpha)|n} =\left(\frac{n!}{m!}\right)^{1/2}\alpha^{m-n}e^{-|\alpha|^2/2}L_n^{(m-n)}(|\alpha|^2)
\end{equation}
 yields 
\begin{multline}
\label{eq:Cond_Exc}
    P(1,n,\tm) =\\ b\sum_m P_m \left|\left(\frac{m!}{n!}\right)^{1/2}\alpha^{n-m}e^{-|\alpha|^2/2}L_m^{(n-m)}\left(|\alpha|^2\right)\right|^2
\end{multline}
where $L_m^{(n-m)}(x)$ are the generalised Laguerre polynomials and the thermal occupation $P_m=e^{-\beta_M\hbar\omega (m+1/2)}/\sum_n e^{-\beta_M\hbar\omega (m+1/2)} = e^{-\beta_M\hbar\omega (m+1/2)}/Z_0 $ with $\beta_\text{M}$ being the inverse temperature of the meter.

In the limit $\hbar\omega\gg k_B \Temp[M]$, we can assume that only the ground state of the oscillator is occupied, that is, $m=0$ and $P_{m=0}=1$. \cref{eq:Cond_Exc} can than be written as
\begin{align}
\label{eq:Cond_Exc2}
    &P(1,n,\tm) =b\left(\frac{1}{n!}\right)|\alpha|^{2n}e^{-|\alpha|^2}\\ \notag
    &=b\left(\frac{1}{n!}\right) \bigg[\frac{g_{\rm eff}^2}{\hbar \omega}\big(1-\cos(\omega \tm)\big)\bigg]^n e^{-\frac{g_{\rm eff}^2}{\hbar \omega}\big(1-\cos(\omega \tm)\big)}
    .
\end{align}

%% file: appendices/zeno.tex
\section{Work Extraction in the Zeno Limit}\label{app:zeno}
The quantum Zeno effect is essentially a freezing of the system dynamics through fast, repeated measurement~\cite{Misra1977,Facchi2008}.
Referring to the engine in the main text, we may then ask how it behaves in the Zeno limit and if net positive work extraction is possible.

The net work output of the engine is $\Wnet = \Wext - \Wmeas$, hence the condition to fulfill is $\Wext > \Wmeas$.
Using \cref{eq:work_extracted_specific,eq:measurement_work_definition}, the condition is
\begin{multline}
    \Delta E \sum_{n=0}^\infty 
    P(n,\tm)\Big[ P(1|n,\tm) - P(0|n,\tm)\Big]\\
    \Theta \Big(P(1|n,\tm) - P(0|n,\tm)\Big)
    > b\geff^2(1-\cos(\omega \tm)\big)
\end{multline}
and we remind ourselves that $a,\,b$ are the initial ground and excited state populations of the \gls{tls}
\begin{align}
    a = \frac{1}{1+e^{-\beta_S\Delta E}} \\
    b = \frac{e^{-\beta_S\Delta E}}{1+e^{-\beta_S\Delta E}}
\end{align}
where $\beta_S$ is the inverse temperature.

When the system evolves unitarily there are essentially two timescales; the time scale of the harmonic oscillator $\omega\tm$ and the time scale of information transfer $\geff\tm$.
We would argue that in the Zeno limit both of these timescales must be small. 
Expanding the measurement work, $\Wmeas$, \cref{eq:measurement_work_specific}, around $\omega\tm=0$ yields
\begin{equation}
    \Wmeas \approx b\geff^2\frac{\omega^2\tm^2}{2}.
\end{equation}

We rewrite the extracted work in the following way: let $n' = \min_n \big(P(1|n\tm) > P(0|n,\tm)\big)$
and find
\begin{equation}
    \Wext = \Delta E \sum_{n=n'}^\infty \Big[P(1,n,\tm)-P(0,n,\tm)\Big].
\end{equation}
where Bayes' rule was used.
The joint probabilities of \gls{tls} state $i$ and \gls{qho} state $n$ at time $\tm$ are given by $\braket{i|\braket{n|\dm(\tm)|n}|i}$.
The density matrix is $\dm(\tm) = e^{-i\tm \Hat{H}/\hbar}\dm[S](0)\otimes\dm[M](0)e^{-i\tm \Hat{H}/\hbar}$ with
\begin{align}
    &\dm[S](0) = a\ket{0}\bra{0}+b\ket{1}\bra{1}\\
    &\dm[M](0) = \sum_{n=0}^\infty \frac{1}{Z_\text{M}}e^{-\beta_M\hbar\omega(n+1/2)}\ket{n}\bra{n}
\end{align}
and $Z_\text{M}$ is the partition function.
We can also diagonalise the Hamiltonian by way of the displacement operator $\hat{D} = e^{iMg\ket{1}\bra{1}\otimes\hat{x}/\hbar}$ to get
\begin{align}
    \dm(\tm) = e^{-it\hat{H}'/\hbar} \hat{D}^\dagger(\tm)\hat{D}\dm(0)\hat{D}^\dagger\hat{D}(\tm)e^{it\hat{H}'/\hbar} \\
    \hat{H}' = \left(\Delta E - \frac{\geff^2}{2}\right)\ket{1}\bra{1} + \frac{1}{2\mathcal{M}}\hat{p}^2  + \frac{\mathcal{M}\omega^2}{2}\hat{x}.
\end{align}
Here we also remind ourselves that $\geff^2 = g^2\mathcal{M}$ where $\mathcal{M}$ is the mass of the harmonic oscillator.

With these preliminaries out of the way we may now write the average work extracted as
\begin{multline}
\label{eq:W_ext_append}
    \Wext = \Delta E \sum_{n=n'}^\infty \Bigg[b \sum_{m=0}^\infty \Big(  \frac{1}{Z_\text{M}}e^{-\beta_\text{M}\hbar\omega(m+1/2)}\\
    \braket{n|\hat{D}^\dagger
    (\tm)\hat{D}|m}\braket{m|\hat{D}^\dagger\hat{D(\tm)}|n} \Big) \\
    -a\frac{1}{Z_\text{M}}e^{-\beta_\text{M}\hbar\omega(n+1/2)}\Bigg]. 
\end{multline}
Using the Zassenhaus formula~\cite{Magnus1954}, evaluating the sum of the second term in $\Wext$, \cref{eq:W_ext_append}, and simplifying now yields the result for the condition $\Wext>\Wmeas$
\begin{multline}
    \sum_{n=n'}^\infty b\sum_{m=0}^\infty e^{-\beta_\text{M}\hbar\omega m}
    \braket{n|e^{-ig\tm\hat{p}/\hbar}|m}
    \braket{m|e^{ig\tm\hat{p}/\hbar}|n} \\
    > \frac{ae^{-\beta_\text{M}\hbar\omega n'}}{1-e^{-\beta_\text{M}\hbar\omega}} +
     \frac{b\geff^2\omega^2\tm^2}{2\Delta E(1-e^{-\beta_\text{M}\hbar\omega})}.
\end{multline}
At this time we make use of the second condition, $gt\ll1$, expand to order 2, and simplify further to get
\begin{multline}\label{eq:condition_zeno}
    \sum_{n=n'}^\infty e^{-\beta_\text{M}\hbar\omega n}
    \Bigg[ -\left(\frac{g\tm}{\hbar}\right)^2 \braket{n|\hat{p}^2(0)|n} \\
    + \sum_{m=0}^\infty e^{-\beta_\text{M}\hbar\omega m}\left(\frac{g\tm}{\hbar}\right)^2\braket{n|\hat{p}(0)|m}\braket{m|\hat{p}(0)|n}\Bigg] \\
    > \frac{(\geff\omega\tm)^2}{2\Delta E (1-e^{-\beta_\text{M}\hbar\omega})} + \frac{(a-b)e^{-\beta_\text{M}\hbar\omega n'}}{b(1-e^{-\beta_\text{M}\hbar\omega})}
\end{multline}
Here, a useful trick is to rewrite the momentum operator in terms of the creation and annihilation operators. 
Doing this and simplifying results in the final condition of \cref{eq:condition_zeno}
\begin{multline}\label{eq:zeno_cond_final}
    4 n'\sinh^2\left(\frac{\beta_\text{M}\hbar\omega}{2}\right) > \\
    \frac{\hbar\omega}{\Delta E}e^{\beta_\text{M}\hbar\omega n'} + \frac{2\hbar(a-b)}{\geff^2\omega\tm^2 b}.
\end{multline}
Now, $n'$ which we may call an activation threshold for when we choose to extract work will be a function of $\tm$ since it depends on where $P(1|n,\tm) > P(0|n,\tm)$. 
Nevertheless, for our model \cref{eq:zeno_cond_final} provides a condition for when positive work can be extracted from the engine.

We may also use \cref{eq:zeno_cond_final} to investigate the behaviour of $\Wext$ as a function of $\tm$.
First we identify that the first term on the right hand side comes from the measurement work, $\Wmeas$, and ignore this term to yield to condition $\Wext > 0$:
\begin{equation}
    4n'\sinh^2\left(\frac{\beta_\text{M}\hbar\omega}{2}\right) > \frac{2\hbar(a-b)}{\geff^2\omega\tm^2 b}.
\end{equation}
Solving for $\omega\tm$ gives 
\begin{equation}\label{eq:zeno_omegatm}
    \omega\tm > \sqrt{ \frac{2\hbar\omega(a-b)}{2  n' \sinh^2\left(\frac{\beta_\text{M}\hbar\omega}{2}\right)\geff^2 b} }, 
\end{equation}
where we see that a higher activation threshold $n'$ allows for shorter measurement times.
Or, conversely, going to shorter measurement times pushes the activation threshold higher, meaning that lower meter states will not contribute to the extracted work $\Wext$,
\begin{equation}\label{eq:zeno_nprime}
    n' > \frac{\hbar\omega(a-b)}{2 \sinh^2\left(\frac{\beta_\text{M}\hbar\omega}{2}\right)\geff^2 b  (\omega\tm)^2}
\end{equation}
where $n' \in \mathbb{N}^{0}$.

%% file: appendices/power_vs_time.tex
\section{Power as a Function of Time}\label{app:power_vs_time}

\begin{figure}[htp]
    \centering
    \includegraphics[]{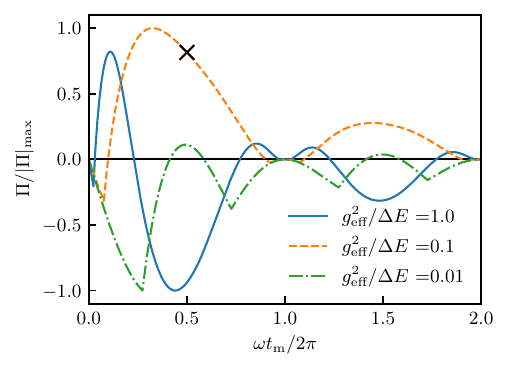}
    \caption{ Power output normalised with its maximal absolute value, see \cref{eq:power}, as a function of time for three values of the effective coupling strength, $\geff^2/\Delta E = 0.01,\,0.1,\,1.0$. 
    The chosen parameters are $\Temp[M]/\Temp[S] = 0.2$, $\Delta E = 4\kb\Temp[S]$, $\hbar\omega = 1.5\kb\Temp[S]$. The $\times$ denotes a specific operating point, serving as reference in \cref{fig:P_eta_both}(a) and \cref{fig:P_eta_info_both}.}
\label{fig:power_vs_time_higher_temp}
\end{figure}

In \cref{fig:power_vs_time}, the power shows an oscillatory behaviour, alternating between positive and negative values over time.
This effect becomes more pronounced as the coupling strength increases.
However, there are other parameters besides measurement time and coupling strength that influence this behaviour.
To illustrate this, \cref{fig:power_vs_time_higher_temp} shows the normalised power as a function of time for a higher temperature ratio $\Temp[M]/\Temp[S]=0.2$, in contrast to the lower temperature ratio $\Temp[M]/\Temp[S]=0.1$ used in \cref{fig:power_vs_time}. With higher meter temperate the non-smooth behaviour due to $\Wext$, see discussion in \cref{sec:results}, becomes more pronounced.
Increasing the meter temperature, $\Temp[M]$, makes the meter less sensitive, thus, reducing the net extracted work. 
As shown in \cref{fig:ergotropy_coupling_heatmap}, when $\Temp[M]/\Temp[S]=0.2$ the engine approaches the boundary of the power-producing region, particularly at both the lowest and highest coupling strengths.
For both the lowest and highest coupling strengths, the information engine oscillates between the two modes of operation---the heat engine and heat valve.
This reiterates the importance of the measurement time, $\tm$, as an engine parameter.

%

%% file: appendices/low_meter_temperature.tex
\section{Thermodynamic efficiency and power at low meter temperatures}
\label{app:low_temp}
In the limit $\hbar\omega\gg k_B \Temp[M]$, we can assume that initially only the ground state of the oscillator is occupied.
The extracted work $\Wext$, \cref{eq:W_ext_append}, together with \cref{eq:Cond_Exc2} and \cref{app:joint_prob_ground_case}
, can be written as
\begin{align}
     W_{\rm ext}=&\Delta E \bigg[ b\sum_{n=n'}^{\infty}\left(\frac{1}{n!}\right) \bigg[\frac{g_{\rm eff}^2}{\hbar \omega}\big(1-\cos(\omega \tm)\big)\bigg]^n \label{eq:work_ext_low}\\  \notag &\times e^{-\frac{g_{\rm eff}^2}{\hbar \omega}\big(1-\cos(\omega \tm)\big)}\bigg]-\Delta Eae^{-\hbar\omega n'/k_B\Temp[M]}.
\end{align}
Note that for low meter temperatures $\hbar\omega\gg k_B \Temp[M]$ one needs to have at least $n'=1$ to have $W_{\rm ext}\geq 0$ while the second term in \cref{eq:work_ext_low} then vanishes. We note that $n'$ depends on the underlying system and meter parameters.  
For $n'=1$, \cref{eq:work_ext_low} reads
\begin{align}
\label{eq:work_ext_low_2}
     W_{\rm ext}=&\Delta E b\bigg[1-e^{-\frac{g_{\rm eff}^2}{\hbar \omega}\big(1-\cos(\omega \tm)\big)}\bigg].
\end{align}
We calculate next the thermodynamic efficiency \cref{eq:efficiency} in this limit. 
As we only consider the heat engine regime, we need to have $\Wext>\Wmeas$. 
Comparing \cref{eq:work_ext_low_2} and \cref{eq:meas_cost} leads to the condition for all $\omega \tm$
\begin{align}
\label{eq:condition_work}
    1-e^{-\frac{g_{\rm eff}^2}{\hbar \omega}\big(1-\cos(\omega \tm)\big)} > \frac{\geff^2}{\Delta E} \big(1-\cos(\omega \tm)\big).
\end{align}
By realising that the argument in \cref{eq:condition_work} $0\leq 1-\cos(\omega \tm) \leq 2$, we can further refine the condition \cref{eq:condition_work} to 
\begin{align}
\label{eq:condition_work_2}
    \frac{2\geff^2}{\hbar \omega}> \ln{\frac{1}{1-2\geff^2/\Delta E}} \geq \frac{2\geff^2}{\Delta E}.
\end{align}

The thermodynamic efficiency \cref{eq:efficiency} can then be written with \cref{eq:work_ext_low_2} and \cref{eq:meas_cost} as
\begin{align}\label{app_eq:low_temp_eff}
    \eta_{\rm HE}&=1-\frac{\Wmeas}{\Wext}\\ \notag
    &=1-\frac{ \geff^2(1-\cos{(\omega \tm))}}{\Delta E \bigg[1-e^{-\frac{g_{\rm eff}^2}{\hbar \omega}\big(1-\cos(\omega \tm)\big)}\bigg]}.
\end{align}

\begin{figure}
    \centering
    \includegraphics[width=0.9\linewidth]{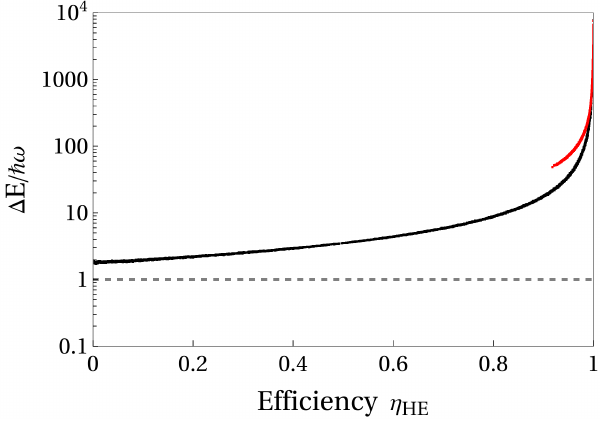}
    \caption{The ratio $\Delta E/\hbar\omega$ along the Power-thermodynamic efficiency Pareto front in Fig.~\ref{fig:P_eta_both} by including the time for application of the $\pi$-pulse (red line) and neglecting it (black line) in the power, plotted as a function of thermodynamic efficiency $\eta_{\rm HE}$. }
    \label{fig:deltaE_vs_efficiency}
\end{figure}

The thermodynamic efficiency $\eta_{\rm HE}$ in \cref{app_eq:low_temp_eff} reaches its maximum value for $\omega \tm \to 0$ which reads
\begin{align}\label{app_eq:low_temp_eff_2}
    \eta_{\rm HE}(\omega \tm \to 0)&=1-\frac{\hbar \omega}{\Delta E},
\end{align}
which together with the condition $\Delta E \geq \hbar \omega$  in \cref{eq:condition_work_2} is bounded by 1. In fact, the condition $\Delta E \geq \hbar \omega$ is fulfilled along the whole Power-thermodynamic efficiency Pareto front as shown in \cref{fig:deltaE_vs_efficiency}.

However, this optimisation must be approached with care. 
The summation in \cref{eq:work_ext_low} begins at $n'$, the smallest value for which $P(1|n,\tm) > P(0|n,\tm)$. Our derivation of \cref{eq:work_ext_low_2,app_eq:low_temp_eff} specifically assumed that $n'=1$. 
As shown in \cref{eq:Cond_Exc2}, $P(1,n,\tm)$ depends on $b = (1+e^{\beta_\text{S}\Delta E})^{-1}$ the excited state population of the \gls{tls}. Here $\beta_\text{S}=(\kb\Temp[S])^{-1}$ is the inverse temperature of the system bath.
If one sends $\Delta E$ to a large number, $b$ approaches zero, which in turn means that $n'$ must tend to infinity for the criterion $P(1|n',\tm) > P(0|n',\tm)$ to hold.
A similar issue of $n'\to\infty$ occurs when taking $\hbar\omega\to\infty$.
Therefore, to preserve the validity of \cref{app_eq:low_temp_eff}, the system temperature $\Temp[S]$ must simultaneously increase with $\Delta E$, thereby keeping the initial state populations $a$ and $b$ constant. As a result, the relative temperature $\Tfrac$ vanishes, $\Tfrac\to 0$ and $\eta_\text{HE} = \eta_\text{CA}=\eta_\text{C}$.

The power, \cref{eq:power}, can be written with \cref{eq:work_ext_low_2} and \cref{eq:meas_cost} as
\begin{align}
\label{eq:power_low_temp}
    \Pi = \frac{b\Delta E \bigg[1-e^{-\frac{g_{\rm eff}^2}{\hbar \omega}\big(1-\cos(\omega \tm)\big)}\bigg]-b\geff^2(1-\cos{(\omega \tm))}}{\tm}.
\end{align}
At maximum efficiency in the limit of $\omega \tm \to 0$, the power in \cref{eq:power_low_temp} vanishes as expected
\begin{align}
\label{eq:power_low_temp_2}
 \lim_{\omega \tm \to 0}   \Pi = b \geff^2 \omega^2 \tm \frac{   \frac{\Delta E}{\hbar\omega}-1}{2}\to 0.
\end{align}

%% file: appendices/Pareto_formalism.tex
\section{Pareto optimization and genetic algorithms}\label{app:Pareto}

Assume we have a system with a compact set of $m$ governing parameters $\mathcal{X} = \{x_1,x_2,\dots,x_m\} \subset \mathbb{R}^m$ that fully describes the system at hand, and a set of $n$ feasible objectives
$\mathcal{Y} = \{y_1,y_2,\dots,y_n\} \subset\mathbb{R}^n$, where each objective is a function $f_i : \mathcal{X}\rightarrow \mathbb{R}$ of the governing parameters, i.e., $y_i = f_i(x_1,x_2,\dots,x_m)$. 
To compute mutually optimal relations between the different objectives we make use of multi-function optimisation to find the parameter values $\mathcal{X}$ that correspond to Pareto-optimal configurations. A configuration is said to be Pareto-optimal if the improvement of any objective is necessarily detrimental to at least one of the others, where we always assume that improving an objective means minimisation; other types of optimisation can always be written in this form. Mathematically, this condition can be encoded through the concept of Pareto dominance~\cite{Emmerich2018}. A vector $\mathbf{y}\in\mathcal{Y}$ is said to \emph{dominate} another vector $\mathbf{y'}\in\mathcal{Y}$, expressed by $\mathbf{y} \succ \mathbf{y'}$, iff
\begin{equation}
    \forall i\in\{1,\dots,n\}: y_i \leq y'_i,\, \text{and }\,\exists j\in\{1,\dots,n\}: y_j < y_j'\,.
\end{equation}
The minimal elements of the Pareto dominance partial ordering are called the Pareto-optimal objective vectors, and the subset of all such non-dominated objective vectors in $\mathcal{Y}$ then constitutes the Pareto front, denoted by $\mathcal{P}(\mathcal{Y}) \subset \mathcal{Y}$.

\begin{figure}[htp]
    \centering
    \includegraphics[width=\linewidth]{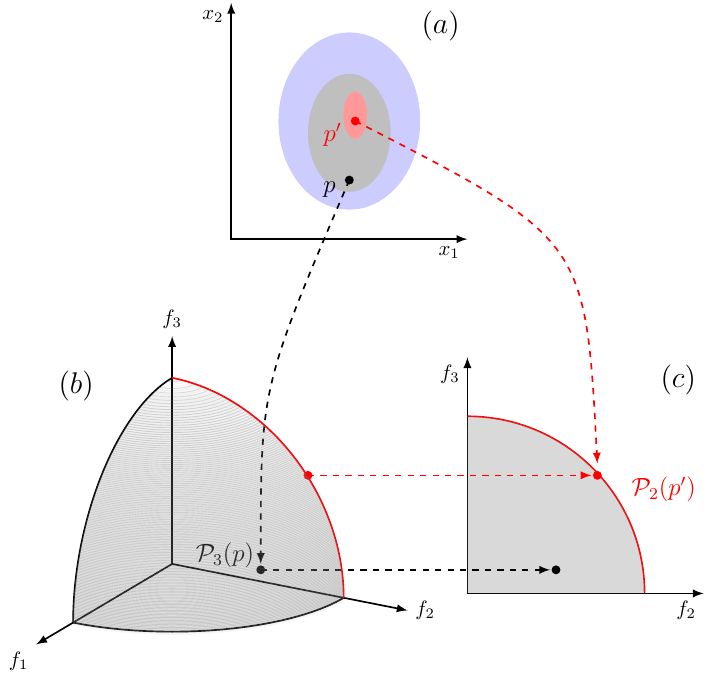}
    \caption{Mappings (dashed lines) from the decision space $\mathcal{X}\subset \mathbb{R}^2$ (panel a, blue region) to the objective spaces (panels b and c). A point $p\in\mathcal{X}$ is mapped to a point $\mathcal{P}_3(p)$ on the two-dimensional Pareto surface $\mathcal{P}_3$ (gray surface in (b)) only if it lies within the gray subset of $\mathcal{X}$ in (a). When projected onto a two-dimensional subspace, e.g., the $f_2-f_3$ plane in panel c, it can become suboptimal with respect to points on the pairwise Pareto front $\mathcal{P}_2$; only points in the red region in (a) are Pareto-optimal in this case.}
    \label{fig:Paretocartoon}
\end{figure}

A simple representation is shown in Fig.~\ref{fig:Paretocartoon}, where the full set of feasible, i.e., physically allowed governing parameters (blue region) is mapped onto a three-dimensional objective space $(m=2,\,n=3)$. Only a particular subset of these parameters (gray region) is associated with the Pareto hypersurface (gray surface in (b)). Note that projecting the Pareto hypersurface down results in the majority of the previously Pareto-optimal points not being optimal any more. However, the pairwise Pareto front can then either be computed through marginalisation or through a new Pareto optimisation directly onto that lower-dimensional objective space.

To compute numerically the Pareto fronts, we use the controlled elitist NSGA-II genetic algorithm~\cite{Deb2008} to perform the optimisation. Unlike, e.g., reinforcement learning approaches for determining Pareto fronts~\cite{Erdman2017}, which often require extensive model tuning and hyperparameter optimization, our genetic algorithms heuristically explore the parameter space without the need for such adjustments.

%% file: appendices/paretoovertime.tex


%% file: apssamp.bbl
\begin{thebibliography}{62}%
\makeatletter
\providecommand \@ifxundefined [1]{%
 \@ifx{#1\undefined}
}%
\providecommand \@ifnum [1]{%
 \ifnum #1\expandafter \@firstoftwo
 \else \expandafter \@secondoftwo
 \fi
}%
\providecommand \@ifx [1]{%
 \ifx #1\expandafter \@firstoftwo
 \else \expandafter \@secondoftwo
 \fi
}%
\providecommand \natexlab [1]{#1}%
\providecommand \enquote  [1]{``#1''}%
\providecommand \bibnamefont  [1]{#1}%
\providecommand \bibfnamefont [1]{#1}%
\providecommand \citenamefont [1]{#1}%
\providecommand \href@noop [0]{\@secondoftwo}%
\providecommand \href [0]{\begingroup \@sanitize@url \@href}%
\providecommand \@href[1]{\@@startlink{#1}\@@href}%
\providecommand \@@href[1]{\endgroup#1\@@endlink}%
\providecommand \@sanitize@url [0]{\catcode `\\12\catcode `\$12\catcode `\&12\catcode `\#12\catcode `\^12\catcode `\_12\catcode `\%12\relax}%
\providecommand \@@startlink[1]{}%
\providecommand \@@endlink[0]{}%
\providecommand \url  [0]{\begingroup\@sanitize@url \@url }%
\providecommand \@url [1]{\endgroup\@href {#1}{\urlprefix }}%
\providecommand \urlprefix  [0]{URL }%
\providecommand \Eprint [0]{\href }%
\providecommand \doibase [0]{https://doi.org/}%
\providecommand \selectlanguage [0]{\@gobble}%
\providecommand \bibinfo  [0]{\@secondoftwo}%
\providecommand \bibfield  [0]{\@secondoftwo}%
\providecommand \translation [1]{[#1]}%
\providecommand \BibitemOpen [0]{}%
\providecommand \bibitemStop [0]{}%
\providecommand \bibitemNoStop [0]{.\EOS\space}%
\providecommand \EOS [0]{\spacefactor3000\relax}%
\providecommand \BibitemShut  [1]{\csname bibitem#1\endcsname}%
\let\auto@bib@innerbib\@empty
\bibitem [{\citenamefont {Benenti}\ \emph {et~al.}(2017)\citenamefont {Benenti}, \citenamefont {Casati}, \citenamefont {Saito},\ and\ \citenamefont {Whitney}}]{Benenti2017Jun}%
  \BibitemOpen
  \bibfield  {author} {\bibinfo {author} {\bibfnamefont {G.}~\bibnamefont {Benenti}}, \bibinfo {author} {\bibfnamefont {G.}~\bibnamefont {Casati}}, \bibinfo {author} {\bibfnamefont {K.}~\bibnamefont {Saito}},\ and\ \bibinfo {author} {\bibfnamefont {R.~S.}\ \bibnamefont {Whitney}},\ }\href {https://doi.org/10.1016/j.physrep.2017.05.008} {\bibfield  {journal} {\bibinfo  {journal} {Phys. Rep.}\ }\textbf {\bibinfo {volume} {694}},\ \bibinfo {pages} {1} (\bibinfo {year} {2017})}\BibitemShut {NoStop}%
\bibitem [{\citenamefont {Binder}\ \emph {et~al.}(2018)\citenamefont {Binder}, \citenamefont {Correa}, \citenamefont {Gogolin}, \citenamefont {Anders},\ and\ \citenamefont {Adesso}}]{QTDBook2018}%
  \BibitemOpen
  \bibinfo {editor} {\bibfnamefont {F.}~\bibnamefont {Binder}}, \bibinfo {editor} {\bibfnamefont {L.~A.}\ \bibnamefont {Correa}}, \bibinfo {editor} {\bibfnamefont {C.}~\bibnamefont {Gogolin}}, \bibinfo {editor} {\bibfnamefont {J.}~\bibnamefont {Anders}},\ and\ \bibinfo {editor} {\bibfnamefont {G.}~\bibnamefont {Adesso}},\ eds.,\ \href {https://link.springer.com/book/10.1007/978-3-319-99046-0} {\emph {\bibinfo {title} {{Thermodynamics in the Quantum Regime}}}}\ (\bibinfo  {publisher} {Springer International Publishing},\ \bibinfo {address} {Cham, Switzerland},\ \bibinfo {year} {2018})\BibitemShut {NoStop}%
\bibitem [{\citenamefont {Kirchberg}\ and\ \citenamefont {Nitzan}(2022)}]{Kirchberg2022}%
  \BibitemOpen
  \bibfield  {author} {\bibinfo {author} {\bibfnamefont {H.}~\bibnamefont {Kirchberg}}\ and\ \bibinfo {author} {\bibfnamefont {A.}~\bibnamefont {Nitzan}},\ }\href {https://doi.org/10.1063/5.0086319} {\bibfield  {journal} {\bibinfo  {journal} {The Journal of Chemical Physics}\ }\textbf {\bibinfo {volume} {156}},\ \bibinfo {pages} {094306} (\bibinfo {year} {2022})}\BibitemShut {NoStop}%
\bibitem [{\citenamefont {Kirchberg}\ and\ \citenamefont {Nitzan}(2023)}]{Kirchberg2023}%
  \BibitemOpen
  \bibfield  {author} {\bibinfo {author} {\bibfnamefont {H.}~\bibnamefont {Kirchberg}}\ and\ \bibinfo {author} {\bibfnamefont {A.}~\bibnamefont {Nitzan}},\ }\bibfield  {journal} {\bibinfo  {journal} {Entropy}\ }\textbf {\bibinfo {volume} {25}},\ \href {https://doi.org/10.3390/e25081218} {10.3390/e25081218} (\bibinfo {year} {2023})\BibitemShut {NoStop}%
\bibitem [{\citenamefont {Campbell}\ \emph {et~al.}(2025)\citenamefont {Campbell}, \citenamefont {D'Amico}, \citenamefont {Ciampini}, \citenamefont {Anders}, \citenamefont {Ares}, \citenamefont {Artini}, \citenamefont {Auff{\ifmmode\grave{e}\else\`{e}\fi}ves}, \citenamefont {Oftelie}, \citenamefont {Bettmann}, \citenamefont {Bonan{\ifmmode\mbox{\c{c}}\else\c{c}\fi}a}, \citenamefont {Busch}, \citenamefont {Campisi}, \citenamefont {Cavalcante}, \citenamefont {Correa}, \citenamefont {Cuestas}, \citenamefont {Dag}, \citenamefont {Dago}, \citenamefont {Deffner}, \citenamefont {Del~Campo}, \citenamefont {Deutschmann-Olek}, \citenamefont {Donadi}, \citenamefont {Doucet}, \citenamefont {Elouard}, \citenamefont {Ensslin}, \citenamefont {Erker}, \citenamefont {Fabbri}, \citenamefont {Fedele}, \citenamefont {Fiusa}, \citenamefont {Fogarty}, \citenamefont {Folk}, \citenamefont {Guarnieri}, \citenamefont {Hegde}, \citenamefont {Hern{\ifmmode\acute{a}\else\'{a}\fi}ndez-G{\ifmmode\acute{o}\else\'{o}\fi}mez}, \citenamefont
  {Hu}, \citenamefont {Iemini}, \citenamefont {Karimi}, \citenamefont {Kiesel}, \citenamefont {Landi}, \citenamefont {Lasek}, \citenamefont {Lemziakov}, \citenamefont {Monaco}, \citenamefont {Lutz}, \citenamefont {Lvov}, \citenamefont {Maillet}, \citenamefont {Mehboudi}, \citenamefont {Mendon{\ifmmode\mbox{\c{c}}\else\c{c}\fi}a}, \citenamefont {Miller}, \citenamefont {Mitchell}, \citenamefont {Mitchison}, \citenamefont {Mukherjee}, \citenamefont {Paternostro}, \citenamefont {Pekola}, \citenamefont {Perarnau-Llobet}, \citenamefont {Poschinger}, \citenamefont {Rolandi}, \citenamefont {Rosa}, \citenamefont {S{\ifmmode\acute{a}\else\'{a}\fi}nchez}, \citenamefont {Santos}, \citenamefont {Sarthour}, \citenamefont {Sela}, \citenamefont {Solfanelli}, \citenamefont {Souza}, \citenamefont {Splettstoesser}, \citenamefont {Tan}, \citenamefont {Tesser}, \citenamefont {Van~Vu}, \citenamefont {Widera}, \citenamefont {Halpern},\ and\ \citenamefont {Zawadzki}}]{Campbell2025Apr}%
  \BibitemOpen
  \bibfield  {author} {\bibinfo {author} {\bibfnamefont {S.}~\bibnamefont {Campbell}}, \bibinfo {author} {\bibfnamefont {I.}~\bibnamefont {D'Amico}}, \bibinfo {author} {\bibfnamefont {M.~A.}\ \bibnamefont {Ciampini}}, \bibinfo {author} {\bibfnamefont {J.}~\bibnamefont {Anders}}, \bibinfo {author} {\bibfnamefont {N.}~\bibnamefont {Ares}}, \bibinfo {author} {\bibfnamefont {S.}~\bibnamefont {Artini}}, \bibinfo {author} {\bibfnamefont {A.}~\bibnamefont {Auff{\ifmmode\grave{e}\else\`{e}\fi}ves}}, \bibinfo {author} {\bibfnamefont {L.~B.}\ \bibnamefont {Oftelie}}, \bibinfo {author} {\bibfnamefont {L.~P.}\ \bibnamefont {Bettmann}}, \bibinfo {author} {\bibfnamefont {M.~V.~S.}\ \bibnamefont {Bonan{\ifmmode\mbox{\c{c}}\else\c{c}\fi}a}}, \bibinfo {author} {\bibfnamefont {T.}~\bibnamefont {Busch}}, \bibinfo {author} {\bibfnamefont {M.}~\bibnamefont {Campisi}}, \bibinfo {author} {\bibfnamefont {M.~F.}\ \bibnamefont {Cavalcante}}, \bibinfo {author} {\bibfnamefont {L.~A.}\ \bibnamefont {Correa}}, \bibinfo {author}
  {\bibfnamefont {E.}~\bibnamefont {Cuestas}}, \bibinfo {author} {\bibfnamefont {C.~B.}\ \bibnamefont {Dag}}, \bibinfo {author} {\bibfnamefont {S.}~\bibnamefont {Dago}}, \bibinfo {author} {\bibfnamefont {S.}~\bibnamefont {Deffner}}, \bibinfo {author} {\bibfnamefont {A.}~\bibnamefont {Del~Campo}}, \bibinfo {author} {\bibfnamefont {A.}~\bibnamefont {Deutschmann-Olek}}, \bibinfo {author} {\bibfnamefont {S.}~\bibnamefont {Donadi}}, \bibinfo {author} {\bibfnamefont {E.}~\bibnamefont {Doucet}}, \bibinfo {author} {\bibfnamefont {C.}~\bibnamefont {Elouard}}, \bibinfo {author} {\bibfnamefont {K.}~\bibnamefont {Ensslin}}, \bibinfo {author} {\bibfnamefont {P.}~\bibnamefont {Erker}}, \bibinfo {author} {\bibfnamefont {N.}~\bibnamefont {Fabbri}}, \bibinfo {author} {\bibfnamefont {F.}~\bibnamefont {Fedele}}, \bibinfo {author} {\bibfnamefont {G.}~\bibnamefont {Fiusa}}, \bibinfo {author} {\bibfnamefont {T.}~\bibnamefont {Fogarty}}, \bibinfo {author} {\bibfnamefont {J.}~\bibnamefont {Folk}}, \bibinfo {author} {\bibfnamefont
  {G.}~\bibnamefont {Guarnieri}}, \bibinfo {author} {\bibfnamefont {A.~S.}\ \bibnamefont {Hegde}}, \bibinfo {author} {\bibfnamefont {S.}~\bibnamefont {Hern{\ifmmode\acute{a}\else\'{a}\fi}ndez-G{\ifmmode\acute{o}\else\'{o}\fi}mez}}, \bibinfo {author} {\bibfnamefont {C.-K.}\ \bibnamefont {Hu}}, \bibinfo {author} {\bibfnamefont {F.}~\bibnamefont {Iemini}}, \bibinfo {author} {\bibfnamefont {B.}~\bibnamefont {Karimi}}, \bibinfo {author} {\bibfnamefont {N.}~\bibnamefont {Kiesel}}, \bibinfo {author} {\bibfnamefont {G.~T.}\ \bibnamefont {Landi}}, \bibinfo {author} {\bibfnamefont {A.}~\bibnamefont {Lasek}}, \bibinfo {author} {\bibfnamefont {S.}~\bibnamefont {Lemziakov}}, \bibinfo {author} {\bibfnamefont {G.~L.}\ \bibnamefont {Monaco}}, \bibinfo {author} {\bibfnamefont {E.}~\bibnamefont {Lutz}}, \bibinfo {author} {\bibfnamefont {D.}~\bibnamefont {Lvov}}, \bibinfo {author} {\bibfnamefont {O.}~\bibnamefont {Maillet}}, \bibinfo {author} {\bibfnamefont {M.}~\bibnamefont {Mehboudi}}, \bibinfo {author} {\bibfnamefont
  {T.~M.}\ \bibnamefont {Mendon{\ifmmode\mbox{\c{c}}\else\c{c}\fi}a}}, \bibinfo {author} {\bibfnamefont {H.~J.~D.}\ \bibnamefont {Miller}}, \bibinfo {author} {\bibfnamefont {A.~K.}\ \bibnamefont {Mitchell}}, \bibinfo {author} {\bibfnamefont {M.~T.}\ \bibnamefont {Mitchison}}, \bibinfo {author} {\bibfnamefont {V.}~\bibnamefont {Mukherjee}}, \bibinfo {author} {\bibfnamefont {M.}~\bibnamefont {Paternostro}}, \bibinfo {author} {\bibfnamefont {J.}~\bibnamefont {Pekola}}, \bibinfo {author} {\bibfnamefont {M.}~\bibnamefont {Perarnau-Llobet}}, \bibinfo {author} {\bibfnamefont {U.}~\bibnamefont {Poschinger}}, \bibinfo {author} {\bibfnamefont {A.}~\bibnamefont {Rolandi}}, \bibinfo {author} {\bibfnamefont {D.}~\bibnamefont {Rosa}}, \bibinfo {author} {\bibfnamefont {R.}~\bibnamefont {S{\ifmmode\acute{a}\else\'{a}\fi}nchez}}, \bibinfo {author} {\bibfnamefont {A.~C.}\ \bibnamefont {Santos}}, \bibinfo {author} {\bibfnamefont {R.~S.}\ \bibnamefont {Sarthour}}, \bibinfo {author} {\bibfnamefont {E.}~\bibnamefont {Sela}},
  \bibinfo {author} {\bibfnamefont {A.}~\bibnamefont {Solfanelli}}, \bibinfo {author} {\bibfnamefont {A.~M.}\ \bibnamefont {Souza}}, \bibinfo {author} {\bibfnamefont {J.}~\bibnamefont {Splettstoesser}}, \bibinfo {author} {\bibfnamefont {D.}~\bibnamefont {Tan}}, \bibinfo {author} {\bibfnamefont {L.}~\bibnamefont {Tesser}}, \bibinfo {author} {\bibfnamefont {T.}~\bibnamefont {Van~Vu}}, \bibinfo {author} {\bibfnamefont {A.}~\bibnamefont {Widera}}, \bibinfo {author} {\bibfnamefont {N.~Y.}\ \bibnamefont {Halpern}},\ and\ \bibinfo {author} {\bibfnamefont {K.}~\bibnamefont {Zawadzki}},\ }\bibfield  {journal} {\bibinfo  {journal} {arXiv}\ }\href {https://doi.org/10.48550/arXiv.2504.20145} {10.48550/arXiv.2504.20145} (\bibinfo {year} {2025}),\ \Eprint {https://arxiv.org/abs/2504.20145} {2504.20145} \BibitemShut {NoStop}%
\bibitem [{\citenamefont {Maruyama}\ \emph {et~al.}(2009)\citenamefont {Maruyama}, \citenamefont {Nori},\ and\ \citenamefont {Vedral}}]{Maruyama2009Jan}%
  \BibitemOpen
  \bibfield  {author} {\bibinfo {author} {\bibfnamefont {K.}~\bibnamefont {Maruyama}}, \bibinfo {author} {\bibfnamefont {F.}~\bibnamefont {Nori}},\ and\ \bibinfo {author} {\bibfnamefont {V.}~\bibnamefont {Vedral}},\ }\href {https://doi.org/10.1103/RevModPhys.81.1} {\bibfield  {journal} {\bibinfo  {journal} {Rev. Mod. Phys.}\ }\textbf {\bibinfo {volume} {81}},\ \bibinfo {pages} {1} (\bibinfo {year} {2009})}\BibitemShut {NoStop}%
\bibitem [{\citenamefont {Sagawa}\ and\ \citenamefont {Ueda}(2010)}]{Sagawa2010}%
  \BibitemOpen
  \bibfield  {author} {\bibinfo {author} {\bibfnamefont {T.}~\bibnamefont {Sagawa}}\ and\ \bibinfo {author} {\bibfnamefont {M.}~\bibnamefont {Ueda}},\ }\href {https://doi.org/10.1103/PhysRevLett.104.090602} {\bibfield  {journal} {\bibinfo  {journal} {Phys. Rev. Lett.}\ }\textbf {\bibinfo {volume} {104}},\ \bibinfo {pages} {090602} (\bibinfo {year} {2010})}\BibitemShut {NoStop}%
\bibitem [{\citenamefont {Mandal}\ \emph {et~al.}(2013)\citenamefont {Mandal}, \citenamefont {Quan},\ and\ \citenamefont {Jarzynski}}]{Mandal2013}%
  \BibitemOpen
  \bibfield  {author} {\bibinfo {author} {\bibfnamefont {D.}~\bibnamefont {Mandal}}, \bibinfo {author} {\bibfnamefont {H.~T.}\ \bibnamefont {Quan}},\ and\ \bibinfo {author} {\bibfnamefont {C.}~\bibnamefont {Jarzynski}},\ }\href {https://doi.org/10.1103/PhysRevLett.111.030602} {\bibfield  {journal} {\bibinfo  {journal} {Phys. Rev. Lett.}\ }\textbf {\bibinfo {volume} {111}},\ \bibinfo {pages} {030602} (\bibinfo {year} {2013})}\BibitemShut {NoStop}%
\bibitem [{\citenamefont {Szilard}(1929)}]{Szilard1929}%
  \BibitemOpen
  \bibfield  {author} {\bibinfo {author} {\bibfnamefont {L.}~\bibnamefont {Szilard}},\ }\bibfield  {journal} {\bibinfo  {journal} {Z. Physik}\ }\href {https://doi.org/10.1007/BF0134128} {10.1007/BF0134128} (\bibinfo {year} {1929})\BibitemShut {NoStop}%
\bibitem [{\citenamefont {Bennett}(1982)}]{Bennett1982}%
  \BibitemOpen
  \bibfield  {author} {\bibinfo {author} {\bibfnamefont {C.}~\bibnamefont {Bennett}},\ }\bibfield  {journal} {\bibinfo  {journal} {Int J Theor Phys}\ }\href {https://doi.org/10.1007/BF02084158} {10.1007/BF02084158} (\bibinfo {year} {1982})\BibitemShut {NoStop}%
\bibitem [{\citenamefont {de~Oliveira~Junior}\ \emph {et~al.}(2025)\citenamefont {de~Oliveira~Junior}, \citenamefont {Brask},\ and\ \citenamefont {Chaves}}]{junior2025}%
  \BibitemOpen
  \bibfield  {author} {\bibinfo {author} {\bibfnamefont {A.}~\bibnamefont {de~Oliveira~Junior}}, \bibinfo {author} {\bibfnamefont {J.~B.}\ \bibnamefont {Brask}},\ and\ \bibinfo {author} {\bibfnamefont {R.}~\bibnamefont {Chaves}},\ }\href {https://arxiv.org/abs/2503.07740} {\bibinfo {title} {A friendly guide to exorcising maxwell's demon}} (\bibinfo {year} {2025}),\ \Eprint {https://arxiv.org/abs/2503.07740} {arXiv:2503.07740 [quant-ph]} \BibitemShut {NoStop}%
\bibitem [{\citenamefont {Bresque}\ \emph {et~al.}(2021)\citenamefont {Bresque}, \citenamefont {Camati}, \citenamefont {Rogers}, \citenamefont {Murch}, \citenamefont {Jordan},\ and\ \citenamefont {Auff\`eves}}]{bresque2021}%
  \BibitemOpen
  \bibfield  {author} {\bibinfo {author} {\bibfnamefont {L.}~\bibnamefont {Bresque}}, \bibinfo {author} {\bibfnamefont {P.~A.}\ \bibnamefont {Camati}}, \bibinfo {author} {\bibfnamefont {S.}~\bibnamefont {Rogers}}, \bibinfo {author} {\bibfnamefont {K.}~\bibnamefont {Murch}}, \bibinfo {author} {\bibfnamefont {A.~N.}\ \bibnamefont {Jordan}},\ and\ \bibinfo {author} {\bibfnamefont {A.}~\bibnamefont {Auff\`eves}},\ }\href {https://doi.org/10.1103/PhysRevLett.126.120605} {\bibfield  {journal} {\bibinfo  {journal} {Phys. Rev. Lett.}\ }\textbf {\bibinfo {volume} {126}},\ \bibinfo {pages} {120605} (\bibinfo {year} {2021})}\BibitemShut {NoStop}%
\bibitem [{\citenamefont {Fadler}\ \emph {et~al.}(2023)\citenamefont {Fadler}, \citenamefont {Friedenberger},\ and\ \citenamefont {Lutz}}]{Fadler2023}%
  \BibitemOpen
  \bibfield  {author} {\bibinfo {author} {\bibfnamefont {P.}~\bibnamefont {Fadler}}, \bibinfo {author} {\bibfnamefont {A.}~\bibnamefont {Friedenberger}},\ and\ \bibinfo {author} {\bibfnamefont {E.}~\bibnamefont {Lutz}},\ }\href {https://doi.org/10.1103/PhysRevLett.130.240401} {\bibfield  {journal} {\bibinfo  {journal} {Phys. Rev. Lett.}\ }\textbf {\bibinfo {volume} {130}},\ \bibinfo {pages} {240401} (\bibinfo {year} {2023})}\BibitemShut {NoStop}%
\bibitem [{\citenamefont {S\'anchez}\ \emph {et~al.}(2019)\citenamefont {S\'anchez}, \citenamefont {Splettstoesser},\ and\ \citenamefont {Whitney}}]{Sanchez2019}%
  \BibitemOpen
  \bibfield  {author} {\bibinfo {author} {\bibfnamefont {R.}~\bibnamefont {S\'anchez}}, \bibinfo {author} {\bibfnamefont {J.}~\bibnamefont {Splettstoesser}},\ and\ \bibinfo {author} {\bibfnamefont {R.~S.}\ \bibnamefont {Whitney}},\ }\href {https://doi.org/10.1103/PhysRevLett.123.216801} {\bibfield  {journal} {\bibinfo  {journal} {Phys. Rev. Lett.}\ }\textbf {\bibinfo {volume} {123}},\ \bibinfo {pages} {216801} (\bibinfo {year} {2019})}\BibitemShut {NoStop}%
\bibitem [{\citenamefont {Raizen}(2009)}]{Raizen2009}%
  \BibitemOpen
  \bibfield  {author} {\bibinfo {author} {\bibfnamefont {M.~G.}\ \bibnamefont {Raizen}},\ }\href {https://doi.org/10.1126/science.1171506} {\bibfield  {journal} {\bibinfo  {journal} {Science}\ }\textbf {\bibinfo {volume} {324}},\ \bibinfo {pages} {1403} (\bibinfo {year} {2009})}\BibitemShut {NoStop}%
\bibitem [{\citenamefont {Toyabe}\ \emph {et~al.}(2010)\citenamefont {Toyabe}, \citenamefont {Sagawa}, \citenamefont {Ueda}, \citenamefont {Muneyuki},\ and\ \citenamefont {Sano}}]{Toyabe2010}%
  \BibitemOpen
  \bibfield  {author} {\bibinfo {author} {\bibfnamefont {S.}~\bibnamefont {Toyabe}}, \bibinfo {author} {\bibfnamefont {T.}~\bibnamefont {Sagawa}}, \bibinfo {author} {\bibfnamefont {M.}~\bibnamefont {Ueda}}, \bibinfo {author} {\bibfnamefont {E.}~\bibnamefont {Muneyuki}},\ and\ \bibinfo {author} {\bibfnamefont {M.}~\bibnamefont {Sano}},\ }\href {https://doi.org/10.1038/nphys1821} {\bibfield  {journal} {\bibinfo  {journal} {Nature Phys}\ }\textbf {\bibinfo {volume} {6}},\ \bibinfo {pages} {988–992} (\bibinfo {year} {2010})}\BibitemShut {NoStop}%
\bibitem [{\citenamefont {Koski}\ \emph {et~al.}(2014)\citenamefont {Koski}, \citenamefont {Maisi}, \citenamefont {Pekola},\ and\ \citenamefont {Averin}}]{Koski2014}%
  \BibitemOpen
  \bibfield  {author} {\bibinfo {author} {\bibfnamefont {J.~V.}\ \bibnamefont {Koski}}, \bibinfo {author} {\bibfnamefont {V.~F.}\ \bibnamefont {Maisi}}, \bibinfo {author} {\bibfnamefont {J.~P.}\ \bibnamefont {Pekola}},\ and\ \bibinfo {author} {\bibfnamefont {D.~V.}\ \bibnamefont {Averin}},\ }\href {https://doi.org/10.1073/pnas.1406966111} {\bibfield  {journal} {\bibinfo  {journal} {Proceedings of the National Academy of Sciences}\ }\textbf {\bibinfo {volume} {111}},\ \bibinfo {pages} {13786} (\bibinfo {year} {2014})}\BibitemShut {NoStop}%
\bibitem [{\citenamefont {Pruchyathamkorn}\ \emph {et~al.}(2024)\citenamefont {Pruchyathamkorn}, \citenamefont {Nguyen}, \citenamefont {Grommet}, \citenamefont {Novoveska}, \citenamefont {Ronson}, \citenamefont {Thoburn},\ and\ \citenamefont {Nitschke}}]{Pruchyathamkorn2024}%
  \BibitemOpen
  \bibfield  {author} {\bibinfo {author} {\bibfnamefont {J.}~\bibnamefont {Pruchyathamkorn}}, \bibinfo {author} {\bibfnamefont {B.-N.~T.}\ \bibnamefont {Nguyen}}, \bibinfo {author} {\bibfnamefont {A.~B.}\ \bibnamefont {Grommet}}, \bibinfo {author} {\bibfnamefont {M.}~\bibnamefont {Novoveska}}, \bibinfo {author} {\bibfnamefont {T.~K.}\ \bibnamefont {Ronson}}, \bibinfo {author} {\bibfnamefont {J.~D.}\ \bibnamefont {Thoburn}},\ and\ \bibinfo {author} {\bibfnamefont {J.~R.}\ \bibnamefont {Nitschke}},\ }\href {https://doi.org/10.1038/s41557-024-01549-2} {\bibfield  {journal} {\bibinfo  {journal} {Nat. Chem.}\ ,\ \bibinfo {pages} {1558–1564}} (\bibinfo {year} {2024})}\BibitemShut {NoStop}%
\bibitem [{\citenamefont {Paneru}\ \emph {et~al.}(2018)\citenamefont {Paneru}, \citenamefont {Lee}, \citenamefont {Tlusty},\ and\ \citenamefont {Pak}}]{Paneru2018Jan}%
  \BibitemOpen
  \bibfield  {author} {\bibinfo {author} {\bibfnamefont {G.}~\bibnamefont {Paneru}}, \bibinfo {author} {\bibfnamefont {D.~Y.}\ \bibnamefont {Lee}}, \bibinfo {author} {\bibfnamefont {T.}~\bibnamefont {Tlusty}},\ and\ \bibinfo {author} {\bibfnamefont {H.~K.}\ \bibnamefont {Pak}},\ }\href {https://doi.org/10.1103/PhysRevLett.120.020601} {\bibfield  {journal} {\bibinfo  {journal} {Phys. Rev. Lett.}\ }\textbf {\bibinfo {volume} {120}},\ \bibinfo {pages} {020601} (\bibinfo {year} {2018})}\BibitemShut {NoStop}%
\bibitem [{\citenamefont {Monsel}\ \emph {et~al.}(2025)\citenamefont {Monsel}, \citenamefont {Acciai}, \citenamefont {S{\ifmmode\acute{a}\else\'{a}\fi}nchez},\ and\ \citenamefont {Splettstoesser}}]{Monsel2025Jan}%
  \BibitemOpen
  \bibfield  {author} {\bibinfo {author} {\bibfnamefont {J.}~\bibnamefont {Monsel}}, \bibinfo {author} {\bibfnamefont {M.}~\bibnamefont {Acciai}}, \bibinfo {author} {\bibfnamefont {R.}~\bibnamefont {S{\ifmmode\acute{a}\else\'{a}\fi}nchez}},\ and\ \bibinfo {author} {\bibfnamefont {J.}~\bibnamefont {Splettstoesser}},\ }\href {https://doi.org/10.1103/PhysRevB.111.045419} {\bibfield  {journal} {\bibinfo  {journal} {Phys. Rev. B}\ }\textbf {\bibinfo {volume} {111}},\ \bibinfo {pages} {045419} (\bibinfo {year} {2025})}\BibitemShut {NoStop}%
\bibitem [{\citenamefont {Sagawa}\ and\ \citenamefont {Ueda}(2012)}]{Sagawa2012Nov}%
  \BibitemOpen
  \bibfield  {author} {\bibinfo {author} {\bibfnamefont {T.}~\bibnamefont {Sagawa}}\ and\ \bibinfo {author} {\bibfnamefont {M.}~\bibnamefont {Ueda}},\ }\href {https://doi.org/10.1103/PhysRevLett.109.180602} {\bibfield  {journal} {\bibinfo  {journal} {Phys. Rev. Lett.}\ }\textbf {\bibinfo {volume} {109}},\ \bibinfo {pages} {180602} (\bibinfo {year} {2012})}\BibitemShut {NoStop}%
\bibitem [{\citenamefont {Potts}\ and\ \citenamefont {Samuelsson}(2018)}]{Potts2018Nov}%
  \BibitemOpen
  \bibfield  {author} {\bibinfo {author} {\bibfnamefont {P.~P.}\ \bibnamefont {Potts}}\ and\ \bibinfo {author} {\bibfnamefont {P.}~\bibnamefont {Samuelsson}},\ }\href {https://doi.org/10.1103/PhysRevLett.121.210603} {\bibfield  {journal} {\bibinfo  {journal} {Phys. Rev. Lett.}\ }\textbf {\bibinfo {volume} {121}},\ \bibinfo {pages} {210603} (\bibinfo {year} {2018})}\BibitemShut {NoStop}%
\bibitem [{\citenamefont {Paneru}\ \emph {et~al.}(2020)\citenamefont {Paneru}, \citenamefont {Dutta}, \citenamefont {Sagawa}, \citenamefont {Tlusty},\ and\ \citenamefont {Pak}}]{Paneru2020Feb}%
  \BibitemOpen
  \bibfield  {author} {\bibinfo {author} {\bibfnamefont {G.}~\bibnamefont {Paneru}}, \bibinfo {author} {\bibfnamefont {S.}~\bibnamefont {Dutta}}, \bibinfo {author} {\bibfnamefont {T.}~\bibnamefont {Sagawa}}, \bibinfo {author} {\bibfnamefont {T.}~\bibnamefont {Tlusty}},\ and\ \bibinfo {author} {\bibfnamefont {H.~K.}\ \bibnamefont {Pak}},\ }\href {https://doi.org/10.1038/s41467-020-14823-x} {\bibfield  {journal} {\bibinfo  {journal} {Nat. Commun.}\ }\textbf {\bibinfo {volume} {11}},\ \bibinfo {pages} {1} (\bibinfo {year} {2020})}\BibitemShut {NoStop}%
\bibitem [{\citenamefont {Bu\ss{}hardt}\ and\ \citenamefont {Freyberger}(2010)}]{busshard2010}%
  \BibitemOpen
  \bibfield  {author} {\bibinfo {author} {\bibfnamefont {M.}~\bibnamefont {Bu\ss{}hardt}}\ and\ \bibinfo {author} {\bibfnamefont {M.}~\bibnamefont {Freyberger}},\ }\href {https://doi.org/10.1103/PhysRevA.82.042117} {\bibfield  {journal} {\bibinfo  {journal} {Phys. Rev. A}\ }\textbf {\bibinfo {volume} {82}},\ \bibinfo {pages} {042117} (\bibinfo {year} {2010})}\BibitemShut {NoStop}%
\bibitem [{\citenamefont {Kirchberg}\ and\ \citenamefont {Nitzan}(2025)}]{Kirchberg2025}%
  \BibitemOpen
  \bibfield  {author} {\bibinfo {author} {\bibfnamefont {H.}~\bibnamefont {Kirchberg}}\ and\ \bibinfo {author} {\bibfnamefont {A.}~\bibnamefont {Nitzan}},\ }\href {https://arxiv.org/abs/2505.00686} {\bibinfo {title} {Quantum information engines: Bounds on performance metrics by measurement time}} (\bibinfo {year} {2025}),\ \Eprint {https://arxiv.org/abs/2505.00686} {arXiv:2505.00686 [quant-ph]} \BibitemShut {NoStop}%
\bibitem [{\citenamefont {Von~Neumann}(1955)}]{VonNeumannBook}%
  \BibitemOpen
  \bibfield  {author} {\bibinfo {author} {\bibfnamefont {J.}~\bibnamefont {Von~Neumann}},\ }\href@noop {} {\emph {\bibinfo {title} {Mathematical Foundations of Quantum Mechanics}}}\ (\bibinfo  {publisher} {Princeton Univ. Press.},\ \bibinfo {address} {Princeton, NJ},\ \bibinfo {year} {1955})\BibitemShut {NoStop}%
\bibitem [{\citenamefont {Korbicz}\ \emph {et~al.}(2017)\citenamefont {Korbicz}, \citenamefont {Aguilar}, \citenamefont {\ifmmode \acute{C}\else \'{C}\fi{}wikli\ifmmode~\acute{n}\else \'{n}\fi{}ski},\ and\ \citenamefont {Horodecki}}]{Korbicz2017}%
  \BibitemOpen
  \bibfield  {author} {\bibinfo {author} {\bibfnamefont {J.~K.}\ \bibnamefont {Korbicz}}, \bibinfo {author} {\bibfnamefont {E.~A.}\ \bibnamefont {Aguilar}}, \bibinfo {author} {\bibfnamefont {P.}~\bibnamefont {\ifmmode \acute{C}\else \'{C}\fi{}wikli\ifmmode~\acute{n}\else \'{n}\fi{}ski}},\ and\ \bibinfo {author} {\bibfnamefont {P.}~\bibnamefont {Horodecki}},\ }\href {https://doi.org/10.1103/PhysRevA.96.032124} {\bibfield  {journal} {\bibinfo  {journal} {Phys. Rev. A}\ }\textbf {\bibinfo {volume} {96}},\ \bibinfo {pages} {032124} (\bibinfo {year} {2017})}\BibitemShut {NoStop}%
\bibitem [{\citenamefont {Arcizet}\ \emph {et~al.}(2011)\citenamefont {Arcizet}, \citenamefont {Jacques}, \citenamefont {Siria}, \citenamefont {Poncharal}, \citenamefont {Vincent},\ and\ \citenamefont {Seidelin}}]{Arcizet2011}%
  \BibitemOpen
  \bibfield  {author} {\bibinfo {author} {\bibfnamefont {O.}~\bibnamefont {Arcizet}}, \bibinfo {author} {\bibfnamefont {V.}~\bibnamefont {Jacques}}, \bibinfo {author} {\bibfnamefont {A.}~\bibnamefont {Siria}}, \bibinfo {author} {\bibfnamefont {P.}~\bibnamefont {Poncharal}}, \bibinfo {author} {\bibfnamefont {P.}~\bibnamefont {Vincent}},\ and\ \bibinfo {author} {\bibfnamefont {S.}~\bibnamefont {Seidelin}},\ }\href {https://doi.org/10.1038/nphys2070} {\bibfield  {journal} {\bibinfo  {journal} {Nature Phys}\ }\textbf {\bibinfo {volume} {7}},\ \bibinfo {pages} {879–883} (\bibinfo {year} {2011})}\BibitemShut {NoStop}%
\bibitem [{\citenamefont {Blais}\ \emph {et~al.}(2004)\citenamefont {Blais}, \citenamefont {Huang}, \citenamefont {Wallraff}, \citenamefont {Girvin},\ and\ \citenamefont {Schoelkopf}}]{Blais2004}%
  \BibitemOpen
  \bibfield  {author} {\bibinfo {author} {\bibfnamefont {A.}~\bibnamefont {Blais}}, \bibinfo {author} {\bibfnamefont {R.-S.}\ \bibnamefont {Huang}}, \bibinfo {author} {\bibfnamefont {A.}~\bibnamefont {Wallraff}}, \bibinfo {author} {\bibfnamefont {S.~M.}\ \bibnamefont {Girvin}},\ and\ \bibinfo {author} {\bibfnamefont {R.~J.}\ \bibnamefont {Schoelkopf}},\ }\href {https://doi.org/10.1103/PhysRevA.69.062320} {\bibfield  {journal} {\bibinfo  {journal} {Phys. Rev. A}\ }\textbf {\bibinfo {volume} {69}},\ \bibinfo {pages} {062320} (\bibinfo {year} {2004})}\BibitemShut {NoStop}%
\bibitem [{\citenamefont {Cottet}\ and\ \citenamefont {Huard}(2018)}]{Cottet2018}%
  \BibitemOpen
  \bibfield  {author} {\bibinfo {author} {\bibfnamefont {N.}~\bibnamefont {Cottet}}\ and\ \bibinfo {author} {\bibfnamefont {B.}~\bibnamefont {Huard}},\ }\bibinfo {title} {Maxwell's demon in superconducting circuits},\ in\ \href {https://doi.org/10.1007/978-3-319-99046-0_40} {\emph {\bibinfo {booktitle} {Thermodynamics in the Quantum Regime: Fundamental Aspects and New Directions}}},\ \bibinfo {editor} {edited by\ \bibinfo {editor} {\bibfnamefont {F.}~\bibnamefont {Binder}}, \bibinfo {editor} {\bibfnamefont {L.~A.}\ \bibnamefont {Correa}}, \bibinfo {editor} {\bibfnamefont {C.}~\bibnamefont {Gogolin}}, \bibinfo {editor} {\bibfnamefont {J.}~\bibnamefont {Anders}},\ and\ \bibinfo {editor} {\bibfnamefont {G.}~\bibnamefont {Adesso}}}\ (\bibinfo  {publisher} {Springer International Publishing},\ \bibinfo {address} {Cham},\ \bibinfo {year} {2018})\ pp.\ \bibinfo {pages} {959--981}\BibitemShut {NoStop}%
\bibitem [{\citenamefont {Emmerich}\ and\ \citenamefont {Deutz}(2018)}]{Emmerich2018}%
  \BibitemOpen
  \bibfield  {author} {\bibinfo {author} {\bibfnamefont {M.~T.~M.}\ \bibnamefont {Emmerich}}\ and\ \bibinfo {author} {\bibfnamefont {A.~H.}\ \bibnamefont {Deutz}},\ }\href {https://doi.org/10.1007/s11047-018-9685-y} {\bibfield  {journal} {\bibinfo  {journal} {Natural Computing}\ }\textbf {\bibinfo {volume} {17}},\ \bibinfo {pages} {585} (\bibinfo {year} {2018})}\BibitemShut {NoStop}%
\bibitem [{\citenamefont {Fudenberg}\ and\ \citenamefont {Tirole}(1991)}]{Fudenberg1991-jb}%
  \BibitemOpen
  \bibfield  {author} {\bibinfo {author} {\bibfnamefont {D.}~\bibnamefont {Fudenberg}}\ and\ \bibinfo {author} {\bibfnamefont {J.}~\bibnamefont {Tirole}},\ }\href@noop {} {\emph {\bibinfo {title} {Game Theory}}},\ The MIT Press\ (\bibinfo  {publisher} {MIT Press},\ \bibinfo {address} {London, England},\ \bibinfo {year} {1991})\BibitemShut {NoStop}%
\bibitem [{\citenamefont {Tomoiag{\u a}}\ \emph {et~al.}(2013)\citenamefont {Tomoiag{\u a}}, \citenamefont {Chindri{\c s}}, \citenamefont {Sumper}, \citenamefont {Sudria-Andreu},\ and\ \citenamefont {Villafafila-Robles}}]{Tomoiaga2013-bi}%
  \BibitemOpen
  \bibfield  {author} {\bibinfo {author} {\bibfnamefont {B.}~\bibnamefont {Tomoiag{\u a}}}, \bibinfo {author} {\bibfnamefont {M.}~\bibnamefont {Chindri{\c s}}}, \bibinfo {author} {\bibfnamefont {A.}~\bibnamefont {Sumper}}, \bibinfo {author} {\bibfnamefont {A.}~\bibnamefont {Sudria-Andreu}},\ and\ \bibinfo {author} {\bibfnamefont {R.}~\bibnamefont {Villafafila-Robles}},\ }\href@noop {} {\bibfield  {journal} {\bibinfo  {journal} {Energies}\ }\textbf {\bibinfo {volume} {6}},\ \bibinfo {pages} {1439} (\bibinfo {year} {2013})}\BibitemShut {NoStop}%
\bibitem [{\citenamefont {Berx}\ and\ \citenamefont {Proesmans}(2024{\natexlab{a}})}]{berx_2024_2}%
  \BibitemOpen
  \bibfield  {author} {\bibinfo {author} {\bibfnamefont {J.}~\bibnamefont {Berx}}\ and\ \bibinfo {author} {\bibfnamefont {K.}~\bibnamefont {Proesmans}},\ }\href {https://doi.org/10.1209/0295-5075/ad2d14} {\bibfield  {journal} {\bibinfo  {journal} {Europhysics Letters}\ }\textbf {\bibinfo {volume} {145}},\ \bibinfo {pages} {51001} (\bibinfo {year} {2024}{\natexlab{a}})}\BibitemShut {NoStop}%
\bibitem [{\citenamefont {Berx}\ and\ \citenamefont {Proesmans}(2024{\natexlab{b}})}]{berx2024}%
  \BibitemOpen
  \bibfield  {author} {\bibinfo {author} {\bibfnamefont {J.}~\bibnamefont {Berx}}\ and\ \bibinfo {author} {\bibfnamefont {K.}~\bibnamefont {Proesmans}},\ }\href {https://doi.org/10.1098/rsif.2024.0232} {\bibfield  {journal} {\bibinfo  {journal} {Journal of The Royal Society Interface}\ }\textbf {\bibinfo {volume} {21}},\ \bibinfo {pages} {20240232} (\bibinfo {year} {2024}{\natexlab{b}})}\BibitemShut {NoStop}%
\bibitem [{\citenamefont {Forão}\ \emph {et~al.}(2025)\citenamefont {Forão}, \citenamefont {Berx},\ and\ \citenamefont {Fiore}}]{forao2025}%
  \BibitemOpen
  \bibfield  {author} {\bibinfo {author} {\bibfnamefont {G.~A.~L.}\ \bibnamefont {Forão}}, \bibinfo {author} {\bibfnamefont {J.}~\bibnamefont {Berx}},\ and\ \bibinfo {author} {\bibfnamefont {C.~E.}\ \bibnamefont {Fiore}},\ }\href {https://arxiv.org/abs/2504.21717} {\bibinfo {title} {Characterization and optimization of heat engines: Pareto-optimal fronts and universal features}} (\bibinfo {year} {2025}),\ \Eprint {https://arxiv.org/abs/2504.21717} {arXiv:2504.21717 [cond-mat.stat-mech]} \BibitemShut {NoStop}%
\bibitem [{\citenamefont {Sheftel}\ \emph {et~al.}(2013)\citenamefont {Sheftel}, \citenamefont {Shoval}, \citenamefont {Mayo},\ and\ \citenamefont {Alon}}]{Sheftel2013}%
  \BibitemOpen
  \bibfield  {author} {\bibinfo {author} {\bibfnamefont {H.}~\bibnamefont {Sheftel}}, \bibinfo {author} {\bibfnamefont {O.}~\bibnamefont {Shoval}}, \bibinfo {author} {\bibfnamefont {A.}~\bibnamefont {Mayo}},\ and\ \bibinfo {author} {\bibfnamefont {U.}~\bibnamefont {Alon}},\ }\href {https://doi.org/https://doi.org/10.1002/ece3.528} {\bibfield  {journal} {\bibinfo  {journal} {Ecology and Evolution}\ }\textbf {\bibinfo {volume} {3}},\ \bibinfo {pages} {1471} (\bibinfo {year} {2013})}\BibitemShut {NoStop}%
\bibitem [{\citenamefont {Berx}\ \emph {et~al.}(2025)\citenamefont {Berx}, \citenamefont {Singh},\ and\ \citenamefont {Proesmans}}]{Berx_2025}%
  \BibitemOpen
  \bibfield  {author} {\bibinfo {author} {\bibfnamefont {J.}~\bibnamefont {Berx}}, \bibinfo {author} {\bibfnamefont {P.}~\bibnamefont {Singh}},\ and\ \bibinfo {author} {\bibfnamefont {K.}~\bibnamefont {Proesmans}},\ }\href {https://doi.org/10.1088/1367-2630/adb7fc} {\bibfield  {journal} {\bibinfo  {journal} {New Journal of Physics}\ }\textbf {\bibinfo {volume} {27}},\ \bibinfo {pages} {023034} (\bibinfo {year} {2025})}\BibitemShut {NoStop}%
\bibitem [{\citenamefont {Taranto}\ \emph {et~al.}(2023)\citenamefont {Taranto}, \citenamefont {Bakhshinezhad}, \citenamefont {Bluhm}, \citenamefont {Silva}, \citenamefont {Friis}, \citenamefont {Lock}, \citenamefont {Vitagliano}, \citenamefont {Binder}, \citenamefont {Debarba}, \citenamefont {Schwarzhans}, \citenamefont {Clivaz},\ and\ \citenamefont {Huber}}]{Taranto2023}%
  \BibitemOpen
  \bibfield  {author} {\bibinfo {author} {\bibfnamefont {P.}~\bibnamefont {Taranto}}, \bibinfo {author} {\bibfnamefont {F.}~\bibnamefont {Bakhshinezhad}}, \bibinfo {author} {\bibfnamefont {A.}~\bibnamefont {Bluhm}}, \bibinfo {author} {\bibfnamefont {R.}~\bibnamefont {Silva}}, \bibinfo {author} {\bibfnamefont {N.}~\bibnamefont {Friis}}, \bibinfo {author} {\bibfnamefont {M.~P.}\ \bibnamefont {Lock}}, \bibinfo {author} {\bibfnamefont {G.}~\bibnamefont {Vitagliano}}, \bibinfo {author} {\bibfnamefont {F.~C.}\ \bibnamefont {Binder}}, \bibinfo {author} {\bibfnamefont {T.}~\bibnamefont {Debarba}}, \bibinfo {author} {\bibfnamefont {E.}~\bibnamefont {Schwarzhans}}, \bibinfo {author} {\bibfnamefont {F.}~\bibnamefont {Clivaz}},\ and\ \bibinfo {author} {\bibfnamefont {M.}~\bibnamefont {Huber}},\ }\href {https://doi.org/10.1103/PRXQuantum.4.010332} {\bibfield  {journal} {\bibinfo  {journal} {PRX Quantum}\ }\textbf {\bibinfo {volume} {4}},\ \bibinfo {pages} {010332} (\bibinfo {year} {2023})}\BibitemShut {NoStop}%
\bibitem [{\citenamefont {Heisenberg}(1949)}]{Heisenberg1949}%
  \BibitemOpen
  \bibfield  {author} {\bibinfo {author} {\bibfnamefont {W.}~\bibnamefont {Heisenberg}},\ }\href {https://research.ebsco.com/linkprocessor/plink?id=718270c1-fd4b-336a-ba2c-bef43c4d70e6} {\emph {\bibinfo {title} {The physical principles of the quantum theory / by Werner Heisenberg ; transl. by Carl Eckart and Frank C. Hoyt.}}}\ (\bibinfo  {publisher} {Dover},\ \bibinfo {year} {1949})\BibitemShut {NoStop}%
\bibitem [{\citenamefont {Atmanspacher}(1997)}]{Atmanspacher01091997}%
  \BibitemOpen
  \bibfield  {author} {\bibinfo {author} {\bibfnamefont {H.}~\bibnamefont {Atmanspacher}},\ }\href {https://doi.org/10.1080/02604027.1997.9972639} {\bibfield  {journal} {\bibinfo  {journal} {World Futures}\ }\textbf {\bibinfo {volume} {49}},\ \bibinfo {pages} {333} (\bibinfo {year} {1997})},\ \Eprint {https://arxiv.org/abs/https://doi.org/10.1080/02604027.1997.9972639} {https://doi.org/10.1080/02604027.1997.9972639} \BibitemShut {NoStop}%
\bibitem [{\citenamefont {Allahverdyan}\ \emph {et~al.}(2004)\citenamefont {Allahverdyan}, \citenamefont {Balian},\ and\ \citenamefont {Nieuwenhuizen}}]{Allahverdyan2004MaximalSystems}%
  \BibitemOpen
  \bibfield  {author} {\bibinfo {author} {\bibfnamefont {A.~E.}\ \bibnamefont {Allahverdyan}}, \bibinfo {author} {\bibfnamefont {R.}~\bibnamefont {Balian}},\ and\ \bibinfo {author} {\bibfnamefont {T.~M.}\ \bibnamefont {Nieuwenhuizen}},\ }\href {https://doi.org/10.1209/epl/i2004-10101-2} {\bibfield  {journal} {\bibinfo  {journal} {Europhysics Letters}\ }\textbf {\bibinfo {volume} {67}},\ \bibinfo {pages} {565} (\bibinfo {year} {2004})}\BibitemShut {NoStop}%
\bibitem [{\citenamefont {Francica}\ \emph {et~al.}(2017)\citenamefont {Francica}, \citenamefont {Goold}, \citenamefont {Plastina},\ and\ \citenamefont {Paternostro}}]{Francica2017DaemonicCorrelations}%
  \BibitemOpen
  \bibfield  {author} {\bibinfo {author} {\bibfnamefont {G.}~\bibnamefont {Francica}}, \bibinfo {author} {\bibfnamefont {J.}~\bibnamefont {Goold}}, \bibinfo {author} {\bibfnamefont {F.}~\bibnamefont {Plastina}},\ and\ \bibinfo {author} {\bibfnamefont {M.}~\bibnamefont {Paternostro}},\ }\href {https://doi.org/10.1038/s41534-017-0012-8} {\bibfield  {journal} {\bibinfo  {journal} {npj Quantum Information}\ }\textbf {\bibinfo {volume} {3}},\ \bibinfo {pages} {1} (\bibinfo {year} {2017})}\BibitemShut {NoStop}%
\bibitem [{\citenamefont {Elouard}\ \emph {et~al.}(2017)\citenamefont {Elouard}, \citenamefont {Herrera-Mart\'{\i}}, \citenamefont {Huard},\ and\ \citenamefont {Auff\`eves}}]{Elouard2017}%
  \BibitemOpen
  \bibfield  {author} {\bibinfo {author} {\bibfnamefont {C.}~\bibnamefont {Elouard}}, \bibinfo {author} {\bibfnamefont {D.}~\bibnamefont {Herrera-Mart\'{\i}}}, \bibinfo {author} {\bibfnamefont {B.}~\bibnamefont {Huard}},\ and\ \bibinfo {author} {\bibfnamefont {A.}~\bibnamefont {Auff\`eves}},\ }\href {https://doi.org/10.1103/PhysRevLett.118.260603} {\bibfield  {journal} {\bibinfo  {journal} {Phys. Rev. Lett.}\ }\textbf {\bibinfo {volume} {118}},\ \bibinfo {pages} {260603} (\bibinfo {year} {2017})}\BibitemShut {NoStop}%
\bibitem [{\citenamefont {Landauer}(1961)}]{Landauer1961}%
  \BibitemOpen
  \bibfield  {author} {\bibinfo {author} {\bibfnamefont {R.}~\bibnamefont {Landauer}},\ }\href {https://doi.org/10.1147/rd.53.0183} {\bibfield  {journal} {\bibinfo  {journal} {IBM Journal of Research and Development}\ }\textbf {\bibinfo {volume} {5}},\ \bibinfo {pages} {183} (\bibinfo {year} {1961})}\BibitemShut {NoStop}%
\bibitem [{\citenamefont {Jordan}\ \emph {et~al.}(2020)\citenamefont {Jordan}, \citenamefont {Elouard},\ and\ \citenamefont {Aufféves}}]{Jordan2020}%
  \BibitemOpen
  \bibfield  {author} {\bibinfo {author} {\bibfnamefont {A.}~\bibnamefont {Jordan}}, \bibinfo {author} {\bibfnamefont {C.}~\bibnamefont {Elouard}},\ and\ \bibinfo {author} {\bibfnamefont {A.}~\bibnamefont {Aufféves}},\ }\href {https://doi.org/10.1007/s40509-019-00217-2} {\bibfield  {journal} {\bibinfo  {journal} {Quantum Studies: Mathematics and Foundations}\ }\textbf {\bibinfo {volume} {7}},\ \bibinfo {pages} {203} (\bibinfo {year} {2020})}\BibitemShut {NoStop}%
\bibitem [{\citenamefont {Mensky}(2000)}]{MenskyBook}%
  \BibitemOpen
  \bibfield  {author} {\bibinfo {author} {\bibfnamefont {M.~M.}\ \bibnamefont {Mensky}},\ }\href@noop {} {\emph {\bibinfo {title} {Quantum Mesurements and Decoherence}}}\ (\bibinfo  {publisher} {Kluwer Academic Publishers},\ \bibinfo {address} {Dordrecht, The Netherlands},\ \bibinfo {year} {2000})\BibitemShut {NoStop}%
\bibitem [{\citenamefont {Zeh}(1970)}]{Zeh1970}%
  \BibitemOpen
  \bibfield  {author} {\bibinfo {author} {\bibfnamefont {H.}~\bibnamefont {Zeh}},\ }\href {https://doi.org/10.1007/BF00708656} {\bibfield  {journal} {\bibinfo  {journal} {Found Phys}\ }\textbf {\bibinfo {volume} {1}},\ \bibinfo {pages} {69–76} (\bibinfo {year} {1970})}\BibitemShut {NoStop}%
\bibitem [{\citenamefont {Shettell}\ \emph {et~al.}(2023)\citenamefont {Shettell}, \citenamefont {Centrone},\ and\ \citenamefont {Garc{\'{i}}a-Pintos}}]{Shettell2023}%
  \BibitemOpen
  \bibfield  {author} {\bibinfo {author} {\bibfnamefont {N.}~\bibnamefont {Shettell}}, \bibinfo {author} {\bibfnamefont {F.}~\bibnamefont {Centrone}},\ and\ \bibinfo {author} {\bibfnamefont {L.~P.}\ \bibnamefont {Garc{\'{i}}a-Pintos}},\ }\href {https://doi.org/10.22331/q-2023-11-14-1182} {\bibfield  {journal} {\bibinfo  {journal} {{Quantum}}\ }\textbf {\bibinfo {volume} {7}},\ \bibinfo {pages} {1182} (\bibinfo {year} {2023})}\BibitemShut {NoStop}%
\bibitem [{\citenamefont {Nitzan}(2024)}]{NitzanBook}%
  \BibitemOpen
  \bibfield  {author} {\bibinfo {author} {\bibfnamefont {A.}~\bibnamefont {Nitzan}},\ }\href@noop {} {\emph {\bibinfo {title} {Chemical Dynamics in Condensed Phases}}},\ \bibinfo {edition} {2nd}\ ed.\ (\bibinfo  {publisher} {Oxford University Press, Inc.},\ \bibinfo {address} {New York},\ \bibinfo {year} {2024})\BibitemShut {NoStop}%
\bibitem [{\citenamefont {Mallet}\ \emph {et~al.}(2009)\citenamefont {Mallet}, \citenamefont {Ong}, \citenamefont {Palacios-Laloy}, \citenamefont {Nguyen}, \citenamefont {Bertet}, \citenamefont {Vion},\ and\ \citenamefont {Esteve}}]{Mallet2009}%
  \BibitemOpen
  \bibfield  {author} {\bibinfo {author} {\bibfnamefont {F.}~\bibnamefont {Mallet}}, \bibinfo {author} {\bibfnamefont {F.~R.}\ \bibnamefont {Ong}}, \bibinfo {author} {\bibfnamefont {A.}~\bibnamefont {Palacios-Laloy}}, \bibinfo {author} {\bibfnamefont {F.}~\bibnamefont {Nguyen}}, \bibinfo {author} {\bibfnamefont {P.}~\bibnamefont {Bertet}}, \bibinfo {author} {\bibfnamefont {D.}~\bibnamefont {Vion}},\ and\ \bibinfo {author} {\bibfnamefont {D.}~\bibnamefont {Esteve}},\ }\bibfield  {journal} {\bibinfo  {journal} {Nature Physics}\ }\textbf {\bibinfo {volume} {5}},\ \href {https://doi.org/10.1038/NPHYS1400} {10.1038/NPHYS1400} (\bibinfo {year} {2009})\BibitemShut {NoStop}%
\bibitem [{\citenamefont {Walter}\ \emph {et~al.}(2017)\citenamefont {Walter}, \citenamefont {Kurpiers}, \citenamefont {Gasparinetti}, \citenamefont {Magnard}, \citenamefont {Poto\ifmmode~\check{c}\else \v{c}\fi{}nik}, \citenamefont {Salath\'e}, \citenamefont {Pechal}, \citenamefont {Mondal}, \citenamefont {Oppliger}, \citenamefont {Eichler},\ and\ \citenamefont {Wallraff}}]{Walter2017}%
  \BibitemOpen
  \bibfield  {author} {\bibinfo {author} {\bibfnamefont {T.}~\bibnamefont {Walter}}, \bibinfo {author} {\bibfnamefont {P.}~\bibnamefont {Kurpiers}}, \bibinfo {author} {\bibfnamefont {S.}~\bibnamefont {Gasparinetti}}, \bibinfo {author} {\bibfnamefont {P.}~\bibnamefont {Magnard}}, \bibinfo {author} {\bibfnamefont {A.}~\bibnamefont {Poto\ifmmode~\check{c}\else \v{c}\fi{}nik}}, \bibinfo {author} {\bibfnamefont {Y.}~\bibnamefont {Salath\'e}}, \bibinfo {author} {\bibfnamefont {M.}~\bibnamefont {Pechal}}, \bibinfo {author} {\bibfnamefont {M.}~\bibnamefont {Mondal}}, \bibinfo {author} {\bibfnamefont {M.}~\bibnamefont {Oppliger}}, \bibinfo {author} {\bibfnamefont {C.}~\bibnamefont {Eichler}},\ and\ \bibinfo {author} {\bibfnamefont {A.}~\bibnamefont {Wallraff}},\ }\href {https://doi.org/10.1103/PhysRevApplied.7.054020} {\bibfield  {journal} {\bibinfo  {journal} {Phys. Rev. Appl.}\ }\textbf {\bibinfo {volume} {7}},\ \bibinfo {pages} {054020} (\bibinfo {year} {2017})}\BibitemShut {NoStop}%
\bibitem [{\citenamefont {Pirkkalainen}\ \emph {et~al.}(2013)\citenamefont {Pirkkalainen}, \citenamefont {Cho}, \citenamefont {Li}, \citenamefont {Paraoanu}, \citenamefont {Hakonen},\ and\ \citenamefont {Sillanpää}}]{Pirkkalainen2013}%
  \BibitemOpen
  \bibfield  {author} {\bibinfo {author} {\bibfnamefont {J.-M.}\ \bibnamefont {Pirkkalainen}}, \bibinfo {author} {\bibfnamefont {S.~U.}\ \bibnamefont {Cho}}, \bibinfo {author} {\bibfnamefont {J.}~\bibnamefont {Li}}, \bibinfo {author} {\bibfnamefont {G.}~\bibnamefont {Paraoanu}}, \bibinfo {author} {\bibfnamefont {P.~J.}\ \bibnamefont {Hakonen}},\ and\ \bibinfo {author} {\bibfnamefont {M.~A.}\ \bibnamefont {Sillanpää}},\ }\bibfield  {journal} {\bibinfo  {journal} {Nature}\ }\textbf {\bibinfo {volume} {494}},\ \href {https://doi.org/10.1038/nature11821} {10.1038/nature11821} (\bibinfo {year} {2013})\BibitemShut {NoStop}%
\bibitem [{\citenamefont {Carnot}(1897)}]{carnot1824reflections}%
  \BibitemOpen
  \bibfield  {author} {\bibinfo {author} {\bibfnamefont {S.}~\bibnamefont {Carnot}},\ }\href@noop {} {\emph {\bibinfo {title} {Reflections on the Motive Power of Fire: And on Machines Fitted to Develop That Power}}},\ \bibinfo {edition} {2nd}\ ed.,\ edited by\ \bibinfo {editor} {\bibfnamefont {R.~H.}\ \bibnamefont {Thurston}}\ (\bibinfo  {publisher} {John Wiley \& Sons},\ \bibinfo {address} {New York},\ \bibinfo {year} {1897})\ \bibinfo {note} {originally published in 1824 in French}\BibitemShut {NoStop}%
\bibitem [{\citenamefont {Callen}(1985)}]{Callen1985thermodynamics}%
  \BibitemOpen
  \bibfield  {author} {\bibinfo {author} {\bibfnamefont {H.~B.}\ \bibnamefont {Callen}},\ }\href@noop {} {\emph {\bibinfo {title} {Thermodynamics and an Introduction to Thermostatistics}}},\ \bibinfo {edition} {2nd}\ ed.\ (\bibinfo  {publisher} {John Wiley \& Sons},\ \bibinfo {address} {New York},\ \bibinfo {year} {1985})\BibitemShut {NoStop}%
\bibitem [{\citenamefont {Curzon}\ and\ \citenamefont {Ahlborn}(1975)}]{Curzon1975}%
  \BibitemOpen
  \bibfield  {author} {\bibinfo {author} {\bibfnamefont {F.~L.}\ \bibnamefont {Curzon}}\ and\ \bibinfo {author} {\bibfnamefont {B.}~\bibnamefont {Ahlborn}},\ }\href {https://doi.org/10.1119/1.10023} {\bibfield  {journal} {\bibinfo  {journal} {American Journal of Physics}\ }\textbf {\bibinfo {volume} {43}},\ \bibinfo {pages} {22} (\bibinfo {year} {1975})}\BibitemShut {NoStop}%
\bibitem [{\citenamefont {Erdman}\ \emph {et~al.}(2023)\citenamefont {Erdman}, \citenamefont {Rolandi}, \citenamefont {Abiuso}, \citenamefont {Perarnau-Llobet},\ and\ \citenamefont {No\'e}}]{Erdman2017}%
  \BibitemOpen
  \bibfield  {author} {\bibinfo {author} {\bibfnamefont {P.~A.}\ \bibnamefont {Erdman}}, \bibinfo {author} {\bibfnamefont {A.}~\bibnamefont {Rolandi}}, \bibinfo {author} {\bibfnamefont {P.}~\bibnamefont {Abiuso}}, \bibinfo {author} {\bibfnamefont {M.}~\bibnamefont {Perarnau-Llobet}},\ and\ \bibinfo {author} {\bibfnamefont {F.}~\bibnamefont {No\'e}},\ }\href {https://doi.org/10.1103/PhysRevResearch.5.L022017} {\bibfield  {journal} {\bibinfo  {journal} {Phys. Rev. Res.}\ }\textbf {\bibinfo {volume} {5}},\ \bibinfo {pages} {L022017} (\bibinfo {year} {2023})}\BibitemShut {NoStop}%
\bibitem [{\citenamefont {Deb}(2008)}]{Deb2008}%
  \BibitemOpen
  \bibfield  {author} {\bibinfo {author} {\bibfnamefont {K.}~\bibnamefont {Deb}},\ }\href@noop {} {\emph {\bibinfo {title} {{Multi-Objective} Optimization using Evolutionary Algorithms}}},\ Wiley Interscience Series in Systems and Optimization\ (\bibinfo  {publisher} {Wiley-Blackwell},\ \bibinfo {address} {Hoboken, NJ},\ \bibinfo {year} {2008})\BibitemShut {NoStop}%
\bibitem [{\citenamefont {Cahill}\ and\ \citenamefont {Glauber}(1969)}]{Glauber1969}%
  \BibitemOpen
  \bibfield  {author} {\bibinfo {author} {\bibfnamefont {K.~E.}\ \bibnamefont {Cahill}}\ and\ \bibinfo {author} {\bibfnamefont {R.~J.}\ \bibnamefont {Glauber}},\ }\href@noop {} {\bibfield  {journal} {\bibinfo  {journal} {Physical Review}\ }\textbf {\bibinfo {volume} {177}} (\bibinfo {year} {1969})}\BibitemShut {NoStop}%
\bibitem [{\citenamefont {Misra}\ and\ \citenamefont {Sudarshan}(1977)}]{Misra1977}%
  \BibitemOpen
  \bibfield  {author} {\bibinfo {author} {\bibfnamefont {B.}~\bibnamefont {Misra}}\ and\ \bibinfo {author} {\bibfnamefont {E.~C.}\ \bibnamefont {Sudarshan}},\ }\href {https://doi.org/10.1063/1.523304} {\bibfield  {journal} {\bibinfo  {journal} {Journal of Mathematical Physics}\ }\textbf {\bibinfo {volume} {18}},\ \bibinfo {pages} {756} (\bibinfo {year} {1977})}\BibitemShut {NoStop}%
\bibitem [{\citenamefont {Facchi}\ and\ \citenamefont {Pascazio}(2008)}]{Facchi2008}%
  \BibitemOpen
  \bibfield  {author} {\bibinfo {author} {\bibfnamefont {P.}~\bibnamefont {Facchi}}\ and\ \bibinfo {author} {\bibfnamefont {S.}~\bibnamefont {Pascazio}},\ }\bibfield  {journal} {\bibinfo  {journal} {Journal of Physics A: Mathematical and Theoretical}\ }\textbf {\bibinfo {volume} {41}},\ \href {https://doi.org/10.1088/1751-8113/41/49/493001} {10.1088/1751-8113/41/49/493001} (\bibinfo {year} {2008})\BibitemShut {NoStop}%
\bibitem [{\citenamefont {Magnus}(1954)}]{Magnus1954}%
  \BibitemOpen
  \bibfield  {author} {\bibinfo {author} {\bibfnamefont {W.}~\bibnamefont {Magnus}},\ }\href {https://doi.org/https://doi.org/10.1002/cpa.3160070404} {\bibfield  {journal} {\bibinfo  {journal} {Communications on Pure and Applied Mathematics}\ }\textbf {\bibinfo {volume} {7}},\ \bibinfo {pages} {649} (\bibinfo {year} {1954})}\BibitemShut {NoStop}%
\end{thebibliography}%
